\documentclass[twocolumn, floatfix]{aastex63}
\usepackage[utf8]{inputenc}
\usepackage[T1]{fontenc}
\usepackage[english]{babel}
\usepackage{graphicx}
\usepackage{color}

\usepackage{multirow, booktabs}  % performance tables
\usepackage[allowspaces]{mathtools}           % paired delimiters, multlined
\usepackage{tensor}              % properly stacked tensor indices
\usepackage{bm}   % \bm for bold Greek symbols in math mode
\DeclareFontFamily{U}{mathb}{\hyphenchar\font45}
\DeclareFontShape{U}{mathb}{m}{n}{
      <5> <6> <7> <8> <9> <10> gen * mathb
      <10.95> mathb10 <12> <14.4> <17.28> <20.74> <24.88> mathb12
      }{}
\DeclareSymbolFont{mathb}{U}{mathb}{m}{n}
\DeclareMathSymbol{\blackdiamond}{2}{mathb}{"0C}

\usepackage{makecell}   %\thead, \makecell, \multirowthead, \multicolumnthead commands
\DeclareMathOperator*{\minmod}{minmod}

\DeclarePairedDelimiter{\paren}{\lparen}{\rparen}
\DeclarePairedDelimiter{\abs}{\lvert}{\rvert}

\newcommand{\pgas}{p_\mathrm{g}}
\newcommand{\pmag}{p_\mathrm{m}}
\newcommand{\rg}{r_\mathrm{g}}
\newcommand{\app}{\texttt{Athena++}}
\newcommand{\athena}{\texttt{Athena}}

\usepackage{varioref}
\usepackage{hyperref}
\shorttitle{Athena++}
\shortauthors{Stone et al.}
\begin{document}

\title{The Athena++ Adaptive Mesh Refinement Framework: Design and Magnetohydrodynamic Solvers}

\author{James M. Stone}
\altaffiliation{current address: School of Natural Sciences, Institute for Advanced Study, Princeton, NJ 08544, USA}
\affiliation{Department of Astrophysical Sciences,
Princeton University, Princeton, NJ 08544, USA}
\affiliation{Program in Applied and Computational Mathematics,
Princeton University, Princeton, NJ 08544, USA}

\author{Kengo Tomida}
\altaffiliation{current address: Astronomical Institute, Tohoku University, Sendai, Miyagi 980-8578, Japan}
\affiliation{Department of Earth and Space Science,
Osaka University, Toyonaka, Osaka, 560-0043, Japan}

\author{Christopher J.\ White}
\affiliation{Kavli Institute for Theoretical Physics,
University of California Santa Barbara, Santa Barbara, CA 93107, USA}

\author[0000-0002-3501-482X]{Kyle G. Felker}
\altaffiliation{current address: Argonne National Laboratory, Lemont, IL 60439, USA}
\affiliation{Program in Applied and Computational Mathematics,
Princeton University, Princeton, NJ 08544, USA}

\correspondingauthor{James M. Stone}
\email{jmstone@ias.edu}

\begin{abstract}
The design and implementation of a new framework for adaptive mesh refinement (AMR) calculations is described.  It is intended primarily for applications in astrophysical fluid dynamics, but its flexible and modular design enables its use for a wide variety of physics.  The framework works with both uniform and nonuniform grids in Cartesian and curvilinear coordinate systems.  It adopts a dynamic execution model based on a simple design called a ``task list'' that improves parallel performance by overlapping communication and computation, simplifies the inclusion of a diverse range of physics, and even enables multiphysics models involving different physics in different regions of the calculation.  We describe physics modules implemented in this framework for both non-relativistic and relativistic magnetohydrodynamics (MHD).  These modules adopt mature and robust algorithms originally developed for the Athena MHD code and incorporate new extensions: support for curvilinear coordinates, higher-order time integrators, more realistic physics such as a general equation of state, and diffusion terms that can be integrated with super-time-stepping algorithms.  The modules show excellent performance and scaling, with well over 80\% parallel efficiency on over half a million threads.  The source code has been made publicly available.
\end{abstract}


\section{Introduction}

Computational methods are now firmly established as essential tools for studying many problems in astrophysical fluid dynamics. A number of publicly-available codes that implement a range of algorithms and features are widely used for such problems.  Examples of widely-used (based on, e.g., citations) grid-based codes include \texttt{ZEUS} \citep{SN92a, SN92b, SN92c, Hayes2006}, \texttt{ART} \citep{Kravtsov1997}, \texttt{FLASH} \citep{Fryxell2000}, \texttt{RAMSES} \citep{Teyssier2002}, \texttt{HARM} \citep{Gammie2003},
\texttt{PLUTO} \citep{Mignone2007a, Mignone2012}, \athena{} \citep{Stone+2008}, and \texttt{Enzo} \citep{Bryan2014}, among others.

There is a common trend amongst modern codes for astrophysical fluid dynamics towards increasingly complexity.  This trend is driven by a number of factors.  Firstly, realistic models of many astrophysical systems require the inclusion of additional physics, such as radiation transfer, self-gravity, chemical or nuclear reaction networks, and (for relativistic flows) dynamical spacetimes.  Secondly, in order to resolve widely disparate length- and time-scales, it is now common for grid based methods to adopt one of several different adaptive mesh refinement (AMR) strategies.  In addition, such codes often implement a variety of algorithmic options, such as different coordinate systems, Riemann solvers (in the case of Godunov schemes), and spatial and temporal approximations of varying formal orders of accuracy.  Supporting all possible combinations of physics and algorithmic options on an AMR mesh is challenging.  Finally, modern high-performance computing systems are becoming increasingly heterogeneous.  Developing portable code that performs well on the wide range of available architectures presents an additional challenge.

The \athena{} code \citep[hereafter S08]{Stone+2008}, written in C, is a prototypical illustration of this evolution towards increasing complexity. The numerical algorithms, based on the extension of unsplit finite-volume methods to magnetohydrodynamics (MHD) using upwind constrained transport (CT), were initially described in \citet{gs05, gs08}.  Subsequently, the code was augmented with different time integrators \citep{sg09}, the shearing-box approximation \citep{SG2010}, cylindrical coordinates \citep{Skinner2010}, special relativity \citep{Beckwith2011}, particles \citep{BaiStone2010}, sink particles \citep{Hao2013}, a total energy conserving formalism for self-gravity \citep{JiangGrav2013}, and radiation transport \citep{Skinner2013,Davis2012,Jiang2012,Jiang2014}, among many other features. Maintaining and updating \athena{} as progressively more physics and algorithms are implemented has become increasingly untenable. Moreover, the AMR strategy adopted in the original code, based on overlapping patches \citep{BergerOliger,BergerColella}, was found not to perform well on modern highly-parallel architectures.

The need to address these issues has led to a complete redesign and rewrite of the code from scratch.  The first and most important aspect of this redesign has been the abstraction of the mesh from the physics modules solved on it.  In the new design, the mesh exists as an independent, abstract framework on which various discretizations of the dependent variables (such as cell-entered volume averages, face-centered area averages, or vertex- or cell-centered pointwise values) are constructed and stored.  Methods for AMR, various boundary conditions, and distributed memory parallelization using domain decomposition are then implemented for these discrete variables, without specific references to any particular physics.  This greatly simplifies the extension of the code to both new coordinates and new physics that are immediately compatible with AMR in any geometry.  Moreover, 
isolating the mesh infrastructure from the physics allows each to be developed independently: for example a performance portable version of the AMR infrastructure based on the \texttt{Kokkos} library \citep{Edwards14} which can be run on heterogeneous architectures (including GPUs) is now under development.

A second important aspect of this redesign has been the adoption of a {\em block-based} AMR design (e.g. \citet{Stout97}), as opposed to the {\em patch-based} AMR in the style of \citet{BergerOliger} implemented in the original version of \athena{} and also used in codes like \texttt{PLUTO} and \texttt{Enzo}.  There are a number of compelling reasons that motivate the adoption of block-based AMR. In patch-based AMR, refined regions are covered by multiple levels of meshes.  Quantities derived from the conserved variables (such as temperature) can therefore possess different values on different levels.  In turn, this can lead to different dynamics on separate levels if, for example, there are source terms such as cooling or chemical or nuclear reaction networks that depend on temperature.  By carefully designing our block-based AMR so that each position in the domain is covered with one and only one mesh level, this complication is eliminated.  Moreover, when patch-based AMR is parallelized using domain decomposition, the overlap between Message Passing Interface (MPI) domains on different levels can become complex.  With our implementation of block-based AMR, different levels communicate only through the boundaries.  This simplifies the implementation, and greatly improves the performance and scaling on parallel architectures.  Perhaps the best known code that uses a block-based AMR strategy is \texttt{FLASH} \citep{Fryxell2000}, which is based on the \texttt{PARAMESH} AMR library \citep{PARAMESH}.  However, rather than using pre-existing libraries we have instead written our own AMR framework in order to support face-centered variables (as required by our implementation of MHD), reduce library dependencies, and improve performance by sacrificing generality.

Finally, a third important aspect of this redesign is the use of dynamic scheduling. Rather than hard-code the order of execution of steps in the numerical algorithms (including MPI send and receives), instead these steps are assembled into lists of encapsulated {\em tasks}. Individual tasks can be executed in any order, provided that the tasks upon which they depend are complete.  The ability to dynamically adapt the order of execution of tasks allows the overlap of parts of the computation with MPI communication (which in turn can improve parallel scaling on very large number of processors).  Moreover, the design enables a wide range of calculations containing different subsets of physics, since it is simple to change the composition of the task list. Even more powerfully, calculations in which different physics is simulated in disjoint regions are enabled simply by constructing separate task lists for each region.  For example, this organization facilitates the straightforward inclusion of a particle-in-cell (PIC) code for modeling the collisionless dynamics of the corona which is formed in the upper regions of an MHD simulation of an accretion disk \citep[e.g.][]{millerstone}.
A variety of sophisticated libraries, such as \texttt{Legion}\footnote{https://legion.stanford.edu/overview/index.html} or \texttt{CHARM++}\footnote{http://charm.cs.illinois.edu/software}, are available which implement dynamic execution using {\em task-based parallelism} (in which a master processes schedules data and tasks to available processors) among many other useful features. Since we only require the ability to schedule tasks dynamically, and in order to reduce dependencies on 
external libraries, we have implemented our own design, requiring a few thousand lines of special-purpose code.

To take advantage of language extensions that improve modularity and organization, we have adopted the C++ language for this framework; therefore we refer to the resulting new code as \app.  This paper provides an introduction to the AMR framework in \app.  We focus on the new features of this design, especially the implementation of block-based AMR with both cell- and face-centered variables (as required for MHD), extension of the design to various coordinate systems, and dynamic execution using task lists.  These features constitute the basic building blocks of the framework, upon which any physics solver can be implemented.

In the interest of providing concrete examples of physics modules within the framework, we also describe the implementation of algorithms for both non-relativistic and relativistic MHD in this framework, based on the methods used in \athena.  Since these algorithms are described in detail in previous papers \citep{gs05,gs08,Stone+2008,Beckwith2011,WhiteStone2016}, we confine the scope of our description to only novel features related to the new framework design.  There are a variety of other physics modules in development within the \app{} framework, and these will be described in future publications.

This paper is organized as follows.  In the following section, we describe the design, implementation, and major features of the AMR framework.  In \S3, we describe the implementation of a solver for non-relativistic MHD in this framework, including tests.  In \S4 we describe a relativistic MHD solver and tests.  Throughout \S5, we discuss other new physics modules under development, and in \S6 we summarize and conclude.

\section{Framework Design}

As mentioned above, the most important design feature in \app{} is the abstraction of the mesh from the physics. In this section, we describe the code framework that achieves this design.

\subsection{The Mesh}

The computational domain in an \app{} calculation is a logically rectangular region whose overall properties are stored within a C++ class called the \texttt{Mesh}.  The domain is further divided into a regular array of sub-volumes whose properties are stored in another class called the \texttt{MeshBlock}.  The latter stores discrete values for the dependent variables in cells as $N$-dimensional arrays, as well as one-dimensional arrays of coordinate positions along each direction.  The number of cells stored in each MeshBlock is arbitrary but it must be identical for all MeshBlocks.  Similarly, the decomposition of the Mesh into MeshBlocks is arbitrary.

In both uniform mesh and AMR calculations, the logical relationship between MeshBlocks is encoded in a tree data structure, either a binary-tree (in one spatial dimension), a quadtree (in two dimensions), or an oct-tree (in three dimensions).  With AMR, the use of a tree is crucial for encoding the relationship between parent and child MeshBlocks, and even with uniform grids it greatly simplifies finding neighboring blocks.  Moreover, it results in the natural assignment of MeshBlocks to processors using Z-ordering, which helps improve locality and speeds up communications.

\subsubsection{Uniform Grids}

For uniform grid calculations, MeshBlocks are used to parallelize the calculation using domain decomposition.  In this case, the total number of MeshBlocks used typically equals the total number of physical processors available (although this is not required).  For serial calculations on a uniform grid, only one MeshBlock is needed.

To construct the tree, the smallest value of $n$ such that $2^n$ exceeds the largest number of MeshBlocks in any dimension is determined.  Different levels $n$ in the tree are referred to as {\em logical levels}.  Thus, the tree is constructed beginning at logical level $0$ and continuing to logical level $n$, and then assign each MeshBlock to the appropriate leaves at level $n$.  Only in the case that the number of MeshBlocks in each dimension is equal and a power of two will every leaf in the tree be assigned a MeshBlock.  In general, there will be both leaves and nodes that are empty.

\begin{figure}[htb!]
\centering
\includegraphics[width=\columnwidth]{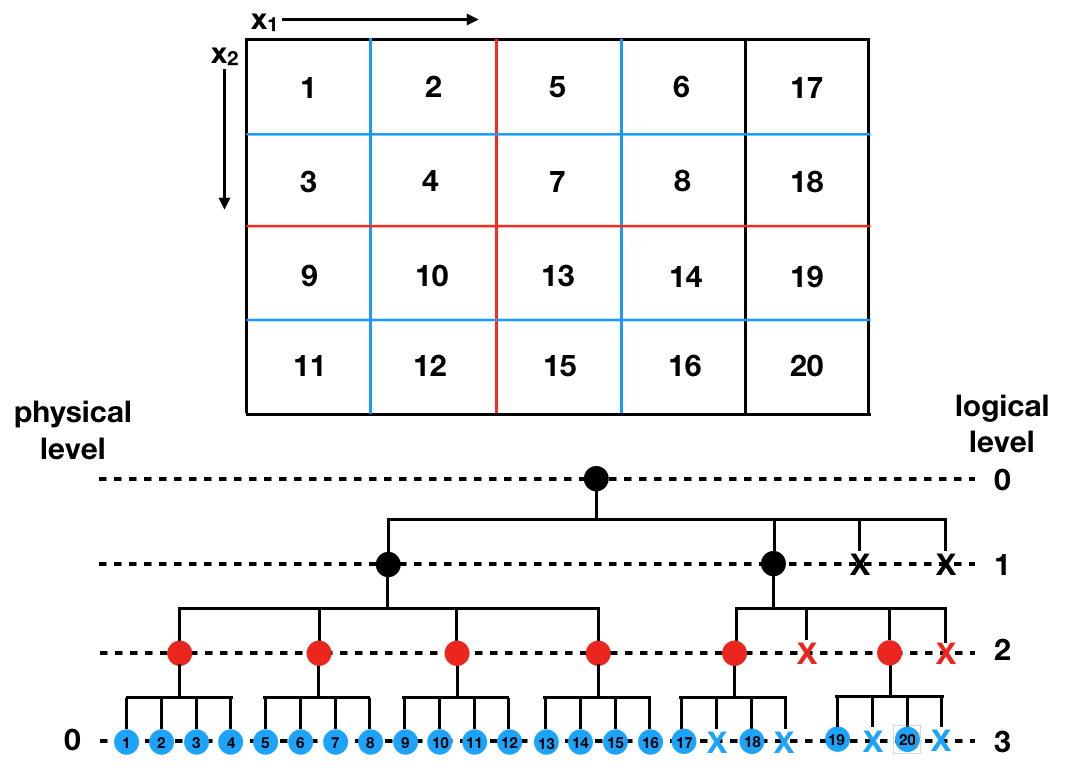}
\caption{Labeling of MeshBlocks (top) and their organization into a quadtree (bottom) for an example uniform grid calculation in two dimensions.  \label{fig:mesh-tree}}
\end{figure}

To illustrate the process, figure~\ref{fig:mesh-tree} diagrams the organization of a uniform grid into MeshBlocks and a quadtree for the specific example of a two-dimensional calculation consisting of $5 \times 4$ MeshBlocks.  In this case the MeshBlocks are stored at logical level 3, and there are empty nodes and leaves at every logical level (except the root, $n=0$).  Note that the {\em physical level} of the grid (which corresponds to the refinement level in AMR) does not equal the logical level, and that the labels of the MeshBlocks are automatically organized into a Z-ordering across the domain (this can be seen by connecting the labels shown in the top panel with a line).  This ordering helps improve locality of communications.

As in the \athena{} code, boundary conditions for the dependent variables stored on each MeshBlock are applied through the use of ghost zones.  The ghost region consists of an extra $N_G$ row of cells added to each array at each boundary.  
Any number of ghost cells are allowed, however for second-order spatial integration algorithms on a uniform mesh for MHD, $N_G=2$, whereas for spatial orders up to four $N_G=3$ for hydrodynamics and $N_G=4$ for MHD.  With AMR, $N_G$ must be an even integer because the restriction step (see section~\ref{subsubsec:restrict-prolong}) reduces the number of cells by a factor of two.
When a calculation contains multiple MeshBlocks, data in the ghost zones may overlap with active cells in adjacent MeshBlocks.  In this case the data must be swapped between MeshBlocks, either via MPI calls if the MeshBlocks are on different processors, or via calls to \texttt{memcpy()} otherwise.  Because \app{} supports cell-, face-, and edge-centered variables, the communication of data between MeshBlocks can become complicated.

\begin{figure}[htb!]
\centering
\includegraphics[width=\columnwidth]{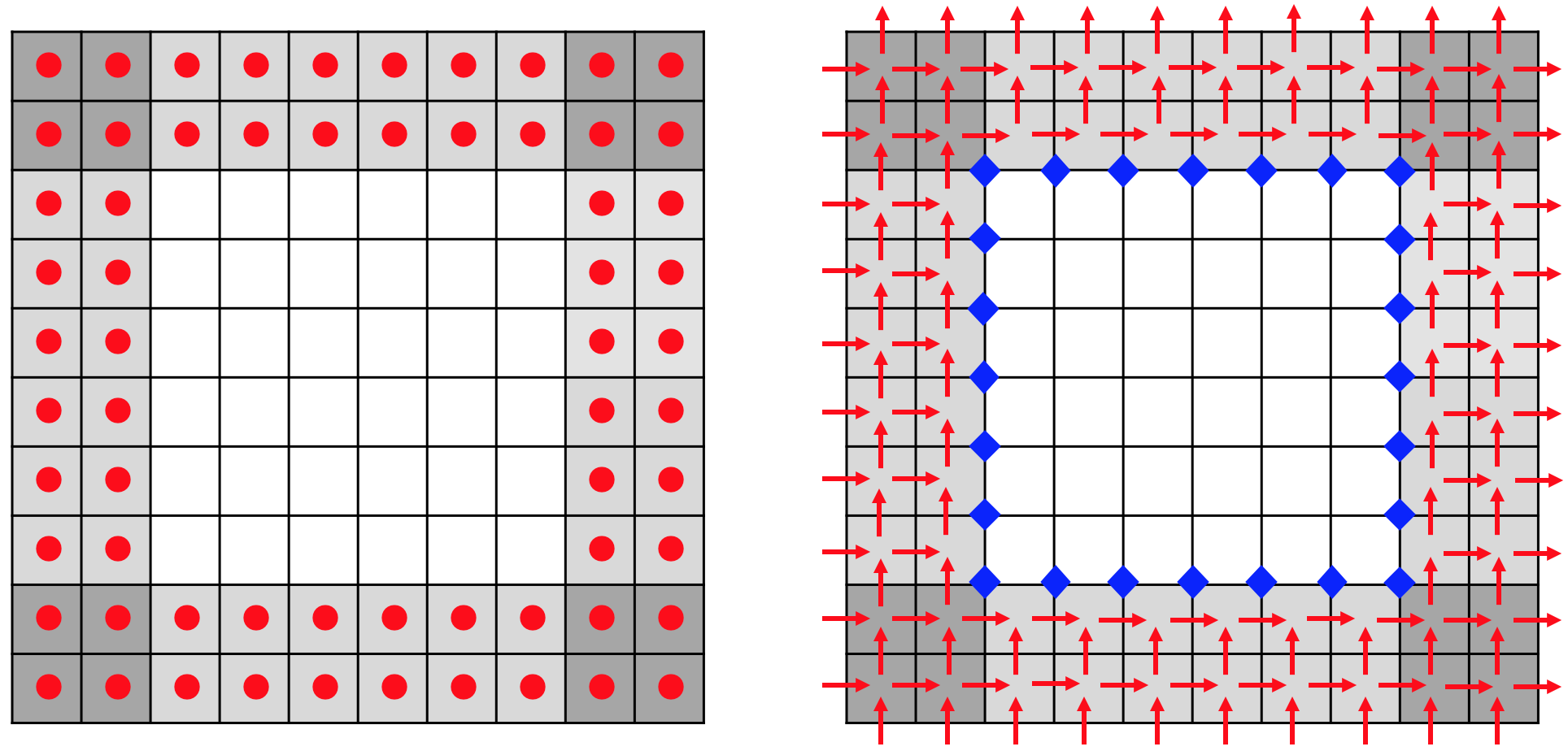}
\caption{{\em Left panel:}\ example of cell-centered data (red dots) that must be communicated to neighboring MeshBlocks in two dimensions.  The shaded cells are ghost cells that overlap with active cells in the eight neighbors.  {\em Right panel:}\ face-centered magnetic fields (red arrows) and edge-centered EMFs (blue dots) that are communicated in this example. \label{fig:uniform-grid-MPI}}
\end{figure}

Figure~\ref{fig:uniform-grid-MPI} diagrams what data must be received from neighbor MeshBlocks for the specific example of a two-dimensional calculation with MeshBlocks of size $6 \times 6$ with two ghost zones.  White cells are ``active cells,'' which are updated in each MeshBlock, while light gray cells are ``face ghost cells'' and dark gray cells are ``edge ghost cells,'' which are filled by data from neighboring MeshBlocks that abut the faces and edges respectively. In 3D, the algorithm also considers ``corner ghost cells'' corresponding to neighboring MeshBlocks that abut the corners.  Note that the ghost cells overlap with as many as eight neighbors in two dimensions (four edges and four corners), and up to 26 neighbors in three dimensions (six faces, twelve edges, and eight corners).  In \app{}, independent communication requests and message buffers are posted for each neighbor.  This differs from the implementation in \texttt{Athena}, in which entire edges in each dimension were communicated sequentially.  The sequential approach introduces dependencies (in the third dimension, ghost cells cannot be communicated until those in the second are finished, which in turn requires those in the first to be finished).  We have found such dependencies can reduce the parallel efficiency on very large numbers of processors.  On the other hand, the large number of communications required per MeshBlock with \app{} can tax some network architectures, thus an additional communication layer that pools messages between MeshBlocks may be a useful feature for future development, at least for some machines.

The left-hand panel in figure~\ref{fig:uniform-grid-MPI} shows the cell-centered data that must be communicated for $N_G=2$, while the right-hand panel shows the same for face-centered and edge-centered data
associated with the constrained transport (CT) algorithm for MHD used in \app{}.  This algorithm requires storing area-averages of the magnetic field on cell faces, and computing line-averages of the electric field (EMF) on cell edges \citep[see][figure 1]{Stone+2008}.
Note that adjacent MeshBlocks share the same face-centered vectors on their surfaces.  In order to enforce the divergence-free constraint, both MeshBlocks must store and evolve the magnetic field components on their surfaces.  However, we have found that in some pathological cases, especially in curvilinear coordinates, round-off error can cause the values for the same magnetic field component stored on different MeshBlocks to diverge in time.  To prevent this, the EMFs computed at cell corners (edges) in two dimensions (three dimensions) are swapped between MeshBlocks, and the average of the values, computed independently on each MeshBlock, is used to update the magnetic fields on the surface.  This adds an additional communication, but ensures consistency (within round-off error) between the field on adjacent MeshBlocks.

While this discussion is motivated by the data associated with the MHD solvers in \app{}, in fact the implementation of communication of ghost cells is highly modular and not specialized to any particular solver.  Communication functions for arbitrary numbers of cell-centered, face-centered, and edge-centered data are provided in separate classes, derived from an abstract base class that implements generic MPI communication patterns.  In turn, these functions can be enrolled using the task list when necessary.

\subsubsection{Static and Adaptive Mesh Refinement}

In the \app{} implementation of AMR, in $n$ dimensions each MeshBlock is refined into $2^n$ finer MeshBlocks, and
the resulting MeshBlock structure is stored in binary-tree ($n=1$), quadtree ($n=2$), or oct-tree ($n=3$). As one cell on a given level corresponds to $2^n$ cells on the next refined level, the number of cells in a MeshBlock in each direction must be even.  In addition, $N_G$ must be even, and only refinement by a factor of two in each dimension simultaneously is allowed. A MeshBlock can contact neighboring MeshBlocks on the same level, one level coarser, or one level finer. Changes in resolution by more than one level at a boundary is not allowed, and this restriction affects which MeshBlocks are flagged for refinement (or derefinement) in addition to the refinement criteria.

\begin{figure}[htb!]
\centering
\includegraphics[width=\columnwidth]{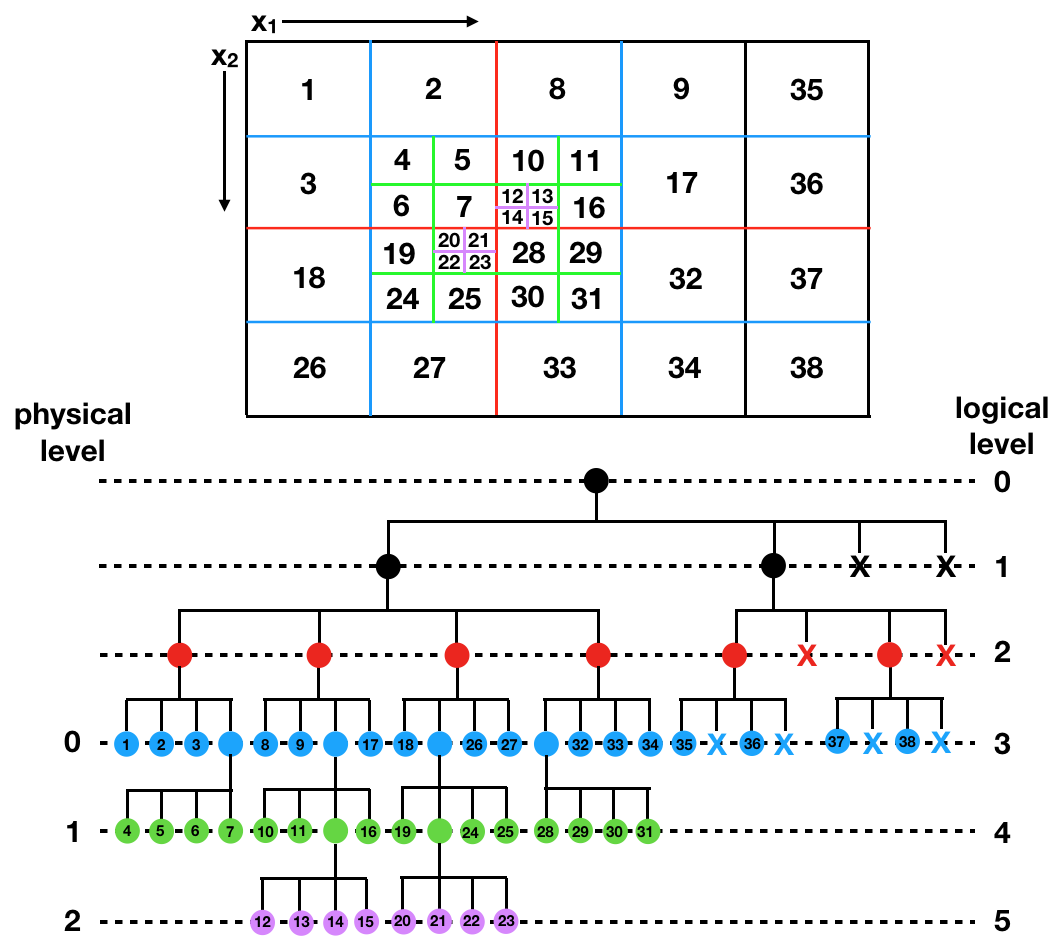}
\caption{Same as figure~\ref{fig:mesh-tree} with AMR. \label{fig:mesh-tree-amr}}
\end{figure}

A driving feature for the tree design of the MeshBlock structure in \app{} is AMR.  Figure~\ref{fig:mesh-tree-amr} shows how the 2D grid shown as an example in figure~\ref{fig:mesh-tree}
might be refined with AMR.  In the example, MeshBlocks 4, 7, 10 and 13 shown in figure~\ref{fig:mesh-tree} have been refined by up to two levels.  This requires inserting additional logical levels (corresponding to extra physical levels) at the appropriate leafs in the tree.  Moreover, the labeling of all subsequent MeshBlocks beyond the first refinement is modified.  The 2D quadtree design is crucial for managing the logical structure of the MeshBlocks, as well as keeping the Z-ordering of labels.  Note that in a parallel calculation, load balancing would be required (see section~\ref{subsubsec:load-balance} below).

In this oct-tree (in 3D) block-based AMR design, the flexibility of the refinement depends on the size of MeshBlocks.
If the root level is tiled with a large number of small MeshBlocks, then smaller volumes can be selected
for refinement, reducing the computational work required.  However, because each MeshBlock contains a fixed number of ghost zones, the fraction of ghost cells compared to active cells is larger for smaller MeshBlocks.  This surface-to-volume effect makes smaller MeshBlocks computationally less efficient. Thus, the best performance requires a careful choice of MeshBlock size in order to balance refinement flexibility (requiring smaller MeshBlocks) and computational efficiency (requiring larger MeshBlocks), see sections~\ref{sec:AMRStrongScaling} and \ref{sec:MeshBlockSize-Performance} for discussion. This is one possible disadvantage of tree block-based AMR. On the other hand, because each MeshBlock has the same logical shape in this design, it is easy to write optimized and flexible code that achieves high performance on modern parallel systems. This is one of the biggest advantages of the oct-tree block-based AMR design.

\subsubsection{Communication Between Different Levels\label{subsec:level_comm}}

The majority of the complexity with block-based AMR is associated with communications between MeshBlocks at different refinement levels.  With AMR, each MeshBlock must communicate with up to 12 neighbors in 2D ($4\times 2$ faces and 4 edges), and up to 56 neighbors in 3D ($6 \times 4$ faces, $12\times 2$ edges, and eight corners). When neighboring MeshBlocks are located on the coarser level, the data is first restricted and then communicated at the lower resolution.  This proceeds through ``coarse buffers'' that
contain copies of the cell-centered and face-centered variables restricted to half the resolution, so that each cell in the coarse buffer (including ghost zones) corresponds to $2^3$ cells (in 3D) in the MeshBlock.

%\kt{When mesh refinement is activated, ``coarse buffers'' for cell-centered variables and face-centered variables are allocated in each MeshBlock, which process communications to and from neighbor MeshBlocks on the coarser level. When a MeshBlock consists of $(N_x+2N_G)\times (N_y+2N_G)\times(N_z+2N_G)$ cells, its coarse buffer contains $(N_x/2+2N'_G)\times (N_y/2+2N'_G)\times(N_z/2+2N'_G)$ cells, where $N'_G=2$ when $N_G=2$ and $N'_G=4$ when $N_G=3$ or $4$. Each cell in the coarse buffer corresponds to $2\times 2\times 2$ cells in the MeshBlock.}

\begin{figure*}[htb!]
\centering
\includegraphics[width=0.8\textwidth]{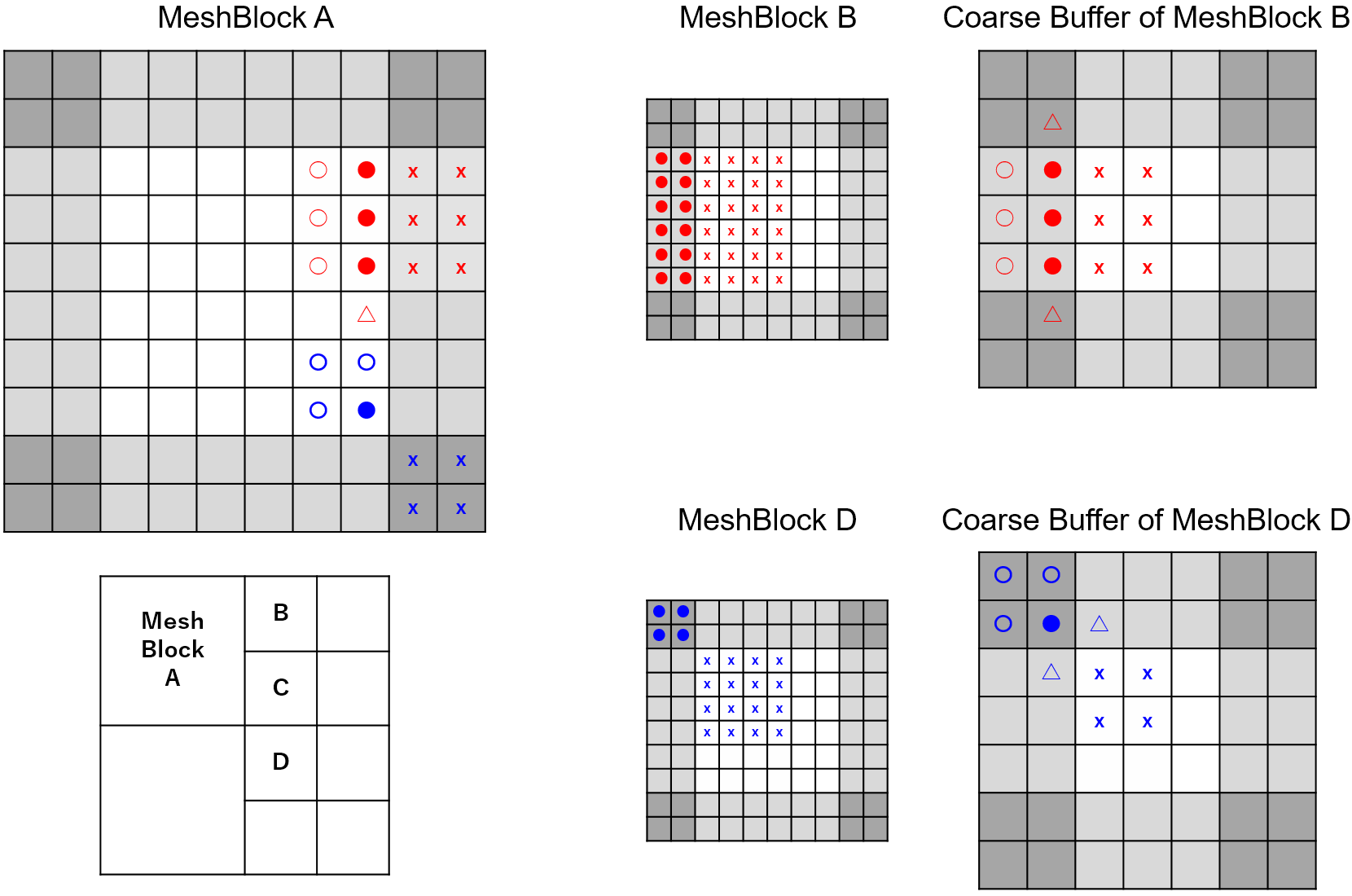}
\caption{Example of communication of cell-centered data between neighboring MeshBlocks at different refinement levels.  The lower left panel shows the configuration of the MeshBlocks in a two-dimensional mesh.  The upper panels show data communicated between MeshBlocks A and B (using red symbols), while the upper left and lower right panel show data communicated between MeshBlocks A and D (using blue symbols).  See the text for a description of the symbols.   \label{fig:amrcomm1}}
\end{figure*}

To illustrate how boundary communications between different levels proceeds, consider an example of a two-dimensional grid in which neighboring MeshBlocks at one face and one edge are at higher resolution.  In the discussion that follows, MeshBlock~A refers to the MeshBlock of interest, MeshBlocks~B and C are neighbors along one face (at higher resolution) and MeshBlock~D is the neighbor at the lower right edge (at higher resolution) (see figure~\ref{fig:amrcomm1}). For simplicity, suppose the MeshBlocks contain $6^2$ cells and two ghost zones, i.e. $N_G=2$. In the figure, red symbols indicate data points communicated between MeshBlocks~A and~B, while blue symbols indicate data communicated between MeshBlocks~A and~D.

From the perspective of MeshBlock~A, the communication procedure for cell-centered variables to and from finer MeshBlocks~B and~D proceeds as follows:
\begin{enumerate}
\item Send active cells overlapping the neighboring MeshBlocks (marked by $\circ$, $\bullet$, and $\triangle$).
\item Receive ghost cells from neighboring MeshBlocks (marked by $\times$).
\end{enumerate}
On MeshBlocks~B and~D, the communication of cell-centered variables to and from the coarser MeshBlock~A are more complicated:
\begin{enumerate}
\item Restrict the active cells overlapping MeshBlock~A (marked by $\times$) to the coarse buffer and send them.
\item Receive the coarse cells from MeshBlock~A (marked by $\circ$, $\bullet$, and $\triangle$) into the coarse buffer.
\item Wait until all the boundary communications (including both cell-centered and face-centered variables) are completed.
\item Fill in the cells adjacent to the cells to be prolongated (marked by $\triangle$ next to $\bullet$). If these cells are on the same level as the MeshBlock, they must be restricted. If they are on the coarser level (i.e. the same level as the coarse buffer), then they have already been received in the coarse buffer.
\item Apply physical boundary conditions on the coarse buffer (if necessary).
\item Perform prolongation and store results into ghost zones overlapping MeshBlock~A (marked by $\bullet$).
\end{enumerate}
The restriction and prolongation algorithms are explained in section~\ref{subsubsec:restrict-prolong}. It is important to note that all the sends, receives, and restriction operations (steps 1 and 2 in the above lists) are independent of each other, while the prolongation can only be performed after the arrival of all the boundary data. Because all the communications are independent, the implementation of the algorithm using the task list is straightforward.

\begin{figure*}[htb!]
\centering
\includegraphics[width=0.8\textwidth]{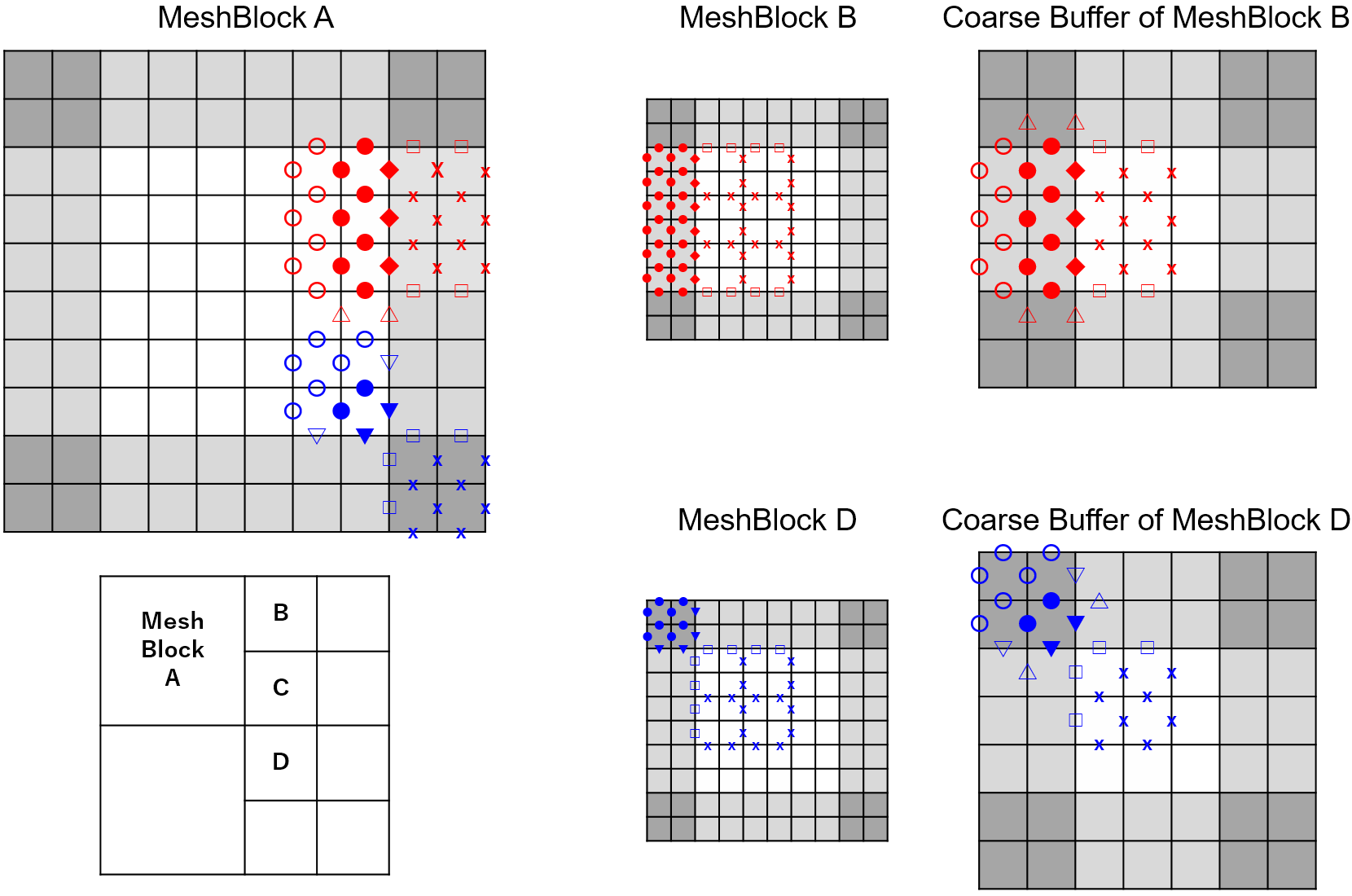}
\caption{Same as figure \ref{fig:amrcomm1} but for face-centered variables. \label{fig:amrcomm2}}
\end{figure*}

For face-centered variables, the communication procedure is slightly more complicated (see figure~\ref{fig:amrcomm2}, note the field component perpendicular to the page in each cell is not shown; it can be transferred in the same way as cell-centered variables discussed above). On MeshBlock~A, the communication procedure to and from finer MeshBlocks~B and~D proceeds as follows:
\begin{enumerate}
\item Send active faces overlapping the neighboring MeshBlocks (marked by $\circ$, $\bullet$, $\blackdiamond$, $\triangle$,$\triangledown$ and $\blacktriangledown$ symbols).
\item Receive ghost faces from neighboring MeshBlocks (marked by $\times$ and $\square$).
\end{enumerate}
Note that the faces marked with the $\blackdiamond$ symbols on MeshBlock A are active faces shared with the neighboring MeshBlocks, and they are not modified by boundary communications as this may cause a violation of the solenoidal constraint (if the data on these faces represent the magnetic field). Instead these faces are sent to the finer MeshBlocks for prolongation. In addition, the faces marked with $\square$ in the ghost zones are also shared by two MeshBlocks. Both MeshBlocks send these faces, and the values that arrive last are stored since (as the restriction operation is conservative) the values should match even if one of the MeshBlocks is on the finer level. Any small differences between the values (at the level of round-off error) are prevented from growing via the flux and EMF correction steps (see section~\ref{subsubsec:flux-emf}), and the error (if any) will not lead to violation of the solenoidal constraint because these values are only used during the reconstruction step and EMF calculation.

On MeshBlocks~B and~D, the exchange of face-centered variables to and from a coarser MeshBlock~A proceeds as follows:
\begin{enumerate}
\item Restrict active faces overlapping MeshBlock~A (marked by $\times$ and $\square$) to the coarse buffer and send them.
\item Receive coarse faces from MeshBlock~A (marked by $\circ$, $\bullet$, $\blackdiamond$, $\triangle$, $\triangledown$, and $\blacktriangledown$ symbols) into the coarse buffer.
\item Wait until all the boundary communications are completed.
\item Fill in the faces adjacent to the faces to be prolongated (marked by $\triangle$ next to $\bullet$ and $\diamond$). If these cells are on the same level as the MeshBlock, they have to be restricted. If they are on the coarser level (i.e. the same level as the coarse buffer), then they have already been received in the coarse buffer.
\item Apply physical boundary conditions on the coarse buffer (if necessary).
\item Perform prolongation and store the results into the ghost zones overlapping MeshBlock~A (marked by $\bullet$ and $\blacktriangledown$)
\end{enumerate}
Again, all the sends, receives, and restriction operations are independent of each other.  Moreover, the communications for cell-centered and face-centered variables are mutually independent. As in the case of MeshBlock~A, the faces marked with $\blackdiamond$ on MeshBlock~B are active and are not modified, and only the faces marked with $\bullet$ are updated by the prolongation operation. On the other hand, the cells marked with $\blacktriangledown$ on MeshBlock~D are in the ghost zone and shared between two MeshBlocks. When both of the MeshBlocks sharing the same face are on the same level (one level coarser than MeshBlock~D), the prolongated values ($\blacktriangledown$ on the horizontal line in this example) are used. If one of them is on the finer level (same as MeshBlock~D), the values from the finer MeshBlock are used because the prolongated values are less accurate ($\blacktriangledown$ on the vertical line).

The communication between MeshBlocks on different levels at the corners in 3D is analogous to the above descriptions.

\subsubsection{Flux and EMF Correction} \label{subsubsec:flux-emf}

In MHD calculations with static and/or adaptive mesh refinement the area integral of the fluxes of the conserved variables on cell faces at the boundaries between MeshBlocks on different levels (as well as the line integral of the EMF along cell edges) must be exactly equal. This requires a special step to correct the coarse cell fluxes with the (generally more accurate) integral of the fine cell fluxes \citep{BergerColella}. The implementation of this correction procedures in \app{} is described below.

\begin{figure}[htb!]
\centering
\includegraphics[width=\columnwidth]{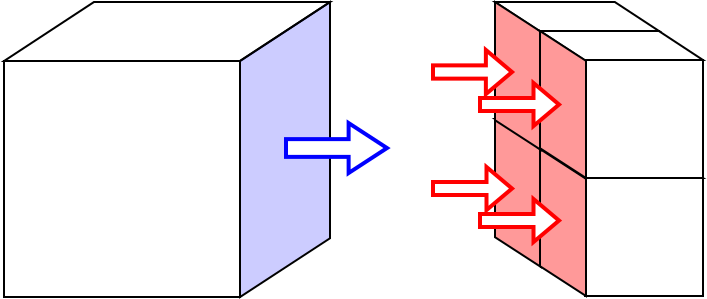}
\caption{Flux correction on cell faces between neighboring MeshBlocks at different refinement levels in 3D. The area-integrated flux on the face of a coarse cell (blue) is replaced by the area-integrated fluxes on the corresponding faces of the fine cells (red). \label{fig:fluxcorr}}
\end{figure}

For the face-centered fluxes of the cell-centered conserved variables, this flux correction step is straightforward (see figure~\ref{fig:fluxcorr}).  In 3D calculations at the interface between different levels one coarse cell abuts four fine cells (two cells in 2D, and one in 1D). The flux used to update the coarse cell on the face that overlaps with the fine cells is simply replaced with the area-weighted sum of the fluxes from these four fine cells.  The step makes use of the communication strategy outlined in the previous section for face-centered data.

\begin{figure}[htb!]
\centering
\includegraphics[width=\columnwidth]{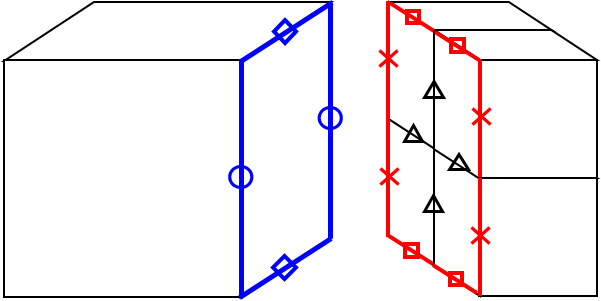}
\caption{EMF correction on cell edges between neighboring MeshBlocks at different refinement levels in 3D. The line-integrated EMFs on the edges of the coarse cell (blue) are replaced by the line-integrated EMFs on the corresponding edges of the fine cells (red). Edges of fine cells that do not overlap any coarse cell edges (marked by $\triangle$) are not used. \label{fig:emfcorrface}}
\end{figure}

\begin{figure*}[htb!]
\centering
\includegraphics[width=\textwidth]{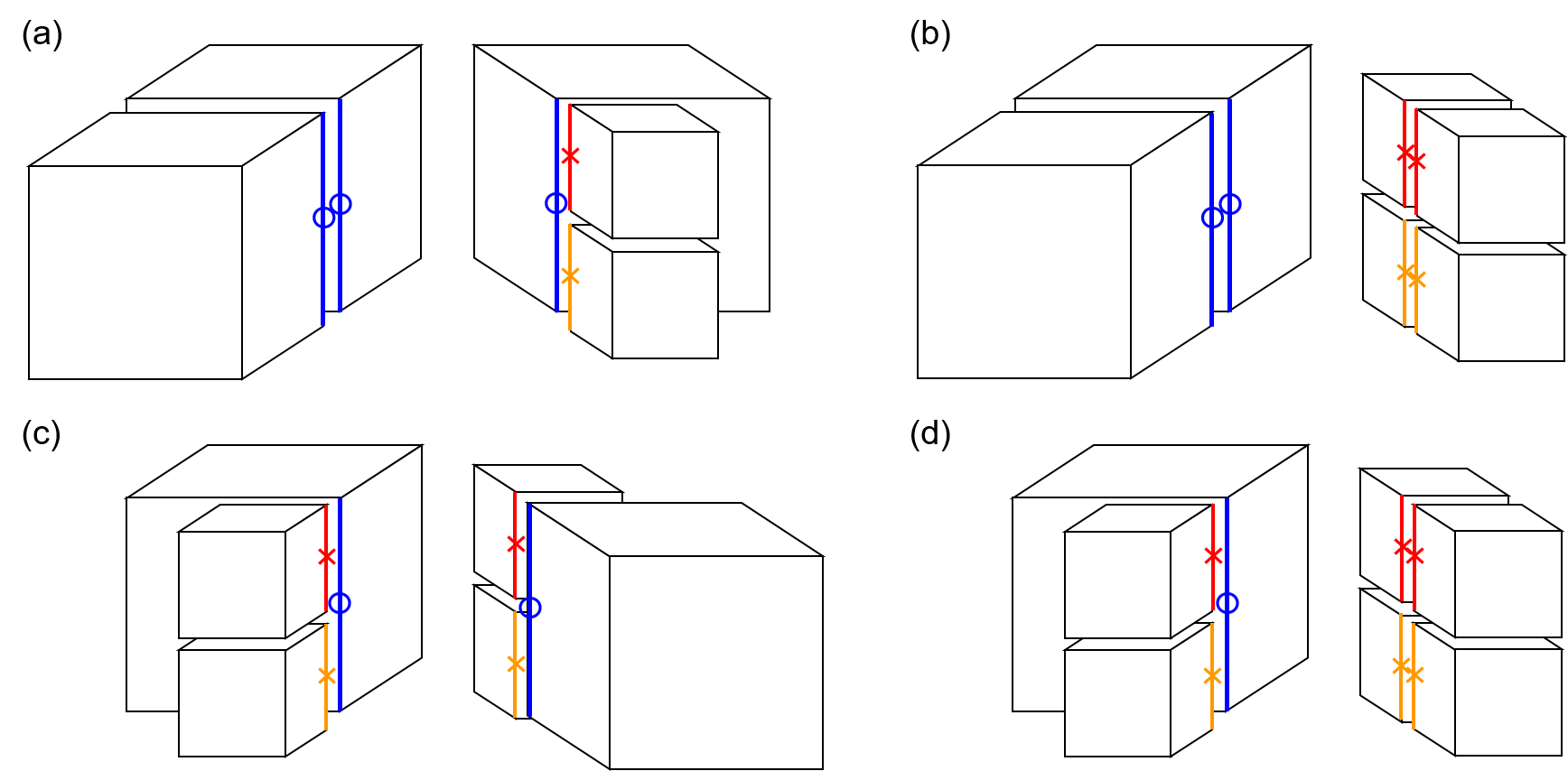}
\caption{Examples of EMF corrections at the edges of cells between MeshBlocks on different refinement levels in various configurations.\label{fig:emfcorredge}}
\end{figure*}

For the edge-centered EMFs needed for the CT algorithm for MHD, this flux correction step is considerably more complicated.  Since the CT schemes preserves the divergence-free constraint to machine precision, it is crucial that the EMFs used to
update the field on overlapping cell edges at different levels be identical, otherwise the magnetic flux at the
faces of the cells will be inconsistent, and the resulting divergence error can grow and cause unphysical dynamics.

When MeshBlocks on different levels share the same face, the EMF on the coarse MeshBlock is replaced with the line-weighted sum over the corresponding fine edges (see figure~\ref{fig:emfcorrface}): 
\begin{equation}
\varepsilon_{\rm coarse}\Delta l_{\rm coarse} = \sum \varepsilon_{\rm fine}\Delta l_{\rm fine} \label{eq:emfcorr}.
\end{equation}
Note that cell edges on the fine MeshBlock that have no corresponding edge on the coarse cell (marked with $\triangle$ in the figure) are not needed. With this correction, the line integral over the coarse cell edges will match those over the corresponding fine faces, which ensures consistent evolution of the magnetic field on the shared face.

This procedure becomes more complicated when MeshBlocks on different levels share an edge rather than a face. Figure~\ref{fig:emfcorredge} shows some representative configurations in this case. In order to satisfy the divergence-free constraint, the line integral of the EMF along the shared edges must match exactly. However, there is no guarantee this will be the case even for shared edges at the same refinement level due to different arrangements of the prolongation operations, nondeterministic ordering of the MPI communications, and round-off error that differs in the calculation of the same EMF on different MeshBlocks. Therefore, both fine and coarse EMFs must be corrected. First, the EMFs on the fine shared edges (marked by red and orange $\times$) are replaced with their average. Then, the EMFs on the coarse shared edges (blue $\circ$) are corrected using the EMFs on the fine edges so that the line integrals of the EMFs match as in equation~\eqref{eq:emfcorr}. The same procedure is applied to edges in the middle of a coarse MeshBlock that overlaps edges of fine MeshBlocks (e.g. the edge shared by MeshBlocks~B and~C facing MeshBlock~A in figure~\ref{fig:amrcomm1}).

Even without mesh refinement, numerical errors can cause a slight mismatch between the EMFs on shared edges between MeshBlocks.  With the constrained transport scheme, such errors never disappear once generated. This problem becomes more prominent when more complex grids with nonuniform mesh spacing and/or curvilinear coordinates are in use. Moreover, aggressive compiler (non-ANSI-conformant) optimizations can introduce and exacerbate differences associated with round-off errors. Therefore, the EMF correction step is applied even when mesh refinement is not used.  In this case, the EMFs on two shared edges are replaced with the arithmetic average of their values.

\subsubsection{Restriction and Prolongation Operators\label{subsubsec:restrict-prolong}}

For simulations with mesh refinement, data on finer MeshBlocks must be mapped onto overlapping cells on coarse MeshBlocks (restriction) and vice versa (prolongation). With our block-based AMR strategy, these interactions occur only at the boundaries between MeshBlocks on different levels, or when MeshBlocks are created or destroyed during refinement or derefinement.

When cell-centered variables are restricted, the volume-weighted average is used:
\begin{equation}
U_{\rm coarse} = \frac{\sum U_{\rm fine} \Delta V_{\rm fine}}{\Delta V_{\rm coarse}}.
\end{equation}
where $U$ denotes the variables being restricted (for MHD the conserved variables are used), and $V$ is the volume of the cells on the fine and course mesh.  For face-centered variables, the area-weighted average is used for quantities defined on the faces shared by MeshBlocks on the fine and coarse levels:
\begin{equation}
F_{\rm coarse} = \frac{\sum F_{\rm fine} \Delta S_{\rm fine}}{\Delta S_{\rm coarse}}.
\end{equation}
Faces on the finer MeshBlock that do not coincide with faces on the coarser MeshBlock are not involved in the restriction step. With MHD, cell-centered magnetic fields and primitive variables are calculated after both cell-centered conservative variables and face-centered fields have been restricted.

For prolongation of cell-centered variables, a multi-dimensional, slope-limited linear reconstruction is used. First, the gradients between neighboring cells in each direction are calculated and slope limiters are applied as in the reconstruction step of the hydrodynamic solver, which is discussed below in section~\ref{subsubsec:reconstruct}. Unlike the limiter used to compute the states at the faces for the Riemann solver, the less aggressive \(\minmod\) slope limiter is used for prolongation.  We have observed that using limiters that are sharper than \(\minmod\) can produce unphysical structures around the level interfaces, as reconstruction during prolongation involves a multi-dimensional profile (unlike the 1D reconstruction during hydrodynamic flux calculations). Then the cell-centered variables are interpolated to the cell-centers on the finer level. For example, to prolongate a cell at $(k,j,i)$, 
\begin{equation}
\begin{aligned}
\frac{\Delta_1 U_{\rm i,j,k}}{\Delta x} &= \minmod\left(\frac{U_{\rm i,j,k}-U_{\rm i-1,j,k}}{\Delta x_{i-1/2}},\frac{U_{\rm i+1,j,k}-U_{\rm i,j,k}}{\Delta x_{i+1/2}}\right), \\
\frac{\Delta_2 U_{\rm i,j,k}}{\Delta y} &= \minmod\left(\frac{U_{\rm i,j,k}-U_{\rm i,j-1,k}}{\Delta y_{j-1/2}},\frac{U_{\rm i,j+1,k}-U_{\rm i,j,k}}{\Delta y_{j+1/2}}\right), \\
\frac{\Delta_3 U_{\rm i,j,k}}{\Delta z} &= \minmod\left(\frac{U_{\rm i,j,k}-U_{\rm i,j,k-1}}{\Delta z_{k-1/2}},\frac{U_{\rm i,j,k+1}-U_{\rm i,j,k}}{\Delta z_{k+1/2}}\right),
\end{aligned}
\end{equation}
\begin{multline}
U_{\rm i\pm1/2,j\pm1/2,k\pm1/2} =  U_{i,j,k} \pm \\ \frac{\Delta_1 U_{\rm i,j,k}}{\Delta x}\Delta x_{\rm f\pm} \pm \frac{\Delta_2 U_{\rm i,j,k}}{\Delta y}\Delta y_{\rm f\pm} \pm \frac{\Delta_3 U_{\rm i,j,k}}{\Delta z}\Delta z_{\rm f\pm},
\end{multline}
where $U$ define at the points with integer indexes are on the coarser level while those with half-integer indices are on the finer level, and $\Delta x_{\rm f\pm}$, $\Delta y_{\rm f\pm}$ and $\Delta z_{\rm f\pm}$ are the distances between the volume-weighted cell-centers of the coarse cell and right/left fine cells in each direction. For the prolongation at interfaces between MeshBlocks on different levels, this prolongation operation is performed using the primitive variables because use of the conservative variables can produce negative pressure. This does not violate the conservation law because the values in the ghost zones are used only through the flux calculation, and conservation in the active zones is ensured by the flux correction procedure. As the communications between MeshBlocks use the conservative variables, they are converted into primitive variables, prolongated, and then converted back to the conservative variables after the prolongation. On the other hand, the conservative variables are used when new MeshBlocks are created by mesh refinement in order to satisfy the conservation law. A pressure floor is applied if negative pressures appear in the refined cells. While the pressure floor violates conservation of the total energy, this method still satisfies conservation of mass and momentum.

For prolongation of face-centered variables, the method of
\citet{TothRoe2002} is adopted, which preserves the divergence of the face-centered fields. First, 2D interpolation on each coarse face is performed with the \(\minmod\) slope limiter to the corresponding fine faces. When fine faces already have values at the fine level (e.g. $\blackdiamond$ on MeshBlock~B in figure~\ref{fig:amrcomm2}), they are not overwritten by the prolongated values; the fine face values are used instead. To determine the field on internal faces on the fine mesh, the method adopted by \citet{TothRoe2002} is adopted, which assumes that the divergence of each fine cell matches the coarse cell (which is zero), while the curl computed at internal edges matches that estimated using the coarse level fields. As pointed out in the original paper, enforcing the curl of the field (currents) to match between levels is an assumption; nevertheless it seems to work well. While this method was originally designed for uniformly-spaced Cartesian grids, it is straightforward to extend it to nonuniform mesh spacing and curvilinear coordinates in the ``finite area'' fashion. For further details, see \citet{TothRoe2002}.

\subsubsection{Load Balancing\label{subsubsec:load-balance}}

In parallel simulations, it is important to keep the computational load balanced among the independent computing elements. By default, \app{} distributes MeshBlocks among computing elements as evenly as possible, assuming each MeshBlock incurs the same computational expense. While this works quite satisfactorily for the hydrodynamic and MHD solvers, calculations involving additional physics can incur uneven computational cost. For example, chemical reactions updated using an iterative solver may require different numbers of iterations on different MeshBlocks. Moreover, when particles such as passive tracers or sink particles are used, they may concentrate in a specific region and increase the load imbalance. 

In order to provide more flexible load balancing, each MeshBlock is given its own ``cost'' parameter and \app{} attempts to redistribute MeshBlocks so that the total cost per process is as even as possible. This cost can be manually set by users or automatically determined by the code using measurements of the compute-time on each MeshBlock gathered from system timing calls. This load balancing is performed periodically, and whenever MeshBlocks are newly created or destroyed. The implementation does have several limitations. Firstly, because a MeshBlock is a unit of both domain decomposition and load balancing, more than one MeshBlock per process is required to adjust the load balance.  Secondly, since the ordering of MeshBlocks cannot be shuffled in the current implementation, certain pathological cases in which the load changes dramatically from one MeshBlock to another can be hard to distribute evenly.  Although more complex load balancing strategies are possible, they lack the simplicity and ease of use of the method implemented in \app{}.

\subsubsection{Time Stepping with AMR\label{subsubsec:timestepping}}

If the maximum signal speed in an MHD calculation ($|v| + C_f$, where $v$ is the fluid velocity and $C_f$ the fast magnetosonic speed) on an AMR mesh is the same on all levels, then the maximum stable time step used to integrate each level will be proportional to the spatial resolution used at each level.  Thus, standard adaptive time stepping can be used, in which each level $l$ uses a timestep that is $2^l$ smaller than that used at the root level $(l=0)$.  Such algorithms require interpolation in both time and space at fine/coarse boundaries to enforce flux conservation \citep[e.g., see][]{Mignone2012}.

In \app{}, we do not use adaptive time stepping, but instead adopt the same fixed time step to integrate all levels.  There are several reasons for this choice.  Firstly, in many MHD applications the maximum signal speed is not constant across all levels.  In fact, it is often the case that the highest speeds (and therefore smallest stable time steps) occur on the root level, where densities may be small and the Alfv\'{e}n speed large.  In this case, by requiring smaller time steps than necessary at the highest refined levels, adaptive time stepping makes the calculation more expensive.  Secondly, the temporal interpolation required by adaptive time stepping introduces additional error to the solution, especially when self-gravity is included.  Finally, the complexity of adaptive time stepping makes the overall calculation more difficult to optimize and load balance; moreover, the cost savings in many cases is not substantial.  For example, if an equal number of cells are being updated at each level (which implies in 3D that roughly 10\% of the volume of the domain is refined at each level), then the reduction in the number of cell updates required is only about $N/2$, where $N$ is the number of levels.  Unless $N$ is large, these savings may be offset by the reduced efficiency of the method on highly parallel systems, making the overall reduction in the amount of CPU time required even smaller.  Moreover, the reduction in work will not decrease the minimum possible wall clock time, which is bounded by the number of time steps needed to update the solution on the finest level.

Recently, several authors have explored the use of variable time-stepping both across AMR levels, and even within MeshBlocks at a given level \citep{DISPATCH,Gnedin2018}.  Tests indicate speed-ups of about an order of magnitude are possible, as well as an increase in accuracy due to the ability to run at close to the maximum stable time step everywhere. Adaptive and/or variable time stepping may be advantageous for very deep AMR hierarchies, or when a very small fraction of the volume is refined, or when the time step varies dramatically within different regions at the same level.   Extending \app{} to enable such capabilities is a topic for future investigation.

\subsection{Comparison to Other AMR Codes\label{subsubsec:other_amr_codes}}

The discussion in the previous sections has focused on the specific implementation of AMR in the \app{} framework.  It is instructive to compare the algorithms we have adopted with those used in other codes.

There are three commonly used algorithms for AMR.  The first is cell-by-cell refinement, as adopted in codes such \texttt{RAMSES} \citep{Teyssier2002} and \texttt{ART} \citep{Kravtsov1997}, in which each individual cell can be refined independently.  The second is patch-based AMR in which refined regions of arbitrary size and shape can be created to cover areas of interest, following the original algorithm of \citet{BergerOliger} and \citet{BergerColella}.  This method is perhaps the most popular, and is implemented in a variety of codes including \texttt{Enzo} \citep{Bryan2014}, \texttt{PLUTO} \citep{Mignone2012}, and \texttt{AMRVAC} \citep{AMRVAC}.  Moreover, sophisticated libraries which implement patch-based AMR for general systems of equations, including Chombo\footnote{Astrophysics Source Code Library, record ascl:1202.008} and AMReX \citep{AMReX_JOSS} are available.  Finally, the third algorithm is block-based AMR in which refinement can occur only in fixed locations using blocks of fixed size.  This is the algorithm adopted in \app{} and described in detail above.  Other codes that adopt this approach include \texttt{FLASH} \citep{Fryxell2000} (which uses an AMR framework implemented in the PARAMESH library \citep{PARAMESH}), the most recent version of \texttt{NIRVANA} \citep{ziegler2008}, and \texttt{DISPATCH} \citep{DISPATCH}.  Another important ingredient to the algorithm is the time-stepping strategy.  Many implementations of AMR use adaptive time-stepping, in which different levels are integrated at different time steps.  As discussed in section~\ref{subsubsec:timestepping}, in \app{} we use a single global time step which is the same for all levels.  This makes our approach less efficient when the grid contains a large number of levels (more than ten) that cover a small fraction of the volume (one percent or less).

Several authors have explored the parallel efficiency of the particular implementation of AMR algorithms in specific codes \citep[e.g.][]{ziegler2008,AMRVAC}.  In sections~\ref{sec:MHD_solver} and \ref{sec:RelativisticMHD_solver} we present similar tests of the efficiency of the AMR algorithms in \app{}.

In fact, the determination of which of the above three approaches for AMR is most efficient is highly application dependent.  The cell-by-cell and patch-based strategies can adapt the mesh to features in the flow more efficiently than the block-based AMR adopted here, mostly because in the latter case refinement can only occur in the predefined locations of MeshBlocks (e.g. see figure \ref{fig:mesh-tree-amr}).  On the other hand, block-based AMR is easier to implement, and therefore easier to optimize on modern highly parallel computing architectures. For example, \citet[][figure 4]{DISPATCH} shows more than an order of magnitude improvement in efficiency using the block-based approach in the \texttt{DISPATCH} code compared to cell-by-cell as in \texttt{RAMSES} on one test.
A comprehensive investigation of the relative merits of each AMR strategy for various applications of interest, including performance and scaling on highly parallel systems, would be extremely instructive, but it is beyond the scope of this paper.

\subsection{Coordinate Systems\label{subsubsec:coord}}

Up to this point, the AMR framework in \app{} has been described without reference to any particular geometry or coordinate system.  Instead, all of the functionality is implemented for logically rectangular arrays of cells.  In principle this enables the code to be used in any coordinate system.

In practice, grid cells stored on the MeshBlocks may have nonuniform spacing, that is the spatial size of the cells may be a smooth function of position in each dimension independently.  Options to create both uniform and logarithmically-spaced cells are provided as built-in features, and there is a simple mechanism to create a custom cell spacing from a user-defined input function.  The physical size, areas, and volumes of cells are constructed and stored in the \texttt{Coordinates} class.  These values are then used whenever needed to construct vector and tensor operators in the specific coordinates.  Currently, \app{} has built-in support for Cartesian, cylindrical, and spherical-polar coordinates for non-relativistic calculations, as well as those using special relativity (SR). General relativity (GR) capabilities support optimized Minkowksi, Schwarzschild, and spherical Kerr-Schild coordinates, as well as any stationary coordinates specified via metric coefficients by the user.  It is straightforward to add new coordinate systems to the code.

Some coordinates systems (for example spherical-polar) introduce coordinate singularities that require special care.  We have implemented ``polar'' boundary conditions on the pole in spherical-polar and spherical-like coordinates. For this boundary condition, the cell-centered and and face-centered variables in the ghost cells are copied from the other side of the pole considering the physical symmetry across the pole. The flux on a face contacting the pole does not have any influence on the active zone because the surface area of the face is zero. On the other hand, the EMFs on the radial edges contacting the pole are replaced with their average because they must have the same value. The robustness of this boundary condition is demonstrated in section~\ref{subsubsec:field-loop-pole}.

\subsection{Hybrid Parallelization Strategy}

Distributed-memory parallelism through domain decomposition is an integral part of the design of the AMR framework in \app{} and has been discussed extensively in the preceding sections.  On some architectures, it is also advantageous to employ shared-memory parallelism based on, e.g., the OpenMP standard.  Because of the significant overhead of launching and terminating threads, we have found that a fine-grained approach to shared-memory parallelism in which parallel regions are forked and joined at the for-loop level is not very efficient. In addition, this approach requires a significant effort to identify and parallelize every region in the code.  Instead, we have found that a coarse-grained approach, in which each MPI rank possesses multiple MeshBlocks that are updated by individual OpenMP threads, is more efficient.  This design does require a thread-safe implementation of the MPI library with support for \texttt{MPI\_THREAD\_MULTIPLE}, which is the fourth and highest level of thread safety defined in MPI. MPI implementations are not required to support this functionality (although most major distributions offer at least partial support), and occasionally users have discovered that the compiled MPI library on their shared cluster was configured with this thread safety disabled. 

\subsection{Dynamic Scheduling via the Task List}

One of the most important capabilities of the \app{} AMR framework is the dynamic execution of tasks.  Similar ideas have been implemented in other codes such as \texttt{DISPATCH} \citep{DISPATCH}, and libraries such as \texttt{CHARM++} and \texttt{Legion} enable task-based parallelism (along with many other features).  We have implemented our own design for task-based dynamic execution in \app{}, which we describe in detail in this section.

Dynamic execution is implemented in a class called the \texttt{TaskList}. Rather than hard-coding the order of execution of functions associated with a physics module, all of the steps in the algorithm are assembled into an array of \texttt{Task} structures. Each \texttt{Task} structure contains a unique \texttt{task\_id}, a \texttt{dependency} encoding of other tasks that must be finished before the current task can be executed, and a pointer to a function that implements the actual work associated with the task. The \texttt{task\_id} and \texttt{dependency} are implemented as bit fields of arbitrary length, and each \texttt{task\_id} has a (different) single bit set to \texttt{1}. Each MeshBlock owns a \texttt{task\_state} to store which tasks are completed, which is also implemented as a bit field.

The key to this implementation is controlling dependencies between Tasks. There are two types of dependencies:\ the first is an internal dependency between Tasks within a single MeshBlock, and the second is an external dependency between different MeshBlocks. The internal dependency controls the ordering of Tasks, and it is implemented using the dependency flag in the \texttt{Task} structure.  The external dependency controls coherency between MeshBlocks associated with boundary communications, and the return value of a Task function implements this control flow.

\begin{figure}[htb!]
\centering
\includegraphics[width=\columnwidth]{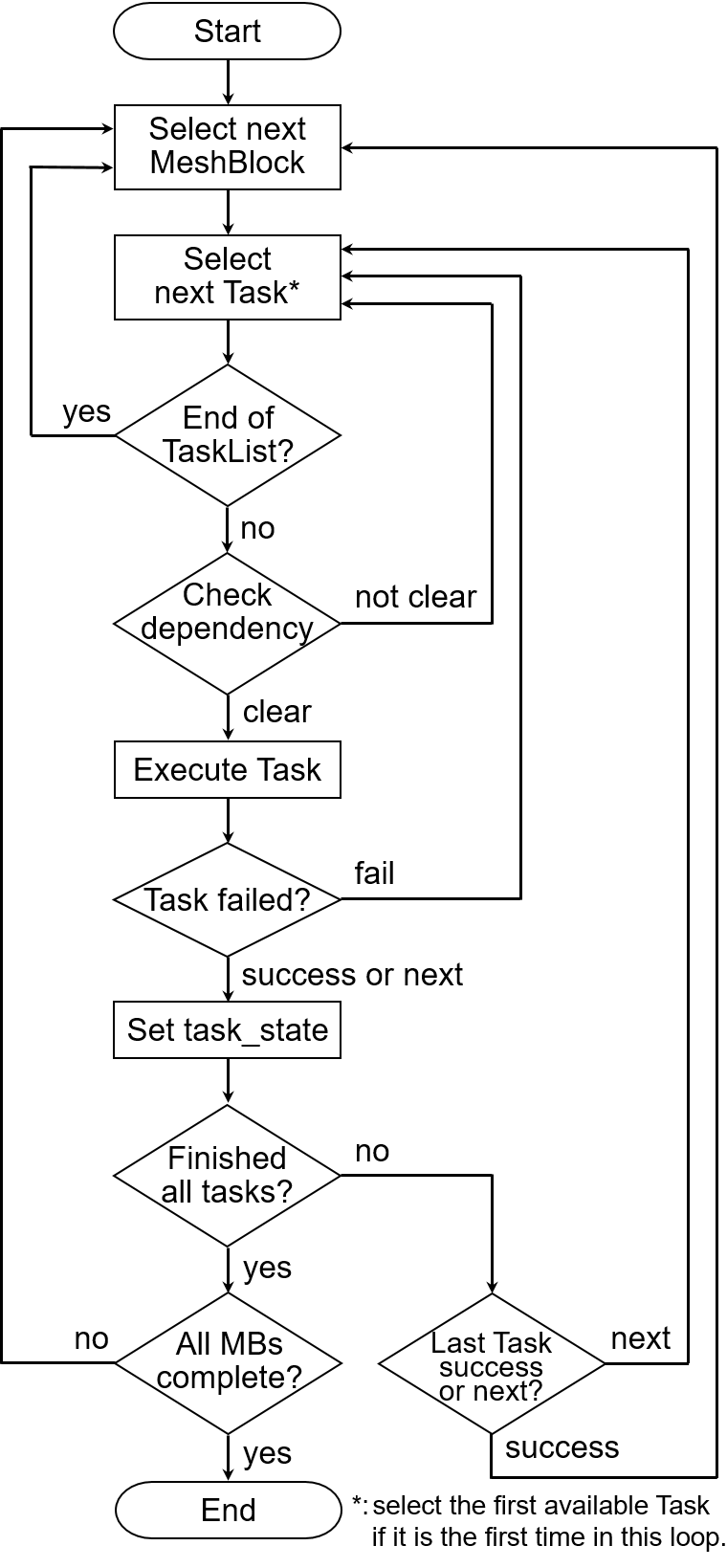}
\caption{A flow chart of dynamic execution using the TaskList. For details see the discussion in the text. \label{fig:taskflow}}
\end{figure}

A flow chart demonstrating how the TaskList is processed is shown in figure~\ref{fig:taskflow}. Execution begins with selection of the first available Task from the TaskList, and a check of its internal dependency (implemented with bitwise operations for efficiency). If the dependency is not cleared, the Task is skipped. If there is no dependency, the Task function is executed. A Task function returns one of three possible results:\ \texttt{success}, \texttt{next}, or \texttt{fail}. When either \texttt{success} or \texttt{next} is returned, the Task is marked as completed and its \texttt{task\_id} is stored in the \texttt{task\_state} by a bitwise \texttt{OR} operation. When the return value is \texttt{success}, the code begins processing another MeshBlock (if any), whereas when \texttt{next} is returned, the subsequent Task on the same MeshBlock is processed. This is used when the ensuing Task should be executed immediately, for example if it involves sending boundary communications. When a Task function returns \texttt{fail}, which typically happens when the function is waiting for MPI communications but one or more messages have not arrived, the \texttt{task\_state} is not updated and the next Task on the same MeshBlock is processed. This procedure is repeated until all the Tasks in all the TaskLists are completed.

\begin{figure}[htb!]
\centering
\includegraphics[width=\columnwidth]{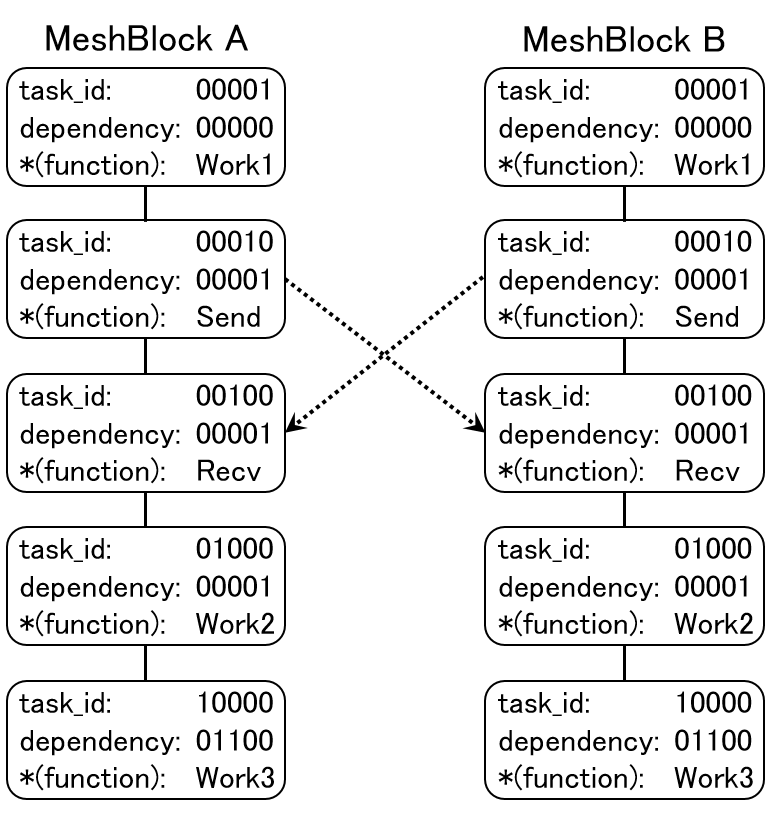}
\caption{Example of a five-step TaskList executed on two MeshBlocks. The dotted arrows indicate communications between MeshBlocks. For details see the discussion in the text. \label{fig:tasklist}}
\end{figure}

To illustrate these concepts further, figure~\ref{fig:tasklist} illustrates an example of two MeshBlocks with very simple five-step TaskLists. These MeshBlocks can be either on the same process or on different processes. Before starting the TaskList, non-blocking MPI receive operations are initiated. When TaskList execution begins, the \texttt{Work1} function referenced in the first Task structure would be called, and provided it completes successfully it will be marked as complete and its \texttt{task\_id} is stored in the \texttt{task\_state} of the MeshBlock. Next, the second task consisting of boundary communications would be executed, as its dependency on the first Task is already cleared. These communications are performed by a standard library \texttt{memcpy()} function call if the neighbor MeshBlock is on the same process, and by non-blocking MPI send operations if it is on a different process. Control will then pass to the third Task in the list. This Task does not depend on the second Task but only on the first Task, which is already cleared. However, this Task also has external dependency on boundary communications from the other MeshBlock. This Task checks completion of the boundary communications using the \texttt{MPI\_Test} functions if the neighbor owned by another process, and returns \texttt{fail} if the messages have not been delivered yet. In this case, the Task is not flagged to be completed, and the next Task in the TaskList is processed. As the fourth Task depends only on the first, this Task is executed even if the third Task is not completed. The fifth Task is then processed, but because it depends on both the third and fourth Tasks, it cannot be executed until those dependencies are cleared. As the execution has now reached the end of the TaskList, control returns to the top of the list and repeats this loop until all of the Tasks are completed.

There are three important reasons why we have found the TaskList to be such a useful design.  The first is that it enables communication to be hidden behind computation.  In the example given in figure~\ref{fig:tasklist}, this is possible because the algorithm contains work (the fourth Task) that does not depend on the completion of some prior communication. Even if this is not the case, by having multiple MeshBlocks on a processor, the communication required by the first and subsequent MeshBlocks can be hidden by the work required at the start of other MeshBlocks.  We have found this feature improves the scaling efficiency of \app{} on very large numbers (millions) of cores.

A second important advantage is that the TaskList provides tremendous flexibility and modularity in incorporating different combinations of physics modules.  In the previous version of the code, different physics algorithms were hard-coded into the main loop and conditionally executed based on a set of nested preprocessor flags. Coding every possible combination of modules in this manner became burdensome. With the TaskList in \app{} physics modules are included at runtime by adding the appropriate steps to the list.  Calculations do not even have to include the MHD modules in order to run.  It is possible to build task lists that simply execute chemistry or radiation transfer modules in a test or post-processing mode.  This makes the code extremely flexible.  Even different numerical algorithms such as higher-order time integrators (see section~\ref{subsubsec:time-int} below) can be constructed simply by encoding them into the task list, rather than hard-coding special purpose functions.

The third advantage of the TaskList is that different MeshBlocks can operate with independent TaskLists and are therefore able to model different physics.  This enables heterogeneous computation in which, for example, some processes solve MHD equations while others solve self-gravity. Heterogeneous parallelization can improve the overall scalability of the code by allocating fewer distributed computing processes for algorithms (e.g. self-gravity) that scale less well. It is even possible to solve different physical models on different MeshBlocks. For example, chemistry or nuclear reaction networks might only be included in certain regions of the flow where they are important, or the general-relativistic MHD equations might be solved only on MeshBlocks near a compact object, while the (much less complex) non-relativistic MHD equations are solved everywhere else.  A final example is that MeshBlocks in regions of very low density may use hybrid particle-in-cell methods to properly capture kinetic physics, while MeshBlocks in denser regions solve the kinetic MHD equations \citep{Garcia1999}.  We will report the usage of the TaskList in such multi-physics applications in the future.

\subsection{Software Design Principles}

\app{} is free-and-open-source software (FOSS). Stable, public releases of the code are hosted on a public GitHub repository\footnote{\url{https://github.com/PrincetonUniversity/athena-public-version}}; however development primarily occurs on a private GitHub repository. Thus the engineering of \app{} is not based on a true \emph{open development model}, although bug reporting, issue tracking, and contributions from the user community are welcomed. Documentation and tutorials are provided on the public GitHub Wiki.  The software is licensed under the permissive 3-clause Berkeley Software Distribution (BSD-3) license, chosen because it has more relaxed rules for redistribution of derivative works than, e.g., a \emph{copyleft} license such as GNU General Public License Version 3 (GPLv3).  This can be an important consideration when integrating \app{} with closed-source software, for example frameworks developed at national laboratories. 

In order to reduce the barriers to entry for using the code, and to maximize the portability of the software (from personal laptops to leadership-class supercomputers and cloud-based containers), \app{} was designed with the smallest number of dependencies possible.  Only a C++ compiler and a Python distribution (versions 2.7+ and 3.4+ both supported) are required in the default configuration. Strict adherence to the C++11 standard is enforced in the source code to ensure compatibility with most modern compilers.  More recent standards are not adopted until all major compilers support new features; to this end migration to the C++14 standard is underway.  To deploy \app{} in parallel, an OpenMP-enabled compiler and/or an MPI library are required. Additional optional functionalities may require linking the solver with compatible FFTw3 and/or HDF5 libraries, although we are working hard to eliminate the latter dependency in the future.  The code has been developed by a core team consisting of the co-authors, with substantial commits from more than a dozen other contributors.  The
first \app{} Developers Meeting and Users Workshop was held in 2019, with 63 attendees and speakers\footnote{http://www.physics.unlv.edu/astro/athena2019/index.html}. 

The decision to not follow an open development model is driven by several factors.  Managing an open development project (including quality control) is more time-consuming and burdensome; the primary focus of the core developers is science applications rather than supporting software development.  Moreover some algorithmic features take years of development and testing before they are generally useful, and granting open access too early seems counterproductive.  \app{}'s current development model strikes a balance between centralizing control over the code's development while also encouraging the dozens of unique clones of the public version that occur per week. However, there are benefits to the open development model \citep{Turk2013}, both for accelerating development of new features and for cultivating a more productive relationship with a self-sustaining community of user-developers who provide valuable contributions.  For this reason, we are actively exploring the re-organization of the \app{} AMR framework and physics modules into separate development repositories.  Since almost all of the factors that drive a private development repository are related to the physics solvers, this would allow the AMR framework to become truly open development.  Moreover, this would enable others to build their own physics solvers on top of the AMR capabilities developed for \app{}.

An important argument in favor of open development models is reproducibility; science applications that use a private development version cannot be easily rerun by the community. However, the ability to reproduce results simply by running the same calculations using the same code does not guarantee those results are correct.  True reproducibility requires results to be checked by an independent implementation of the same algorithms, or even more importantly by running different algorithms as implemented in different codes to solve the same mathematical model.  Open-source software and open development are useful instruments for supporting reproducibility, but they are not sufficient to guarantee it on their own \citep{Stodden2014}. Nevertheless, we support such efforts by bundling input files and validation test scripts with the source code distribution. The analysis and plotting scripts used to produce many of the published results from \app{} are also included; this is an increasingly popular best practice that many other projects have adopted \citep[for example][]{Oishi2018}.

Perhaps the most important ingredient for reproducibility is validation and verification.  In this paper, and in
\citet{Stone+2008}, we provide a comprehensive series of test problems based on known analytic solutions and comparison
of results computed by \app{} with those from other codes (see especially section~\ref{subsubsec:khi-benchmark}).  Another crucial component for promoting computational reproducibility and manageability in a codebase the size of \app{} is automated testing. A regression test suite written in Python is distributed with the source code.  It consists of more than 60 separate tests ranging from simple compilation checks to multiphysics benchmark problems.  Whenever possible, such tests involve comparison to analytic solutions (such as linear wave convergence, or planar shock tube problems) to avoid issues related to numerical precision.  In addition, style checks and code linting of C++ and Python source are provided by Google's open-source \texttt{cpplint.py} static code checker and the \texttt{Flake8} tool, respectively. Every pull request and change to the repository's main branch are automatically tested using continuous integration (CI). A local Jenkins\footnote{https://jenkins.io/} server and the cloud-based Travis CI\footnote{https://travis-ci.org/} service independently execute every available test. We have found that it is valuable to repeat the tests with multiple combinations of compilers, target architectures, and dependency library versions in order to catch subtle bugs that may only emerge in certain programming environments. Code coverage analysis is provided by GCC's \texttt{gcov} utility combined with the Linux Testing Project's graphical front-end \texttt{lcov}\footnote{http://ltp.sourceforge.net/coverage/lcov.php}. The testing regime currently achieves approximately 65\% of C++ line coverage. The important role that CI and regression testing has played in the development of \app{} cannot be overemphasized.

\section{A Non-Relativistic MHD Solver\label{sec:MHD_solver}}

As we have previously highlighted, the AMR framework described in the preceding section can be used with any grid-based physics solver.   In order to provide a concrete example of the (most popular) use of the \app{} AMR framework, in this section we describe the implementation of a module to solve the equations of non-relativistic hydrodynamics and MHD. 

The underlying algorithms implemented in this module are nearly identical to those used in the original C-version of \athena, and are described in detail in \citep{Stone+2008}.  Therefore we only provide an overview of the method in this section with particular focus on any changes we have made in re-implementing the methods in \app{}.

\subsection{Equations and Discretization}

The module solves the equations of non-ideal MHD
\begin{subequations} \label{eq:cons}
\begin{align}
\frac{\partial \rho}{\partial t} +
\mathbf{\nabla\cdot} (\rho\mathbf{ v}) &= 0,
\label{eq:cons_mass} \\
\frac{\partial \rho \mathbf{ v}}{\partial t} +
\mathbf{\nabla\cdot} \left(\rho\mathbf{ vv} - \mathbf{ BB} + {\sf P^*} + \mathbf{ \Pi}\right)  &= 0,
\label{eq:cons_momentum} \\
\!\begin{multlined}[b]
\frac{\partial E}{\partial t} +
\nabla\cdot \bigl[(E + P^*) \mathbf{ v} - \mathbf{ B} (\mathbf{ B \cdot v}) \\ 
+ \mathbf{ \Pi}\cdot\mathbf{ v} + \eta\mathbf{ J}\times\mathbf{ B} \qquad \\ 
+ \frac{\eta_{\rm AD}}{|\mathbf{ B}|^2}\left\{\mathbf{B}\times(\mathbf{J}\times\mathbf{B})\right\}\times\mathbf{B}+\mathbf{ Q}\bigr]
\label{eq:cons_energy}
\end{multlined} &= 0, \\
\!\begin{multlined}[b]
\frac{\partial \mathbf{ B}}{\partial t} -
\mathbf{\nabla} \times \bigl[\left(\mathbf{ v} \times \mathbf{ B}\right) - \eta\mathbf{ J} \qquad \\
- \frac{\eta_{\rm AD}}{|\mathbf{B}|^2}\mathbf{ B}\times\left(\mathbf{ J}\times\mathbf{ B}\right) % \\ - \eta\mathbf{ J} 
\bigr]
\label{eq:induction} 
\end{multlined} &= 0,
\end{align}
\end{subequations}
where ${\sf P^*}$ is a diagonal tensor with components
$P^* = P + B^{2}/2$ (with $P$ being the gas pressure), $\mathbf{ \Pi}$ is the viscous stress tensor
\begin{equation}
\label{eq:viscosity}
\Pi_{ij} = \rho\nu\left(\frac{\partial v_i}{\partial x_j}+\frac{\partial v_j}{\partial x_j}
-\frac{2}{3}\delta_{ij}\nabla\cdot\mathbf{v}\right) \,,
\end{equation}
and $\nu$ is the coefficient of kinematic viscosity.
$E$ is the total energy density
\begin{equation}
  E = e + \frac{1}{2}\rho v^{2} + \frac{B^{2}}{2},
\label{eq:total_energy}
\end{equation}
with $e$ as the internal energy density;
$\mathbf{ Q}$ is the heat flux
\begin{equation}
\mathbf{ Q} = \kappa \nabla T
\label{eq:heat-flux}
\end{equation}
with thermal conductivity $\kappa$ and temperature $T$; and $\mathbf{ J} = \nabla \times B$ the current density.  These equations are written in units such that the magnetic permeability $\mu=1$.  

These equations include terms for isotropic viscosity and thermal conduction, as well as Ohmic resistivity and ambipolar diffusion in the strong coupling limit. The coefficients of kinematic viscosity $\nu$, thermal conductivity $\kappa$, and Ohmic resistivity $\eta$ are constants by default; however it is straightforward to extend them to be functions of position and the dynamical variables.  There is no single form for the conductivity $\eta_\mathrm{AD}$ needed with ambipolar diffusion as this depends on the ionization, recombination, and collision rates in the plasma. Therefore, no default form is provided.  Instead, a function to compute $\eta_\mathrm{AD}$ must be implemented as part of the calculation, and a simple mechanism is provided to users in order to do this.

An equation of state (EOS) is needed to compute the pressure $P$ and temperature $T$ from the total energy and other conserved quantities. In \app{} any general EOS can be used.  This includes both an ideal gas law (in which case $P=(\gamma-1)e$), or a barotropic EOS (for example isothermal, in which case $P=c_s^2 \rho$, where $c_s$ is the isothermal sound speed). Any general EOS that provides $P = P(\rho, e)$ and $a^2 = a^2(\rho, p)$ (where $a$ is the sound speed), either as an analytic function or through interpolation of tabular data, can be used.  A complete description of the implementation of the general EOS functionality in \app{} is provided in \citet{coleman19}. This functionality is validated using tests from \citet{Chen2019}.

Equations \eqref{eq:cons_mass} through \eqref{eq:cons_energy} are discretized using a finite-volume approach, with the cell-averaged conserved variables stored at the volume centers of cells.  Note that in curvilinear coordinates, it is important to distinguish volume centers from geometric centers, especially for algorithms with formal spatial accuracy higher than second order \citep{BlondinLufkin}.  The induction equation~\eqref{eq:induction} is discretized using the upwind constrained transport algorithm developed in \citet{gs05,gs08}, and therefore the components of the magnetic field are area averages stored at cell faces. See \citet[][Section 3]{Stone+2008} for details.

\subsection{Numerical Algorithm}

To provide robust and accurate shock capturing, the MHD module in \app{} is based on a Godunov-type method.  The major components of such algorithms for ideal hydrodynamics are (1) a method for the non-oscillatory spatial reconstruction of the fluid variables to compute interface states, (2) a Riemann solver to compute upwind fluxes and electric fields at cell faces, and (3) a time-integration algorithm to advance the solution. Each of these steps is described in subsections below.

In order to preserve the divergence-free constraint on the magnetic field at every substep, a dimensionally-unsplit algorithm is required.  The most accurate unsplit algorithm used in \athena{}, the corner transport upwind (CTU) method (\citet{Colella1990}, described in detail in \citet{Stone+2008}) requires a characteristic projection of the interface states during the reconstruction phase.  For relativistic MHD, such projections are very complex, and for that reason in \app{} the CTU integrator is not used but instead simpler unsplit integration algorithms are adopted (see \S3.2.3).  Of course, it would still be possible to implement the CTU algorithm in \app{} provided its use is  restricted to non-relativistic MHD.

\subsubsection{Spatial Reconstruction Methods\label{subsubsec:reconstruct}}

As in \athena, three different spatial reconstruction methods are implemented in the \app{} MHD module:\ (1) a first-order donor cell (DC) method, (2) a second-order piecewise linear method (PLM), and (3) a fourth-order piecewise parabolic method (PPM). 

Variable reconstruction is performed on either the primitive variables $\mathbf{W} = (\rho, \mathbf{v}, P, \mathbf{B})$, or (for non-relativistic, ideal EOS problems) on the characteristic variables $\mathbf{C} = {\sf L}\cdot \mathbf{W}$ , where ${\sf L}$ is the matrix of left-eigenvectors of the system of equations \citep[see][Appendix A]{Stone+2008}. The latter approach can help reduce oscillations in the solutions, especially for MHD problems, as we demonstrate in \S3.3 below. However, the projection procedure is different from the approach used in \athena{} and described in \citet[][Section 4.2.2]{Stone+2008}. The reader is referred to \citet[][Section 2.2.2]{FelkerStone2018} for a detailed description of the characteristic reconstruction steps used in \app{}. 

There are several other important changes to the reconstruction algorithms implemented in \app{} compared to those in the original version of \athena{} and described in \S4.2 in \citet{Stone+2008}. Firstly, the characteristic tracing performed in step~7 of  \S4.2.2 and step~10 in \S4.2.3 of \citet{Stone+2008} is no longer required because the CTU integrator is not implemented.  Secondly, the reconstruction stencils and slope limiters are modified to ensure the reconstruction remains total variation diminishing (TVD) with both nonuniform and curvilinear meshes. For PLM reconstruction, \app{} uses the original van~Leer limiter \citep{vanLeer1974} when the grid is uniformly spaced and there is no geometric factor (e.g. uniform Cartesian grids and uniformly spaced $\phi$ direction in cylindrical/spherical coordinates), and the modified van~Leer limiter described in \citet{Mignone2014} for nonuniform and/or curvilinear meshes. The weights for the smooth reconstruction stencil are automatically modified for nonuniform and/or curvilinear grids if the backwards and forwards difference approximations to the derivative are divided by the distance to the centroid of volume.

The PPM reconstruction algorithm in \app{} has also been significantly modified to improve accuracy on curvilinear and nonuniform meshes. We again refer the reader to \citet[][Section 2.2.2]{FelkerStone2018} for a complete description of the five PPM limiter formulations that were considered during the development of \app{} and a summary of the errata in the original references for each limiter.

The primary PPM limiter is a smooth extrema preserving limiter described in \citep{McCorquodale2015a} which extends the work of \citet{ColellaSekora2008}; it is used for all problems on Cartesian meshes in \app{}. For curvilinear grids, the steps of the original PPM limiter of \citet{ColellaWoodward1984} are modified in \S3.3 of \citet[][]{Mignone2014} to account for the difference between the geometric and volumetric centers of the cells. 

For nonuniform, Cartesian-like grid directions, equation 1.6 of the original PPM publication \citep{ColellaWoodward1984} provides the reconstruction stencil for the initialization of the variable face-averages at fourth-order spatial
accuracy. For uniform grids, it reduces to the well-known weights of \citep[][equation 1.9]{ColellaWoodward1984}:
\begin{equation}
 Q_{i-1/2} = \frac{7}{12}(Q_{i-1} + Q_i) -\frac{1}{12}(Q_{i-2} + Q_{i+1})\, .\label{eq:ppm-uniform-weights}
\end{equation}
The procedure outlined in \citet[][Section 2.2]{Mignone2014} is followed for computing the curvilinear counterparts to the weights in equation~\eqref{eq:ppm-uniform-weights} along the radial direction in spherical-polar and cylindrical coordinates, and along the meridional direction in spherical-polar coordinates.

By default, \app{} uses a second-order accurate constrained transport solver for MHD problems. With this configuration, the overall accuracy of the MHD solver remains formally \(\mathcal{O}(\Delta x^2)\) even when a higher order reconstruction method is employed. However, the use of higher-order algorithmic components often still significantly improves the accuracy of solutions (see section~\ref{subsubsec:linwave} for a demonstration). Extension to a fully fourth-order accurate scheme has already been implemented in \app{} and published in \citet{FelkerStone2018}.

\subsubsection{Riemann Solvers \label{subsubsec:riemann}}

As in \athena, the HLLE, HLLC, and HLLD approximate Riemann solvers are implemented in \app{}, as well as Roe's linearized solver.  We find exact solvers do not provide any significant increase in accuracy for most problems (although they may make the algorithms more robust on problems involving strong rarefactions), so currently none are implemented. 

The HLLE, HLLC, and HLLD solvers have been extended to be compatible with a general EOS.  This requires the sound speed $a$ be provided either as an analytic function, or through interpolation of tabular data.  A complete description of the changes to these solvers for a general EOS is provided in \citet{coleman19}.

\subsubsection{Time Integrators \label{subsubsec:time-int}}

The final major component of the main MHD algorithm concerns the temporal evolution of the fluid variables. A method of lines formulation is adopted, in which the spatial discretization steps in sections~\ref{subsubsec:reconstruct} and \ref{subsubsec:riemann} provide an estimate of the flux divergence of the system of conservation equations at a single time \(t\). When combined with a suitable method for integrating the time-dependent system of ordinary differential equations (ODEs), a complete scheme with formal \(\mathcal{O}(\Delta x^n, \Delta t^m)\) accuracy is constructed.
It is important that dimensionally-unsplit integrators are used for MHD so that the divergence-free constraint
applies at every substep.  In \athena{}, both the \(\mathcal{O}(\Delta t^2)\) accurate van~Leer (VL2) predictor-corrector integrator described in \citet{sg09} and the CTU method of \citet{Colella1990} are implemented.  However, as discussed earlier, the characteristic projection method required by the CTU integrator makes it difficult to use for relativistic flows.  Thus, in \app{} the VL2 integrator is implemented along with several strong-stability preserving (SSP) and/or low-storage Runge-Kutta (RK) methods.

In \app{}, the 2S class of low-storage RK methods discussed in \citet{Ketcheson2010} is adopted. Let \(u^{(0)}, u^{(1)}\) refer to the two registers in memory for storing the conserved fluid variables defined across the mesh at different time abscissae within a single timestep. We now describe our implementation of Algorithm 3 of \citet{Ketcheson2010}. The notation is modified to use 0-based indexing for the variable registers and the integrator stages, and the relative index of \(\delta_i \equiv \delta_{j-1}\) increased by 1 from the original \(\delta_j\).

At every cycle, \(u^{(0)} = u^n\), \(u^{(1)} = 0\) is assigned before the first stage of the integrator. While \(u^{(0)} = u^n\) is already implicitly guaranteed from the output of the 2S algorithm in the previous timestep, these two-register integrators typically require explicit assignment operations in order to clear the cached data in \(u^{(1)}\). Then, for \(s=0\ldots N_{\mathrm{stages}} - 1\):
\begin{equation}
\begin{aligned}
    u^{(1)} &\leftarrow u^{(1)} + \delta_{s} u^{(0)} \\
    u^{(0)} &\leftarrow \gamma_{s0}u^{(0)} + \gamma_{s1}u^{(1)} + \beta_{s, s-1} \Delta t F(u^{(0)})
\end{aligned} \label{eq:rk-2s}
\end{equation}
where \(u^{n+1} \equiv u^{(0)}\) after the final stage in the cycle. In all cases, \(\delta_0 = 1\) and \(\gamma_{01} = 1\) since the first stage is always a forward Euler step using data from the previous cycle. 

A wide range of integrators of varying orders of accuracy, number of stages, and stability properties can be represented within this framework. For completeness, the coefficients of the most commonly-used (and simplest) selections available in \app{} are documented below. All of the following integrators are defined with \(\delta_i = 0\) for \( i > 0 \); however, the generality of \eqref{eq:rk-2s} enables the trivial implementation of more advanced limiters such as the non-SSP RK4()4[2S] \citep[see][table 2]{Ketcheson2010}. Furthermore, it is straightforward to extend the framework to three-register 3S* methods which are useful for high-order schemes \citep{FelkerStone2018}.

The integrators available in \app{} are:

\underline{RK1}:\ forward Euler method.
\begin{equation}
\begin{aligned}
    \mathbf{\gamma}_0 &= \{0, 1\} &\beta_{0,-1} &= 1
\end{aligned}
\end{equation}

\underline{VL2}:\ (default) predictor-corrector midpoint method. The predictor step must always compute and apply diffusive first-order accurate fluxes that are produced by donor cell reconstruction.
\begin{equation}
\begin{aligned}
    \mathbf{\gamma}_0 &= \{0, 1\} &\beta_{0,-1} &= 1/2 \\
    \mathbf{\gamma}_1 &= \{0, 1\} &\beta_{1,0} &= 1
\end{aligned}
\end{equation}

\underline{RK2}:\ \citep[][equation 3.1]{GottliebKetchesonShu2009} also known as SSPRK(2,2) and Heun's second-order method. Optimal (in error bounds) explicit two-stage, second-order SSPRK method.
\begin{equation}
\begin{aligned}
    \mathbf{\gamma}_0 &= \{0, 1\} &\beta_{0,-1} &= 1 \\
    \mathbf{\gamma}_1 &= \{1/2, 1/2\} &\beta_{1,0} &= 1/2
\end{aligned}
\end{equation}

\underline{RK3}:\ \citep[][equation 3.2]{GottliebKetchesonShu2009} also known as SSPRK(3,3). Optimal explicit three-stage, third-order SSPRK method. 
\begin{equation}
\begin{aligned}
    \mathbf{\gamma}_0 &= \{0, 1\} &\beta_{0,-1} &= 1 \\
    \mathbf{\gamma}_1 &= \{1/4, 3/4\} &\beta_{1,0} &= 1/4 \\
    \mathbf{\gamma}_2 &= \{2/3, 1/3\} &\beta_{2,1} &= 2/3 
\end{aligned}
\end{equation}
Note, the RK2 and RK3 methods each have an SSP coefficient of \(c=1\), which implies that their CFL constraint \(C_0=1\), the same as the stability limit for RK1. In practice, the RK1 integrator is only stable with first-order (DC) fluxes.  The stability of RK2 and RK3 is hard to prove with high-order fluxes, but in practice the limit \(C_0=1\) seems to work for both PLM and PPM reconstruction for most
problems.  In 1D, VL2 is stable up to \(C_0=1\), while in 2D and 3D VL2 \(C_0 = 1/2\).  Moreover, the method is positive-definite for \(C_0 \leq 1/3\) when first-order fluxes are used in both the predictor and corrector steps \citep{sg09}. In our experience, the most useful combinations of integrators and reconstruction algorithms are RK1+DC (for testing), VL2 or RK2 with either PLM or PPM, and RK3+PPM.

% Cannot use this SSP guarantee in theory, since both PLM nor PPM fluxes are unconditionally unstable under RK1 integration. 
% In practice, the stability limit of RK2/3 appears to be 1 in 1D, 2D, 3D regardless of spatial discretization choices?? TODO: check this
%For comparison, we have also tested the five-stage, fourth-order, three-register SSPRK(5,4) method, which has an SSP coefficient of \(c=1.508\) \citep{GottliebKetchesonShu2009}.

% This is more restrictive than the limits for the CTU integrator used in \athena: \(C_0 = 1\) in 2D and 3D (6 Riemann solve, one predictor step variant) \citep{Stone+2008}. In 1D, both VL2 and CTU are stable up to \(C_0=1\). 

\subsubsection{Discretization of the Momentum Equation in Curvilinear Coordinates \label{MomEqCurvilinear}}

Equations \ref{eq:cons} are written in conservative form, enabling numerical algorithms that exactly preserve
the integrals of the dependent variables over the domain.  However, in general curvilinear coordinates, the tensor
operators associated with the flux divergence lead to geometrical factors that usually are written as source terms.
For example, in cylindrical coordinates ($R, \phi, Z$), the $\phi-$component of the momentum equation can be written as
\begin{equation}
\frac{\partial \rho {v}_{\phi}}{\partial t} +
\frac{1}{R}\frac{\partial (R M_{R\phi})}{\partial R} + 
\frac{1}{R}\frac{\partial M_{\phi\phi}}{\partial \phi} + 
\frac{\partial M_{Z\phi}}{\partial Z} = - \frac{M_{R\phi}}{R},
\label{eq:phi_mom_cyl}
\end{equation}
while in spherical polar coordinates ($r,\theta,\phi$) it can be written as
\begin{multline}
\frac{\partial \rho {v}_{\phi}}{\partial t} +
\frac{1}{r^2}\frac{\partial (r^2 M_{r\phi})}{\partial r} + 
\frac{1}{r\sin\theta}\frac{\partial \sin\theta M_{\theta\phi}}{\partial \theta} + 
\frac{1}{r\sin\theta}\frac{\partial M_{\phi\phi}}{\partial \phi} \\
=  - \left( \frac{M_{\phi r} + \cot\theta M_{\phi\theta}}{r} \right),
\label{eq:phi_mom_sph}
\end{multline}
where the $M_{ii}$ are components of the total stress tensor.
However, when these equations are written using the angular momentum (for example $R\rho V_{\phi}$ in cylindrical coordinates), they again can be expressed in conservation form, with the geometrical factors
embedded in the divergence of the fluxes of angular momentum.

It is possible to express the source terms that appear in the $\phi-$component of the momentum equation in
cylindrical and spherical polar coordinates in a discrete form that also guarantee conservation of the angular momentum to machine precision.  In particular, the term on the RHS of equation~\ref{eq:phi_mom_cyl} must be written as
\begin{multline}
\frac{M_{R\phi}}{R} \approx \frac{(R_{i+1/2} - R_{i-1/2})}{(R_{i-1/2} + R_{i+1/2})V_R} \times\\
  \left( R_{i+1/2}M_{R\phi,i+1/2} + R_{i-1/2}M_{R\phi,i-1/2} \right),
  \label{eq:cyl_geometric_srcterm}
\end{multline}
where the half integer indices denote quantities at radial cell faces, $V_R = (R_{i+1/2}^2 - R_{i-1/2}^2)/2$, and the 
components of the stress tensor at radial cell faces are the fluxes of momentum given by the solution to the Riemann problem that are used to update the cell.  When the source term in equation~\ref{eq:phi_mom_cyl} is written in this form, it can be shown that the discrete form of the full equation (including the flux-divergence terms) is algebraically identical to the conservative difference formula for the angular momentum equation in cylindrical coordinates.  Thus, by using this form for the ``geometric source term,'' it is possible to conserve angular momentum to machine precision.  This discretization of the momentum equation is adopted in \app{} in cylindrical coordinates.

Similarly, in spherical polar coordinates, the angular momentum can be conserved to machine precision if the source
terms on the RHS of equation~\ref{eq:phi_mom_sph} are discretized appropriately.  The first term can be written in a form similar to equation~\ref{eq:cyl_geometric_srcterm}, but using $V_R=(r_{i+1/2}^3 - r_{i-1/2}^3)/3$.  The second term must be approximated as
\begin{multline}
\frac{\cot\theta M_{\phi\theta}}{r} \approx \frac{(S_{j+1/2} - S_{j-1/2})}{r_{i}(S_{j-1/2} + S_{j+1/2})V_{\theta}} \times\\
  \left( S_{j+1/2}M_{\phi\theta,j+1/2} + S_{j-1/2}M_{\phi\theta,j-1/2} \right),
  \label{eq:sph_geometric_srcterm}
\end{multline}
where $S=\sin\theta$, $V_{\theta}=(\cos\theta_{j+1/2} - \cos\theta_{j-1/2})/2$, and once again the components of the stress
tensor at cell faces in the $\theta-$direction are the momentum fluxes returned by the Riemann solver and used to update the cell.

Of course there are also similar terms that appear in the other components of the momentum equation.  For these terms, the appropriate volume average can be used.  In addition, a variety of coordinate source terms appear in
the momentum equation in general relativistic calculations, depending on the choice of variables.  A discrete form
that conserves the $z-$angular momentum is possible; refer to section~\ref{subsec:GR_equations} for additional details.

\subsubsection{Diffusion Terms \label{subsubsec:diffusion}}

The MHD module includes terms for modeling many different diffusion processes, for example isotropic viscosity, resistivity, thermal conduction, and ambipolar diffusion.  These terms can be included as an explicit update in each step of the time integrator in a fully unsplit fashion.  This is the most accurate formulation for the terms, as it ensures they are evolved at the same temporal order of accuracy as the main, non-diffusive integration algorithm.

To guarantee conservation of momentum, energy, and magnetic flux, the diffusion terms are added as the divergence of
the respective fluxes (see equation~\ref{eq:cons}).  Second-order finite differencing is used to compute the components of the viscous stress tensor, heat flux, or EMF as appropriate.  For higher-order algorithms, higher-order difference approximations for these fluxes may be required.

Explicit integration of diffusive physics requires a very restrictive timestep stability limit that is inversely proportional to the square of the spatial resolution.  When the diffusive terms are relatively large (for example at low Reynolds number), or at very high resolution, this timestep limit can severely restrict the calculation.  Therefore, a Runge--Kutta--Legendre (RKL) super-time-stepping (STS) module \citep{Meyer12,Meyer14} has been implemented (P. Mullen, private communication) which includes both the RKL1 (temporally first-order accurate) and RKL2 (temporally second-order accurate) schemes.  When STS is enabled, diffusive physics is advanced forward in time by a separate super-time-step in an operator-split update.  Each super-time-step is comprised of $s$ stages and is equivalent to $O(s^2)$ times the explicit diffusive timestep.   The super-time-step size is set to be equal to the full (M)HD timestep for the RKL1 algorithm, or half the (M)HD timestep for the RKL2 algorithm.  Two operator-split super-time-steps are required in a single (M)HD update for the second-order accurate RKL2 scheme. All schemes have been shown to (1) produce errors that converge at the appropriate rate for smooth flows, and (2) yield the expected speedup (roughly $\propto s$).  The algorithm has been parallelized and employs the same task-based execution strategy discussed in the previous sections. 

\subsubsection{Additional Physics\label{subsubsec:additional-phys}}

There are a number of extensions to the basic algorithms for non-relativistic MHD that have been implemented in \app{}, in addition to the general EOS and diffusion terms for non-ideal MHD described above.  We describe three such extensions below.

{\em Passive Scalars.}  An arbitrary number of passive scalars that are advected with the fluid flow can be added to the MHD solver.  These quantities independently obey a simple conservative transport equation
\begin{equation}
\frac{\partial \rho C_i}{\partial t} +
\mathbf{\nabla\cdot} \left[\rho\mathbf{ v}C_i\right] = 0 \, ,
\label{eq:passive_scalars} 
\end{equation}
where $C_i$ (primitive variable) is the specific density of each scalar and \((\rho C_i)\) is the mass of each scalar species (conserved variable).  These quantities provide useful flow diagnostics for following transport and mixing, and they are also necessary for coupling chemical or nuclear reactions to the MHD.  In the latter case, source terms representing the net reaction rates are added to the right hand side of each transport equation, typically via an operator-split method. A complete description of the implementation of chemical networks in \app{} will be given in a future publication.

{\em Shearing Box Approximation.} For the purposes of studying the dynamics of an accretion disk in a locally rotating frame, the shearing box approximation is a valuable tool in astrophysical fluid dynamics. A complete description of the implementation of the local shearing box approximation in the \athena{} code was presented in \citep{SG2010}. This feature has also been implemented in \app{} using the same algorithm.

{\em Orbital Advection.}  In order to speed up and improve accuracy of calculations in the local shearing box, an orbital advection algorithm has been implemented in \app{}, following the methods described in \citep{SG2010}.  
The method was developed for hydrodynamics by \citet{Masset}, implemented in the FARGO code, and later extended to MHD \citep{JGG,FARGO3D}.  Orbital advection algorithms have also been implemented in the PLUTO code \citep{PLUTO-Fargo}.
The algorithm in \app{} also can be employed in global calculations of accretion disk dynamics in cylindrical and spherical-polar coordinates.

\subsection{Tests of Non-Relativistic MHD Algorithms}

A comprehensive test suite of the MHD algorithms in \app{} is presented in S08 and will not be repeated here.  In this section we present test results only to demonstrate the properties of new algorithmic features in the code, such as the new reconstruction algorithms and time integrators.  We emphasize that whenever values for the magnetic field are listed, they are given in code units with magnetic permeability $\mu=1$.

\subsubsection{Linear Wave Convergence Test} \label{subsubsec:linwave}

\begin{figure*}[htb!]
\centering
\includegraphics[width=2.0\columnwidth]{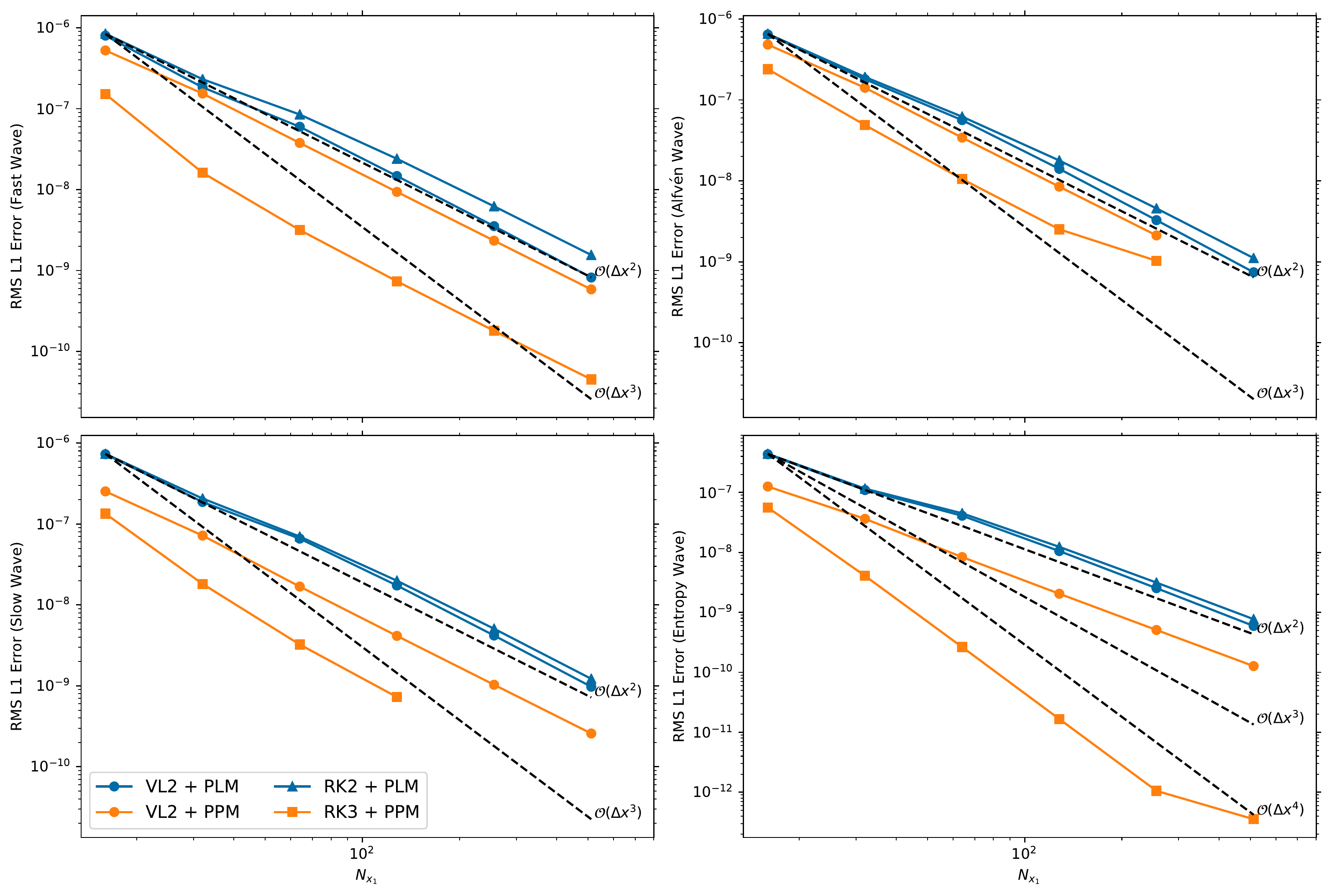}
\caption{MHD linear wave convergence plots produced using a variety of temporal integrators and variable reconstruction methods without mesh refinement. Clockwise from top-left:\ fast magnetosonic, Alfv\'{e}n, entropy, and slow magnetosonic wave modes of the linearized system.\label{fig:linear-waves}}
\end{figure*}

Measuring the convergence of linear waves provides a quantitative test of errors in the algorithm.  For this test, parameters similar to those used in the original \athena{} paper \citep{Stone+2008,gs08} are adopted. The box size is \((L_x, L_y, L_z) = (3.0, 1.5, 1.5)\), and a grid of \(2N \times N \times N\) cells is used with periodic boundary conditions. A plane wave with a perturbation wavelength \(\lambda = 1\) and amplitude \(A=10^{-6}\) is initialized propagating along the diagonal of the mesh.  Uniform grid resolutions ranging from \(N=16\) to \(N=256\) are adopted, and the error at each resolution is measured by the root mean square of the volume-weighted L1 norms of each variable as
\begin{equation}
    \langle E\rangle = \left[\sum_n \left(\frac{\sum |U_n - U_{n,{\rm exact}}|\Delta V}{\sum \Delta V}\right)^2\right]^{1/2},
    \label{eq:RMS-error}
\end{equation}
where \(U_n\) and \(U_{n,{\rm exact}}\) are the numerical and exact solutions of the \(n\)-th variable and \(\Delta V\) is the volume of a cell.

Figure~\ref{fig:linear-waves} displays the results for each different wave family (slow and fast magnetosonic, Alfv\'{e}n, and entropy waves) computed using different time integrators (both VL2 and RK3), and different spatial reconstruction algorithms (both PLM and PPM). In all cases the HLLD approximate Riemann solver is used.  As expected, strict second-order overall convergence is observed when either the VL2 time integrator or the PLM reconstruction method is used.  The error amplitudes are somewhat lower for each wave when the more accurate PPM reconstruction is used with the VL2 time integrator, although the convergence rate is still exactly second-order.  The most accurate combination of algorithms is clearly RK3+PPM.  Errors in the solution computed with this choice can be an order of magnitude or more lower than those produced by VL2 and PLM.  Moreover, for some wave families the convergence rate of the error is higher across a significant range of resolutions (close to third-order). Since the method does not possess formal third-order spatial accuracy in multidimensional problems, this likely indicates that temporal errors dominate in these cases.

\subsubsection{Linear Waves in Non-Ideal MHD} \label{subsubsec:linwave-damped}

\begin{figure}[htb!]
\centering
\includegraphics[width=\columnwidth]{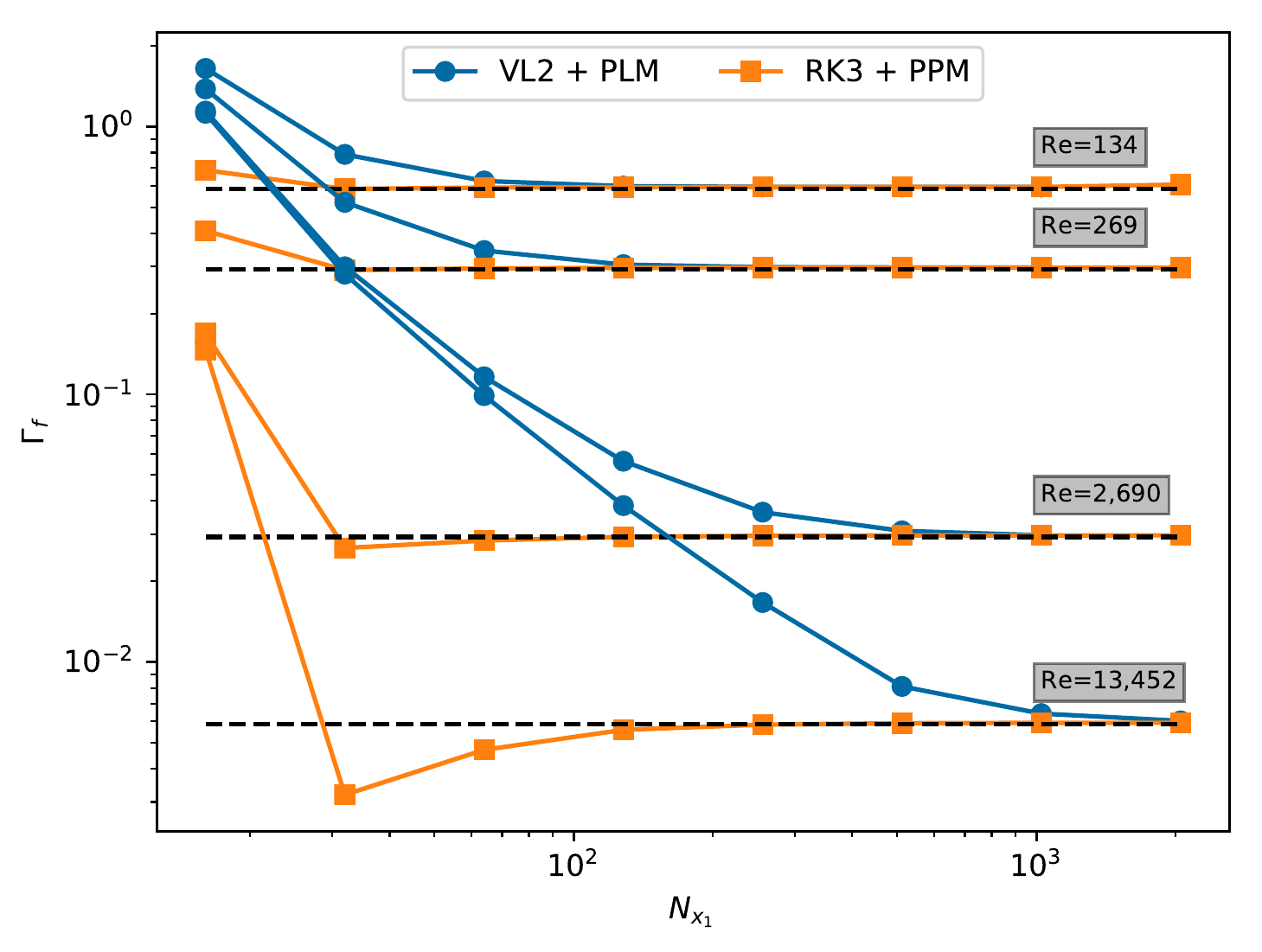}
\caption{Convergence to the analytic fast-mode decay rates (dashed black lines) for a range of Reynolds numbers spanning two orders of magnitude. The more accurate RK3+PPM solver produces linear wave decay rates significantly closer to the predicted values than VL2+PLM solutions at higher resolutions. \label{fig:fast-wave-decay}}
\end{figure}

\begin{figure}[htb!]
\centering
\includegraphics[width=\columnwidth]{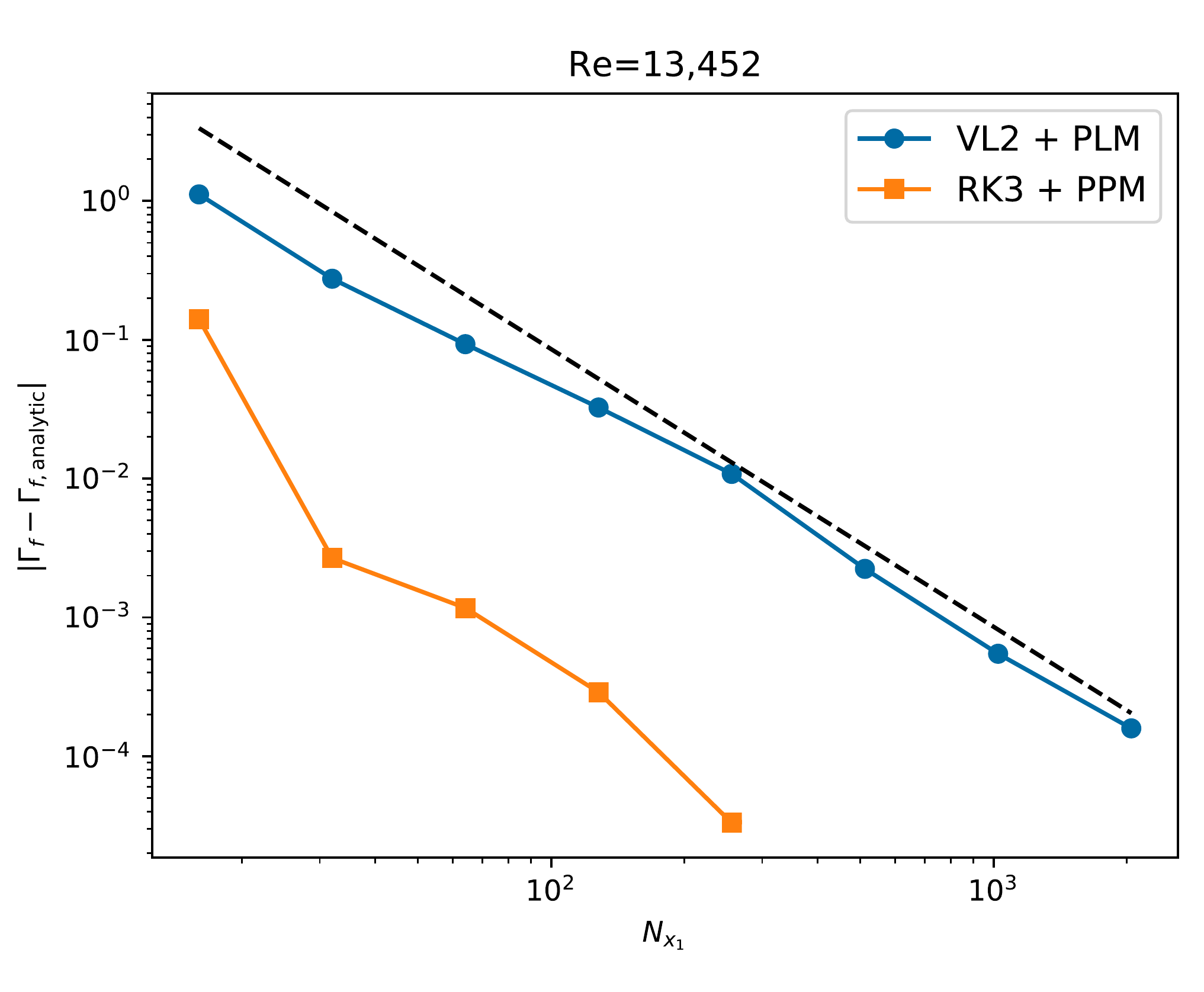}
\caption{Convergence of the \(L_1\) error of the fast-mode decay rate for the largest Reynolds number case shown in figure~\ref{fig:fast-wave-decay}. \label{fig:fast-wave-decay-convergence}}
\end{figure}

The MHD module in \app{} includes terms to model diffusive processes such as isotropic viscosity, resistivity, and thermal conduction.  To test these terms, a 2D variant of the linear wave problem described in the previous subsection is considered. The domain size is \( (2/\sqrt{5}) \times  (1/\sqrt{5}) \), and a linearized fast mode wave is initialized with \(\lambda = 1 \), at an angle \(\theta = \tan^{-1}(2) \approx 63.43^\circ \) inclined with respect to the \(x_1\) axis, and with a perturbation amplitude \( A = 10^{-4} \). The CFL number is set to 0.4, and the fast wave (with wavespeed \(c_f=2\)) is evolved to \(t=0.75\).  The explicit, unsplit algorithm for the diffusion terms is used.  Because an exact eigenmode of the non-ideal MHD wave equation is not initialized, at very high resolutions the error is limited by errors in the initial data.  We discuss this further below.

Kinematic viscosities ranging from $\nu = 10^{-2}$ to $10^{-4}$ are considered. The standard and magnetic Prandtl numbers are fixed to be \( \text{Pr} = \text{Pr}_\text{m}= 1/2 \) for all tests; that is, \(\kappa = \eta = \nu/2 \). The decay rate of the wave in each simulation is measured by applying weighted least-squares (WLS) fitting to the time series of \(\ln(\max(|v_2|)) \) in the solution.

Following \citet[][see equation 3.13]{RyuJonesFrank1995}, the decay rate of the fast wave (including thermal conduction) is
\begin{equation}
\Gamma_{f, \mathrm{ analytic}} = \left( \frac{19\nu}{4} + 3\eta + \frac{3\kappa(\gamma-1)^2}{4\gamma}\right) \frac{2k^2}{15}.
\end{equation}
As the authors note, this expression is applicable only up to first order in the diffusion coefficients, and in the limits
\begin{equation}
\nu k\text{ and } \eta k \ll c_f, c_A, c_s, \text{ or } a,
\end{equation}
where \(k = 2\pi\) here. For these parameters, the Reynolds number is defined as \citep[][equation 3.15]{RyuJonesFrank1995} 
\begin{equation}
R_f \equiv \frac{4\pi^2 c_f}{\lambda\Gamma_f} = \frac{8\pi^2}{\Gamma_f} 
\end{equation}

Figure~\ref{fig:fast-wave-decay} shows the convergence of the numerically measured decay rates to these analytic values over a wide range of Reynolds numbers as the spatial resolution of the mesh is increased. The analytic rates given by the above equation are juxtaposed as dashed black lines in all four cases. Excellent agreement is observed.

%For small Reynolds number flows, the treatment of the viscosity, thermal conduction, and resistivity terms (at second-order accuracy) dominates the timestep restriction. As discussed in section~\ref{subsubsec:diffusion}, \app{} offers super-time-stepping for these parabolic terms in order to avoid the need for an excessive number of small timesteps in these cases. For large Reynolds number flows, a fine spatial resolution is required to resolve the slow diffusion processes and correctly model the predicted decay rate. 

Figure~\ref{fig:fast-wave-decay-convergence} demonstrates second-order convergence with mesh resolution of the decay rate at a single fixed Reynolds number (the largest value considered in the previous plot). As is evident in the previous figure, the RK3+PPM configuration initially converges to the analytic decay rate much more quickly than the formally second-order accurate VL2+PLM solver.  Below values of about $10^{-4}$, the error is dominated by the initial conditions since a wave solution for the ideal (rather than non-ideal) MHD equations is used.  Thus, the errors stop converging beyond the values shown in the plot.  

\subsubsection{Riemann Problems}

In order to test the MHD algorithms with nonlinear solutions, we present the results from multiple shock tube (Riemann) problems.  In all cases, the problems are calculated in one dimension along the $x_1$ axis (we have tested the code generates identical solutions when the tests are run along the $x_2$ or $x_3$ axes, and in \citet{Stone+2008} we have shown the results for shock tubes run along a grid diagonal in multidimensions).  While a huge range of such Riemann problems are available for testing, we focus on two that demonstrate key features of the algorithms: the Shu-Osher problem in hydrodynamics and the Brio-Wu problem in MHD.

\begin{figure}[htb!]
\centering
\includegraphics[width=\columnwidth]{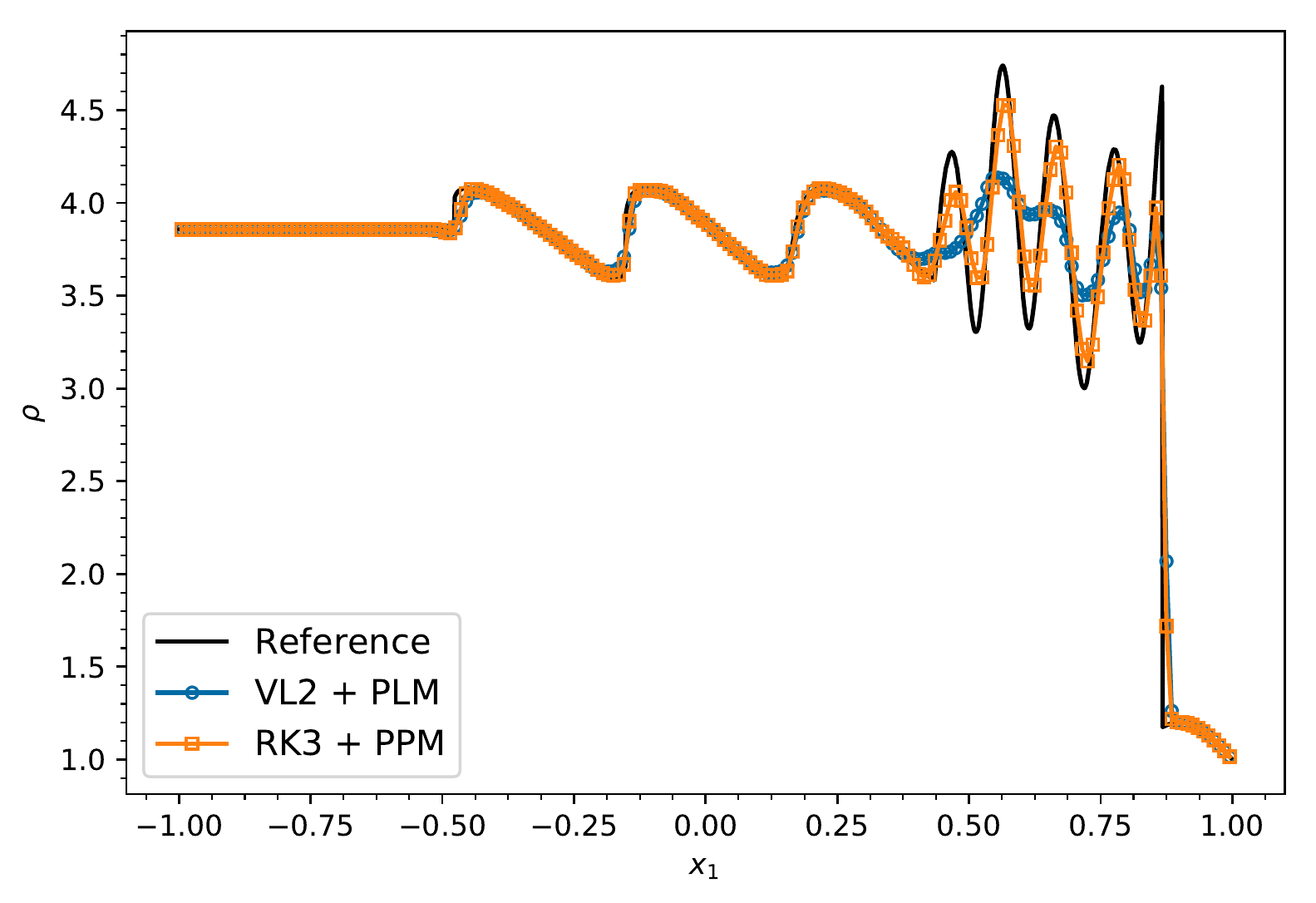}
\caption{Density in the Shu-Osher hydrodynamic shock tube test at \(t=0.47\) for \(N=200\) cells. The reference solution was computed using RK3+PPM with \(N=8192\) cells. \label{fig:shu_osher}}
\end{figure}

Figure~\ref{fig:shu_osher} shows the density at $t=0.47$ for the Shu-Osher shock tube problem \citep{ShuOsher}, which involves the interaction of a shock with a smoothly varying background medium. The ability to resolve the sharp features formed by shock compression is a measure of the numerical diffusion in the scheme.  Results for both the VL2+PPM and RK3+PPM solver configurations using the HLLC Riemann solver and $N=200$ cells are shown along with a reference solution computed using $N=8192$ cells and the RK3+PPM algorithm. It is clear that the RK3+PPM method is able to capture the short wavelength oscillations present in the density profile using only a few cells, and it significantly outperforms the VL2+PLM method on this test.

\begin{figure*}[htb!]
\centering 
\includegraphics[width=2.0\columnwidth]{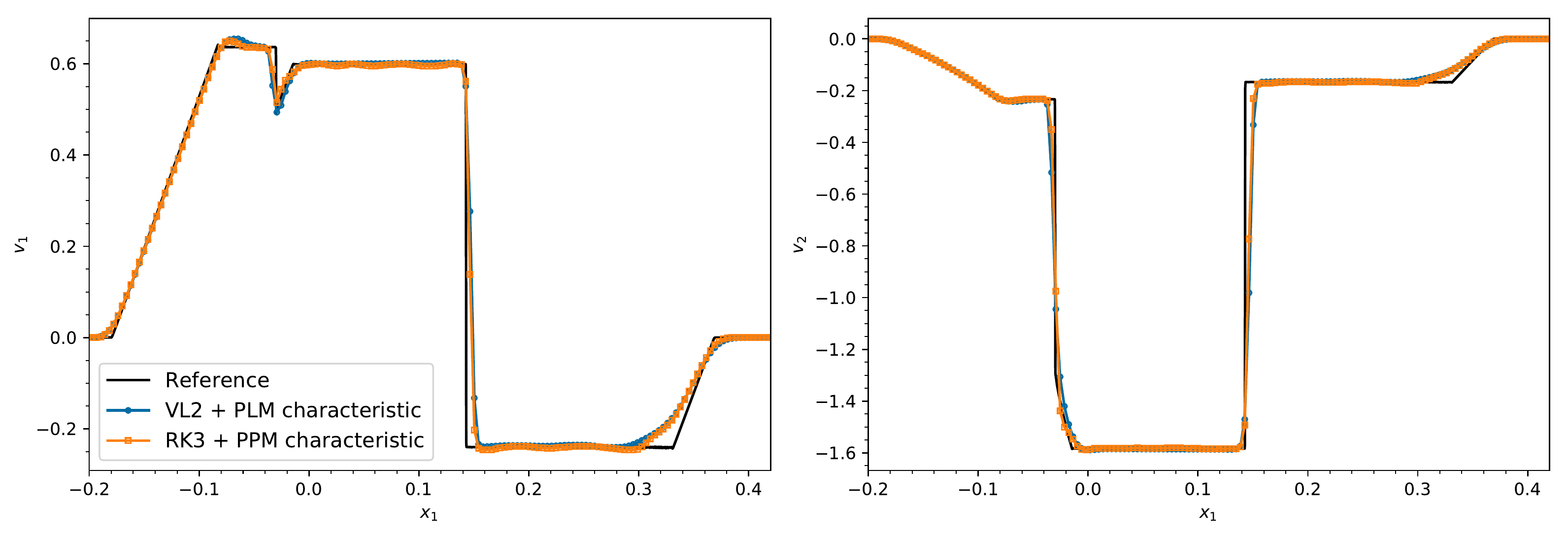}
\caption{Longitudinal and transverse velocity solutions in the Brio-Wu MHD shock tube test at \(t=0.1\) for \(N=256\) cells. The reference solution was computed using RK3+PPM reconstruction on characteristic variables with \(N=8192\) cells. \label{fig:bw_shock}}
\end{figure*}

Figure~\ref{fig:bw_shock} compares the results of the Brio-Wu MHD shock tube test \citep{BrioWu} at time $t=0.1$ for the same two algorithms but with the HLLD Riemann solver.  In this test, reconstruction is performed using the characteristic rather than the primitive variables.  If the latter approach is used, the solver produces significant oscillations behind the right-moving fast rarefaction.  For this reason, this test is an important demonstration of the need for characteristic reconstruction for certain problems.  The results show little difference between the two algorithms:\ RK3+PPM captures the head and foot of rarefactions slightly more accurately.  However, both methods perform well for solutions involving MHD shocks and rarefactions in each wave family.

\subsubsection{Oblique C-Shock with Ambipolar Diffusion}

\begin{figure}[htb!]
    \centering
    \includegraphics[width=0.8\columnwidth]{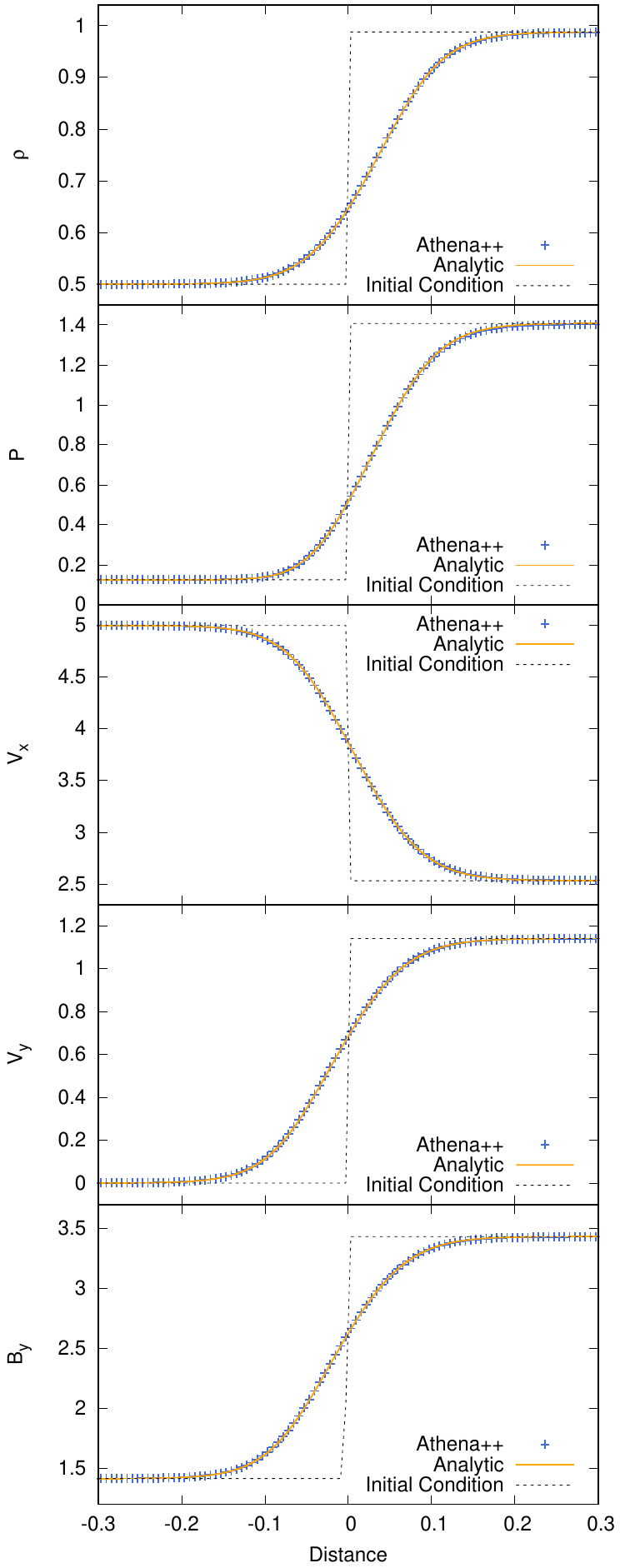}
    \caption{Steady-state solution in the adiabatic oblique C-shock test. \label{fig:adcshock}}
\end{figure}

In addition to Ohmic resistivity, the core MHD module in \app{} includes terms to model ambipolar diffusion.  To test this term, results for an oblique C-shock test are presented.  The test is identical to the problem described in \citet{Masson2012} \citep[see also][]{Wardle1999}. An adiabatic EOS is used in order to test the heating and energy flux associated with ambipolar diffusion as well. Like the shock-tube test, the left state is \((\rho, v_x, v_y, B_x, B_y, P)\) \(=(0.5, 5, 0, \sqrt{2}, \sqrt{2}, 0.125)\) and the right state is \((\rho, v_x, v_y, B_x, B_y, P) = (0.9880,\allowbreak 2.5303,\allowbreak 1.1415,\allowbreak \sqrt{2},\allowbreak 3.4327,\allowbreak 1.4075)\). A density-dependent ambipolar diffusion coefficient $\eta_{\rm AD}=1/(75\rho)$ is used. In order to reduce the symmetry in the problem, a two-dimensional domain of $[-0.5, 0.5] \times [-0.0078125, 0.0078125]$ with resolution of 1/128 is used, with the initial interface rotated by an angle $\theta=\tan^{-1}(3/4)$ using shifted-periodic boundary condition in the \(y\)-direction \citep[see][Appendix A.6]{Tomida2015}. The boundary conditions in the \(x\)-direction are both outflow. Starting from the initial discontinuity, the problem is run until $t=10$ so that the shock profile reaches a steady state.  The profile along the shock propagation direction is shown in figure~\ref{fig:adcshock}.  Even at this relatively low resolution, \app{} successfully reproduces the analytic solution.

\subsubsection{Liska--Wendroff Implosion} \label{subsubsec:lw-implode}

The implosion test discussed in \S 4.7 of \citet{LiskaWendroff2003} and first introduced in \citet{Hui1999} is an extraordinarily sensitive test of the directional symmetry-preserving abilities of a hydrodynamics code. 
The initial condition consists of two uniform states separated by a diagonal discontinuity near the bottom left corner of the domain, with the jumps in the variables identical to those in the familiar Sod shock tube test \citep{Sod1978}; see table 1 in \citet{Stone+2008} for the precise values.  Reflecting boundary conditions are used on all four sides.
A shock wave launched by the high pressure region is reflected by the bottom and left boundaries, generating narrow jets of gas characteristic of double Mach reflections \citep{wc84}. Refer to section~\ref{subsubsec:dmr} for the full double Mach relection test. The resulting two jets collide at the lower-left corner, and launch two vortices and a single, narrow jet of low density gas along the grid diagonal.  As the evolution progresses, reflected shocks interact with the contact discontinuity and seed the growth of fingers via the Richtmeyer--Meshkov instability.  The key ingredient of the test is that the jet will not propagate
exactly along the domain diagonal unless the solver maintains reflective symmetry to machine precision across this plane.

\begin{figure}[htb!]
    \centering
    \includegraphics[width=\columnwidth]{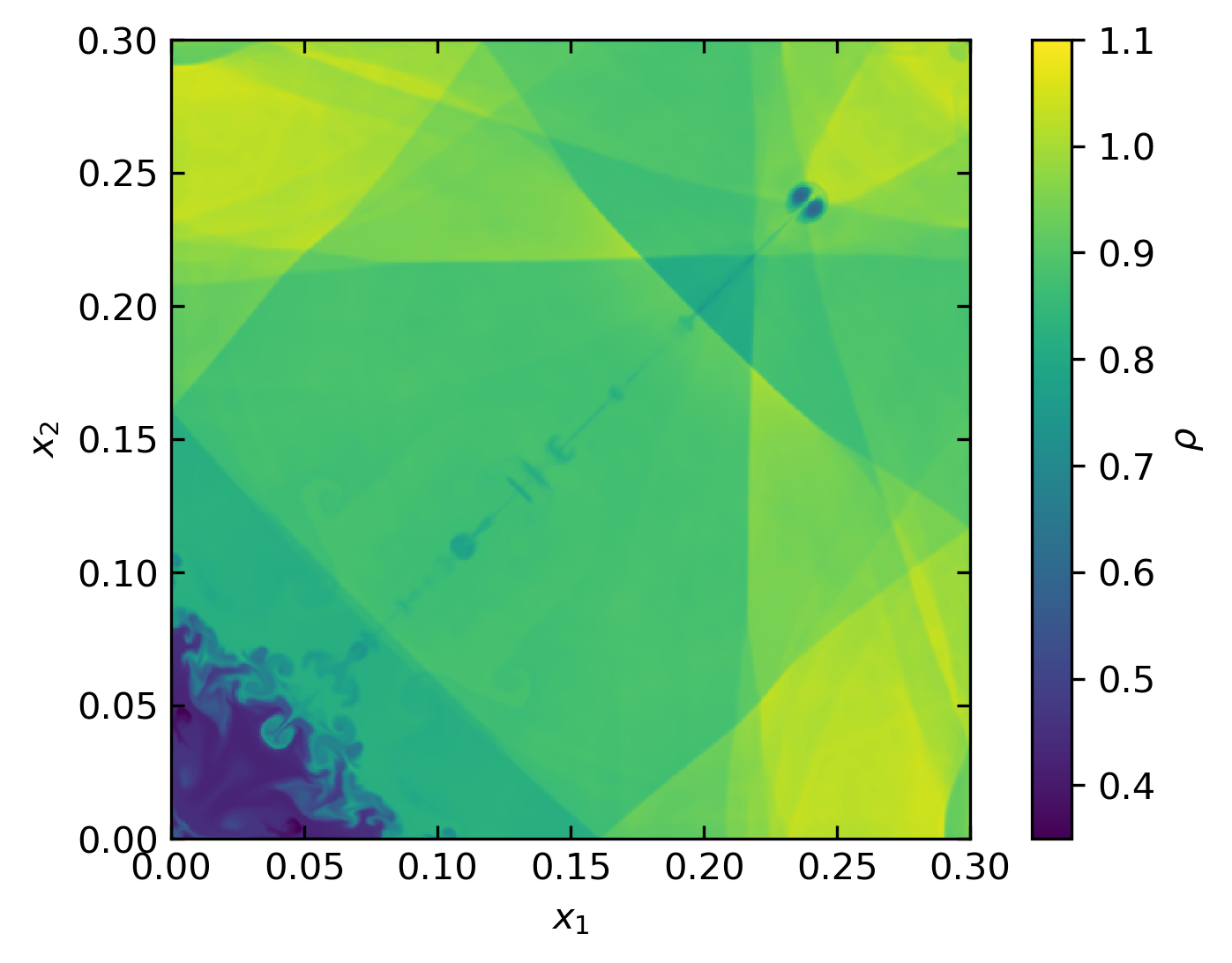}
    \caption{Density in the LW implosion test at \(t=2.5\) using RK3+PPM. Exact symmetry is maintained along the diagonal, and a low density jet is produced there as a consequence.
    % a linear color map between \(\rho=0.35\) and \(
    \label{fig:lw-implode}}
\end{figure}

Figure~\ref{fig:lw-implode} shows the density at \(t=2.5\) for a \(512 \times 512\) mesh. PPM reconstruction of the characteristic variables was used in conjunction with the HLLC Riemann solver and the RK3 timestepper. The results are perfectly symmetric to double precision machine epsilon for all output variables. Symmetry is maintained for all resolutions and solver permutations that we applied to this problem.

% https://www.astro.princeton.edu/~jstone/Athena/tests/implode/Implode.html

Achieving exact symmetry on this problem is in fact extremely challenging.  The PPM reconstruction algorithm is particularly sensitive; the non-associativity of floating-point arithmetic necessitates that the stencils are written in the C++ source code such that they are calculated without a directional bias.
The use of MPI or compiler options that do not guarantee a value-safe floating-point arithmetic mode break \app's ability to preserve directional symmetry in this test. In order to produce the results shown in figure~\ref{fig:lw-implode} with the Intel C++ compiler, options to disable reassociation of operands and contractions of expressions into fused multiply--add (FMA) operations were both required.

Finally, we have also confirmed that when run with AMR, exact symmetry is preserved for this problem, although the details of the solution (for example the strength of the vortices that produce the jet) depend on the refinement condition adopted.  This is similar to the behavior on a uniform grid; lower resolution produces a weaker vortices and a shorter jet.

\subsubsection{Kelvin--Helmholtz Accuracy Benchmark} \label{subsubsec:khi-benchmark}

Finally, to benchmark these algorithms for hydrodynamics against known reference solutions, we consider the Kelvin--Helmholtz (KH) instability test described in \citet{Lecoanet2015}.  This paper described a well-posed benchmark problem, presented resolved reference solutions computed using the pseudo-spectral code \texttt{Dedalus} \citep{Burns2019}, and compared these results to those produced by the original C-version of \athena.  In this section, we reproduce the analysis of \citet{Lecoanet2015}, and compare the results from \app{} to \texttt{Dedalus} (and therefore \athena, as well).

The stratified variant of the problem considers an initial condition with a smooth transition of \(\Delta\rho/\rho_0 = 1\) between the two shearing layers and results in behavior that is challenging for a numerical method to resolve with respect to instabilities and small-scale structure. The authors found that \athena{} (C) required a resolution of \(16384 \times 32768\) cells in order to converge to the same solution that \texttt{Dedalus} achieved with \(2048 \times 4096\) Fourier modes.  

When repeating the tests and comparing the results from \athena{} (C) with those from \app{}, it is worth keeping in mind several key algorithmic differences between the two codes. The results presented in \citet{Lecoanet2015} were generated by the \athena{} (C) code using the CTU integrator combined with PPM reconstruction of the characteristic variables (although the authors found that other algorithmic options produced similar results), and the diffusion terms were applied at first-order accuracy in time using operator-splitting. In contrast, \app{} computes the diffusion processes in an \emph{unsplit} fashion, and does not implement the CTU integrator. All of the \app{} results shown in this section were produced with reconstruction on primitive hydrodynamic variables and the HLLC Riemann solver. 

Explicit diffusion is added via isotropic fluid viscosity \(\nu\), thermal conduction \(\kappa\), and a separate passive dye diffusion process \(\nu_{\text{dye}}\). For the test shown in this section, \(\nu = \kappa = \nu_{\text{dye}} = 2\times 10^{-5}\) corresponding to a Reynolds number \(\mathrm{Re} = 10^5\). The CFL number used for the \app{} tests is \(C_0 = 0.4\). 

\begin{figure*}[htb!]
\centering
% 4x3 plot with just RK3+PPM at A4096 and A8192 vs. D4096
%\includegraphics[width=15cm]{figures/stratified_khi_snapshot_grid_RK3_PPM_D4096.png}
% 4x5 plot with VL2+PLM added to the above plot
\includegraphics[width=1.0\textwidth]{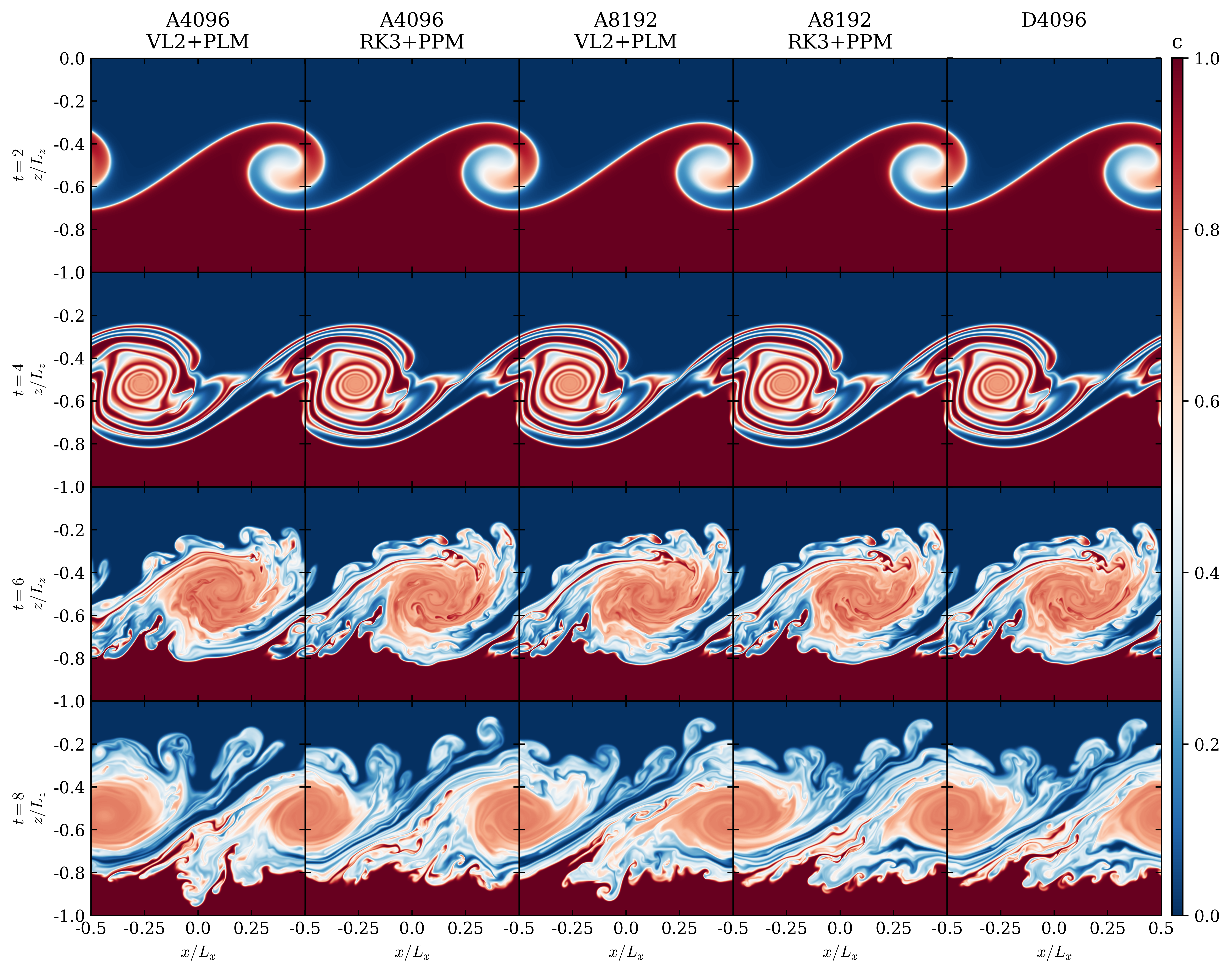}
\caption{Snapshots of solution to KH instability problem with \app{} and \texttt{Dedalus} at various resolutions and times. Compare to figure~8 of \citet{Lecoanet2015}. \label{fig:khi-lecoanet}}
\end{figure*}

Figure~\ref{fig:khi-lecoanet} plots the dye field of the lower half of the domain at \(t=2,4,6,8\) for \app{} VL2+PLM and RK3+PPM at various resolutions. As in \citet{Lecoanet2015}, the columns are labeled with `A' for \app{} or `D' for \texttt{Dedalus} and the \(N\) degrees of freedom in the horizontal direction. The \texttt{Dedalus} results shown were produced from same data as the original study, which was furnished by the authors of \citep{Lecoanet2015}.

The results in figure~\ref{fig:khi-lecoanet} compare very favorably to the original \athena{} (C) results. Note that only \(8192\times 16384\) cells are required to converge to the \texttt{Dedalus} reference solution when RK3+PPM is used with \app{}, which is half the resolution required in the original study.  An important contribution to this improvement is the use of an unsplit algorithm for the diffusion terms. The \(4096\times 8192\) second-order VL2+PLM solution suffers from the onset of the inner vortex instability (IVI) at \(t=4\), albeit at a much smaller amplitude than the A4096 CTU+PPM results from \citet{Lecoanet2015} figure~8. Both A4096 RK3+PPM and A8192 VL2+PLM avoid the onset of the IVI, although these solutions still exhibit visible differences from D4096 in the filament structure at \(t=6\). However, the A4096 RK3+PPM solution qualitatively appears very close to the converged solution. A detailed comparison of the results, along with quantitative study of the errors between solutions is provided in \citet{Felker2019-PhD}.  A further notable result is that due to the much higher computational performance of finite volume compared to spectral methods, the A8192 solution took only one half of the time required to compute the D4096 solution.  Thus, \app{} achieves spectral accuracy for this problem at less cost.

\subsection{Tests of AMR with MHD}

Next, we present the results for a series of test problems that demonstrate the accuracy of our AMR methods.

\subsubsection{Linear Wave Convergence with AMR}

Locally refined grids should produce more accurate solutions than a uniform resolution root grid, and the global convergence rate on AMR grids should be second-order. To test these expectations, the MHD linear convergence test can be used to provide quantitative measures of the errors and convergence rate of solutions on an AMR grid in \app{}.

The test is identical to that already presented in section~\ref{subsubsec:linwave-damped} for a uniform grid.  Results with the same range of resolutions from \(N=16\) to \(N=256\) are presented; however with AMR one additional finer level (at twice higher resolution per dimension) is introduced in regions where the density is within 90\% of the peak value.  Note that this refinement condition was chosen solely for demonstrating the behavior of the AMR, and it was not motivated by any physical requirements. Each MeshBlock consists of \(4^3\) cells for the lowest resolution run and \(64^3\) cells for the highest, so that the refined regions occupy the same volume.  As before, errors are measured using the root mean square of the volume-weighted L1 norms of each variable (equation \ref{eq:RMS-error}).
The VL2 time integrator and both the PLM and PPM reconstruction algorithms are used for comparison.  

\begin{figure}[htb!]
\centering
\includegraphics[width=\columnwidth]{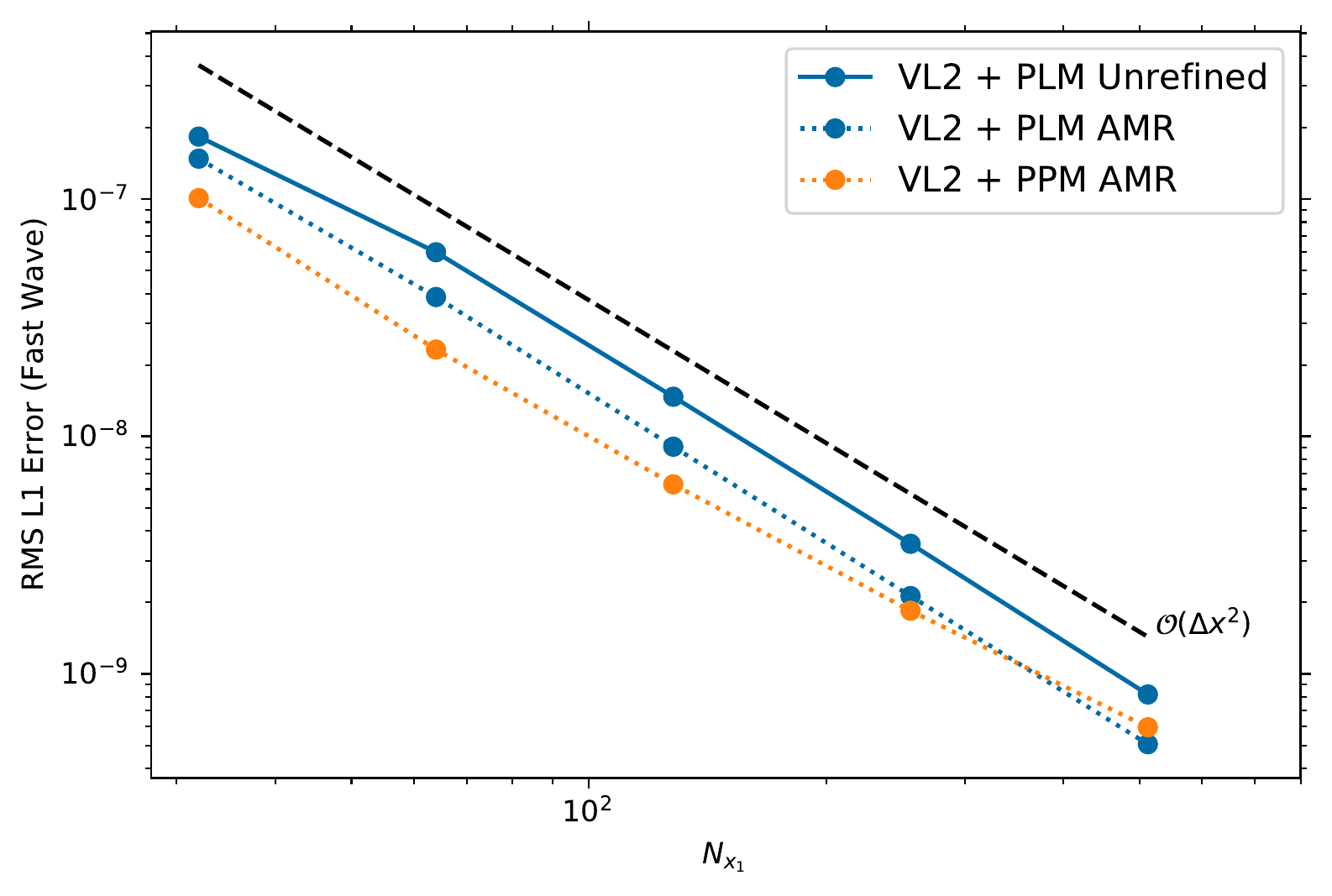}
\caption{Errors in a linear wave convergence test with and without AMR.  Results are shown for the fast wave, but other modes show similar trends.  Second-order convergence is achieved in all cases. \label{fig:linear-wave-amr}}
\end{figure}

The results for the fast wave are shown in figure~\ref{fig:linear-wave-amr}; the other waves behave similarly. As expected, both unrefined and AMR simulations achieve global second-order accuracy, and the AMR simulations exhibit slightly better error than the unrefined grid simulations with the same root grid resolution. As in figure~\ref{fig:linear-waves}, using PPM with the van~Leer integrator yields a lower spatial error for most of the resolutions. However, at the largest resolution, the second-order accurate truncation errors of the AMR prolongation and restriction operators dominate the higher-order terms associated with the PPM reconstruction.  This plot is extremely informative, and clearly demonstrates second-order convergence of global errors is achieved with the AMR algorithm in \app{}.

Additional computational expense is incurred when enabling AMR due to the addition of refinement, derefinement, prolongation, and restriction operations. Although more cells are added when the grid is refined, the overall efficiency of calculating the solution for a single cell remains high. At the highest tested root grid resolution of \(512\times 256 \times 256\) cells, the second-order solver advances the unrefined mesh at 132.5 million zone-cycles/second when deployed with MPI on four dual-socket nodes of an Intel Skylake system. When AMR capabilities are enabled, the solver slows to 106.0 million zone-cycles/second, representing a performance overhead of about 20\%. This relative performance cost is larger at lower resolutions, but it is quickly amortized by increasing the size of the blocks. Furthermore, the efficiency does not decline as more levels are added or as the refined region grows. However, we emphasize that the overhead of AMR depends on many factors such as the volume of the refined region, frequency of refinement operations, and the size of MeshBlocks, and therefore it is highly problem dependent (see discussion of results from other tests below). 

Note that the linear wave convergence test is not only simple but also highly sensitive to most subtle defects in the code. For example, if boundary communication between levels is implemented incorrectly, the AMR calculation will have a larger error than a uniform grid at the resolution of the root level.  Moreover, even if only one boundary cell is communicated incorrectly (for example at the edge or the corner of the MeshBlock, see section~\ref{subsec:level_comm}), this will be evident through the lack of convergence of the $L_{\infty}$ error.

\subsubsection{Double Mach Reflection Test\label{subsubsec:dmr}}

The double Mach reflection problem \citep{wc84} is a standard test for hydrodynamics codes. It involves a Mach 10 shock which reflects from an inclined plane.  This interaction produces complex structures such as discontinuities, a triple point, and a jet. Therefore, this is a good problem for evaluating the correctness of the AMR implementation and the robustness of the code with shocks.

For this test, characteristic reconstruction is used in order to suppress numerical oscillations produced at the strong shock, and the HLLE approximate Riemann solver is chosen in order to suppress the Carbuncle-like instability at the head of the jet \citep{rage}. The H-correction scheme in the \athena{} code \citep{Stone+2008} suppresses these instabilities; however, it has not yet been implemented in \app{}. The initial and boundary conditions are given in \citet{wc84}. For the uniform grid simulation, the resolution is \(\Delta x = 1/120\). For the AMR simulation, the root grid is set to be four times coarser (\(\Delta x = 1/30\)) and up to two finer levels are used so that the finest structures are captured with the same resolution as the uniform grid. Each MeshBlock has \(6\times6\) cells. A refinement condition based on the second spatial derivatives (i.e. curvature) is used, as in \citet{sfumato}:
\begin{equation} \label{eq:curvature}
    \epsilon=\max\left(\frac{|\partial_x^2q_{i,j}+\partial_y^2q_{i,j}|\Delta x^2}{q_{i,j}}\right),
\end{equation}
where \(q_{i,j}\) is a quantity such as density or pressure. A MeshBlock is flagged to be refined when \(\epsilon\) exceeds 0.01 and derefined when \(\epsilon\) falls below 0.005. 

\begin{figure}[htb!]
\includegraphics[width=\columnwidth]{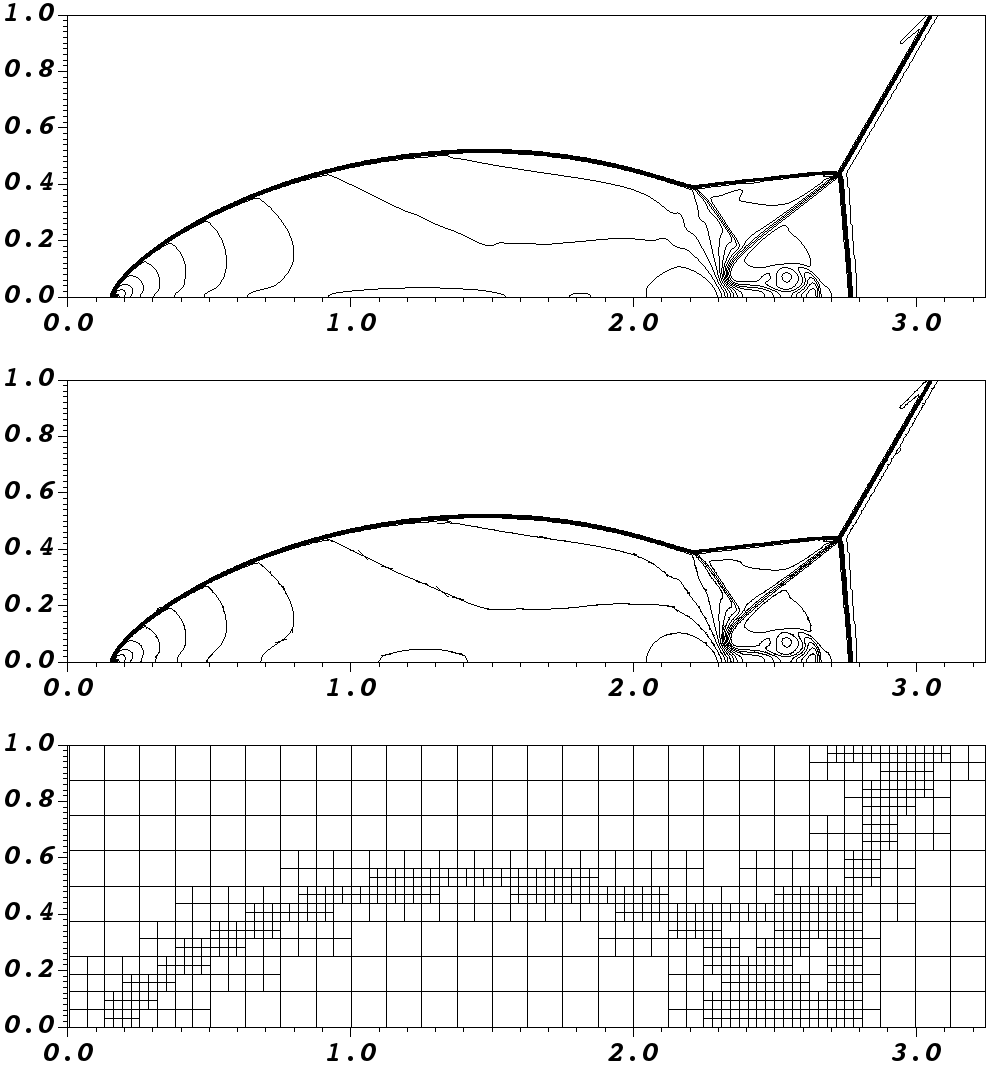}
\caption{Double Mach reflection test with a uniform grid (top) and AMR (middle) using the same effective resolution. The density at \(t=0.2\) is shown with 30 levels of contours. The bottom panel shows the distribution of MeshBlocks. \label{fig:dmr}}
\end{figure}

The results are shown in figure~\ref{fig:dmr}. The AMR grid produces results that are essentially indistinguishable from those on a uniform mesh.  The lower panel shows the distribution of MeshBlocks in the AMR calculation.  A relatively small volume of the domain requires the finest resolution, so our block-based AMR algorithm remains fairly efficient for this problem.  In particular, a uniform grid tiled with $6^2$ MeshBlocks requires 279 CPU seconds on a single core of a Skylake 6148 processor, whereas the AMR run using the same sized MeshBlocks required only 63.2 CPU seconds, for a speedup of 4.4.  It must be noted, however, that performance depends strongly on MeshBlock size (see sections~\ref{sec:AMRStrongScaling} and \ref{sec:MeshBlockSize-Performance}).  For example, doubling the size of the MeshBlocks to $12^2$ decreases the run times to 127 and 31.8 CPU seconds for the uniform mesh and AMR runs respectively, and on a uniform grid with a single MeshBlock the run time is only 42.8 CPU seconds.  Even though with AMR using larger MeshBlocks results in the finest level covering a larger fraction of the domain (making the calculation less efficient by this measure), nevertheless the time to solution is decreased.

\subsubsection{Kelvin--Helmholtz Tests}\label{subsubsec:khi-tests}

To further compare the accuracy and costs of solutions computed with an AMR grid with those using a uniform mesh, results from a series of KHI tests in both hydrodynamics and MHD are presented. The computational domain is chosen to span \(-0.5<x<0.5\) and \(-0.5<y<0.5\) with periodic boundary conditions in both \(x\)- and \(y\)- directions. A shear flow with a density contrast of two and a velocity jump of one is initialized, using a smooth (resolved) interface so that the profiles of density and velocity follow
\begin{subequations} \begin{gather}
    \rho=1.5-0.5\tanh\left(\frac{|y-0.25|}{L}\right), \\
    v_x=0.5\tanh\left(\frac{|y-0.25|}{L}\right), \\
    v_y=A\cos(4\pi x)\exp\left[-\frac{(y-0.25)^2}{\sigma^2}\right].
\end{gather} \end{subequations}
Here, \(L = 0.01\) is the thickness of the shearing layer, \(A = 0.01\) is the amplitude of the initial perturbation with a wavelength of 0.5, and \(\sigma = 0.2\) is the thickness of the perturbed layer.
The total pressure is constant everywhere and equal to \(p=2.5\), with adiabatic index $\gamma=1.4$.  This gives a sound speed \(C^2 = 3.5\) in the lowest density region.  The use of a smooth initial profile for the interface rather than a discontinuity is crucial for obtaining a well-posed problem that converges with resolution \citep[e.g.][]{McNally12}.
 
For the MHD test, a uniform horizontal field of \(B_{x}=0.1\) is added. The HLLD flux for MHD, HLLC flux for hydrodynamics, PLM reconstruction, and VL2 integrator are all used. The problem is run first with a uniform grid of \(2048\times 2048\), and then the calculation is repeated with AMR using four levels so that the same maximum resolution is achieved as the uniform mesh when the root grid resolution is \(256\times 256\). MeshBlocks of size \(8^2\) and \(16^2\) are used with a refinement condition based on the velocity shear
\begin{equation}
    g=h\times \max \left(\partial_x v_y, \partial_y v_x \right) \, .
\end{equation}
A MeshBlock is refined if \(g\) is larger than 0.01 or derefined if \(g\) is smaller than 0.005.

% KT: the figures might be too large
\begin{figure*}[htb!]
\centering
\includegraphics[width=15cm]{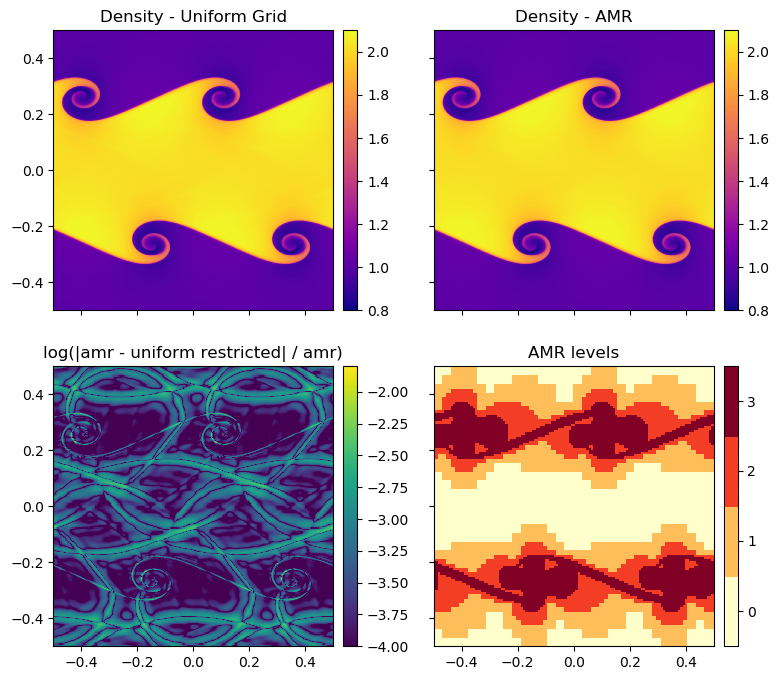}
\caption{Hydrodynamic Kelvin--Helmholtz instability test with AMR. The top left panel shows the density at \(t=1.2\) using a \(2048\times 2048\) uniform grid, while the top right panel is the result with the same effective resolution using 4 levels of AMR with MeshBlocks of \(8^2\). The two are indistinguishable. The bottom left panel shows the fractional difference in density between the uniform grid and AMR runs. The bottom right panel shows the distribution of MeshBlocks in the AMR run. \label{fig:hd_kh}}
\end{figure*}

The results for the hydrodynamic test are shown in figure~\ref{fig:hd_kh}.  The density (shown in the top panels) in the AMR and uniform grid runs is indistinguishable.  The fractional difference in the density between the two calculations, shown in the lower left panel, is more illsutrative.  It is dominated by short wavelength sound waves that
are damped in the low resolution (coarse mesh) regions of the AMR calculation.  Very narrow features that follow the cat's eye rolls produced by the KHI are barely discernible.  They are associated with slight (less than one grid cell) differences in the positions of the interfaces in the two calculations.  It is likely that such differences are unavoidable, since the interaction of the sound waves that cannot be represented in the AMR calculation (but are present on the uniform grid) with the interfaces can produce differences of the observed magnitude.  This is an interesting lesson on the limitations of AMR.  If the dynamics of these waves are important (for example, body modes of the KHI in astrophysical jets, e.g. \citet{hardee79}), then AMR cannot be used for the problem in this way.  The lower left panel in figures~\ref{fig:hd_kh} shows that the volume filling factor of MeshBlocks at the finest level in the AMR solution is relatively small.

\begin{figure*}[htb!]
\centering
\includegraphics[width=15cm]{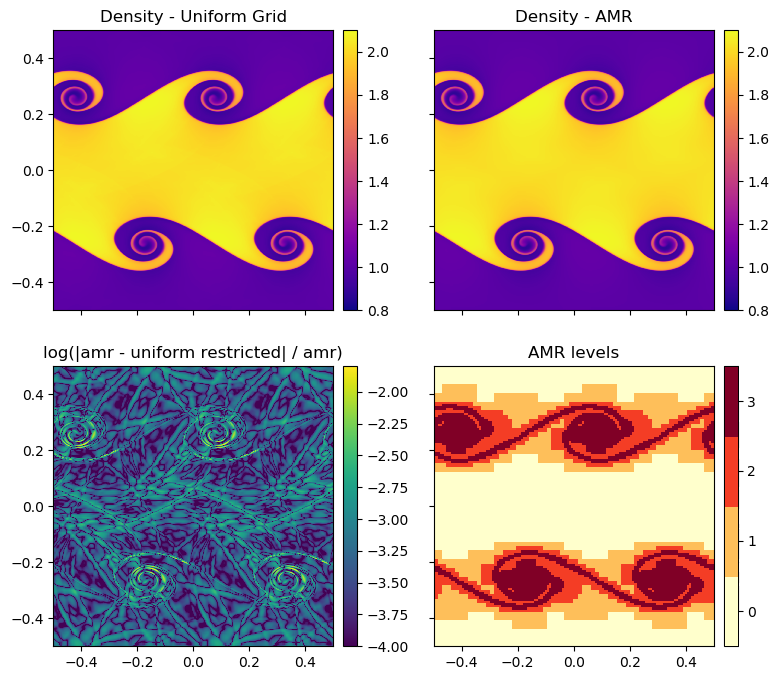}
\caption{Same as figure~\ref{fig:hd_kh} but for the MHD Kelvin--Helmholtz instability at \(t=1.5\). \label{fig:mhd_kh}}
\end{figure*}

The results for the MHD test are shown in figure~\ref{fig:mhd_kh}. The results and conclusions are nearly identical to those for the hydrodynamic version of this problem.  The fractional density difference shown in the lower left panel reveals a more intricate pattern in the MHD calculation because it consists of both fast and slow modes that are both damped.  Taken together, figures~\ref{fig:hd_kh} and \ref{fig:mhd_kh} show that AMR is able to capture the dynamics of the KHI on isolated interfaces very successfully.

The fractional difference in the density between uniform grid and AMR calculations, computational time, performance, and number of cells per calculation are summarized in table~\ref{tbl:khsummary}. These performance measurements include file outputs every \(\Delta t=0.01\). The largest differences emerge mainly at the discontinuities, because even a tiny phase error can produce large pointwise differences. The computational time and performance are measured using 32 cores (16 cores per socket) of an Intel Skylake Xeon 6148 node. While the computational throughput considerably degrades when AMR is in use (\(\sim 40\%\) with \(16^2\) and \(\sim 18\%\) with \(8^2\)), it still reduces the overall computational cost and data size. For this specific test, MeshBlocks of \(16^2\) are optimal, but in other problems this size should be chosen carefully based on the required accuracy and efficiency.

\begin{deluxetable*}{ccccccc}  % does not accept p{} alignment
% deluxetable preamble commands:
\tablewidth{\textwidth}
\tabletypesize{\small}
\tablecolumns{7}
 \tablecaption{Summary of AMR KHI test accuracy and efficiency\label{tbl:khsummary}}
 \tablehead{ & \colhead{\thead{Grid}} & \colhead{\thead{Maximum density \\difference (\%)}} & \colhead{\thead{Mean density \\difference (\%)}} & \colhead{\thead{Time \\ (s)}} & \colhead{\thead{Performance \\(MZone-cycles/s)}} & \colhead{\thead{Number of Cells \\($10^6$)}}}

\startdata
\multirow{3}{*}{Hydro} & Uniform & --- & --- & 319 & 270.0 & 4.19\\
                       & AMR $8^2$ & 0.70 & 0.046 & 319 & 46.7 & 1.09\\
                       & AMR $16^2$ & 0.80 & 0.044 & 152 & 108.5 & 1.25\\
\cmidrule(lr){1-7}
\multirow{3}{*}{MHD} & Uniform & --- & --- & 883 & 118.7 & 4.19\\
                     & AMR $8^2$ & 1.59 & 0.060 & 688 & 21.4 & 0.90\\
                     & AMR $16^2$ & 1.27 & 0.056 & 387 & 49.1 & 1.16\\
\enddata
\end{deluxetable*}

\subsubsection{3D Blast Wave Tests\label{subsubsec:cartesian-blast}}

The Sedov-Taylor solution \citep{sedov1946,taylor1950} provides the basis for useful quantitative tests involving the propagation of blast waves. In order to demonstrate the AMR capabilities of \app{} in 3D, we have performed 3D blast wave tests with and without magnetic fields.

For both non-magnetized and magnetized models, the same initial condition (apart from the magnetic field) is used. The computational domain spans a cubic region with edge length \(L=1\) and periodic boundary conditions on all faces. The initial density is set to one, while the pressure is 0.001 everywhere. To initialize the blast wave, the total internal energy in a region a radius of 0.01 at the center of the domain $E_{\rm tot} = \int dV P/(\gamma-1)=1$ with $\gamma=5/3$, giving a pressure of $1.6\times10^5$ in this region.  For the MHD version of the problem, the magnetic field is uniform and inclined to the grid:\ \(B_x=\sqrt{3}\) and \(B_y=1\).

The VL2+PLM algorithm is used for both models, along with the HLLE solver for hydrodynamics (to suppress the Carbuncle instability) and the HLLD solver for the MHD simulation. A refinement condition based on the pressure jump is used:
\begin{equation}
    g=h\times \max\left(\frac{|\nabla p|}{p}\right).
\end{equation}
A MeshBlock is refined when \(g\) exceeds a threshold value (0.1 in hydrodynamics and 0.2 for MHD) and is flagged for derefinement if \(g\) is smaller than 1/4 of this value.  The root grid consists of \(128^3\) cells with two additional levels of refinement, resulting in an effective resolution of \(512^3\). For comparison, the result from a uniform grid calculation using \(512^3\) cells is also shown.

\begin{figure*}[htb!]
\centering
\includegraphics[width=\textwidth]{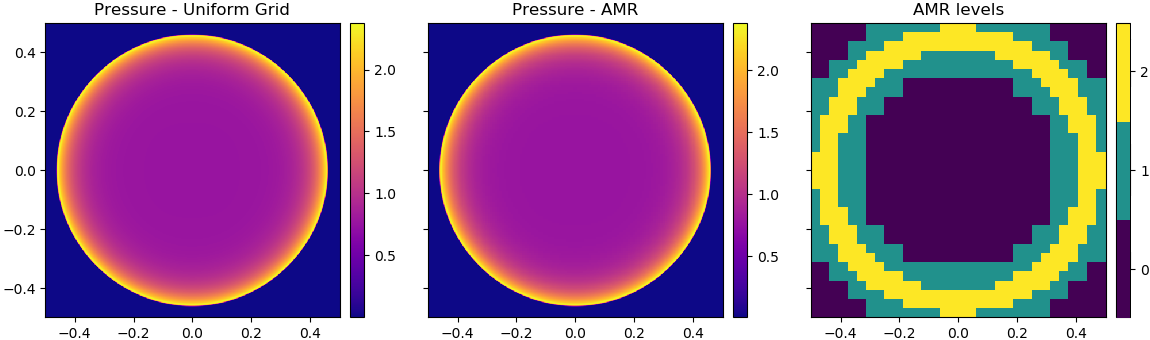}
\caption{Two-dimensional slice of the pressure in a hydrodynamic blast wave test at \(t=0.1\) with a uniform grid (left) and AMR (middle) using the same effective resolution. The right panel shows the distribution of MeshBlocks. \label{fig:hd_blast}}
\end{figure*}

\begin{figure*}[htb!]
\centering
\includegraphics[width=\textwidth]{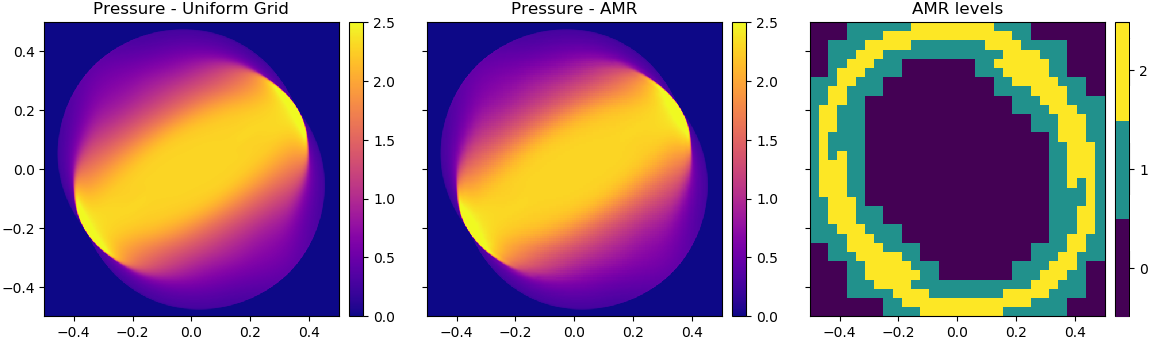}
\caption{The same as figure~\ref{fig:hd_blast} but for the MHD blast wave test at \(t=0.08\). \label{fig:mhd_blast}}
\end{figure*}

Figure~\ref{fig:hd_blast} shows the distributions of the pressure and MeshBlocks at the end of the hydrodynamic calculation.  Comparison of the solutions on the uniform and AMR grids shows essentially no difference.  Excellent spherical symmetry is maintained.  MeshBlocks at the finest level fill only a small fraction of the domain.  Figure~\ref{fig:mhd_blast} shows the same plots for the MHD calculation.  Again the solutions on the uniform and AMR mesh are visually identical.  Although the magnetic field breaks the spherical symmetry of the problem, reflection symmetry perpendicular to the field direction is maintained.   

\begin{figure}[htb!]
\centering
\includegraphics[width=\columnwidth]{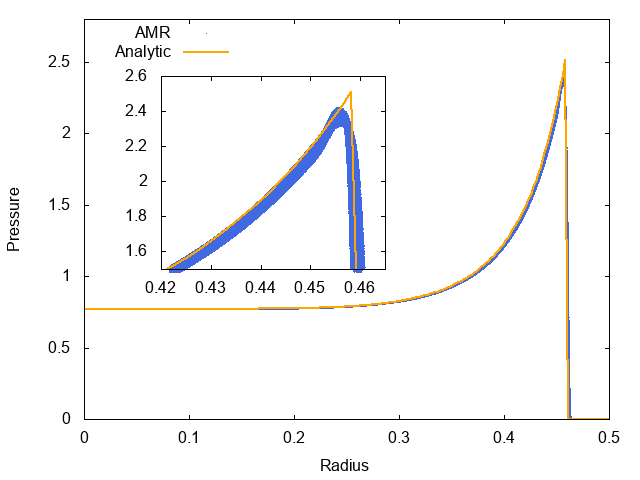}
\caption{Radial profile of the gas pressure in the hydrodynamic blast wave test with AMR. The pressure in each cell as a function of distance from the center is shown with a blue dot, and the orange line indicates the analytic solution.\label{fig:sedov}}
\end{figure}

A more quantitative comparison of the hydrodynamic solution is shown in figure~\ref{fig:sedov}.  The pressure in each grid point in the AMR solution is plotted as a function of radial distance from the center, and this is compared to the analytic Sedov-Taylor blast solution obtained using \texttt{sedov3.f} developed by F. X. Timmes\footnote{
http://cococubed.asu.edu/research\_pages/sedov.shtml}. Note the excellent agreement.  The finite width of the points from the numerical solution is in part an unavoidable consequence of the representation of a sphere on a Cartesian mesh. These figures demonstrate that AMR can reproduce both the uniform grid and analytic solutions very well.

\subsection{Tests of Curvilinear Coordinates and AMR}

Finally, we show results for test problems in curvilinear coordinates (a new capability in \app{}), both with and without AMR.

\subsubsection{Advection Tests in Curvilinear Coordinates}

\begin{figure*}[htb!]
\centering 
\includegraphics[width=2.0\columnwidth]{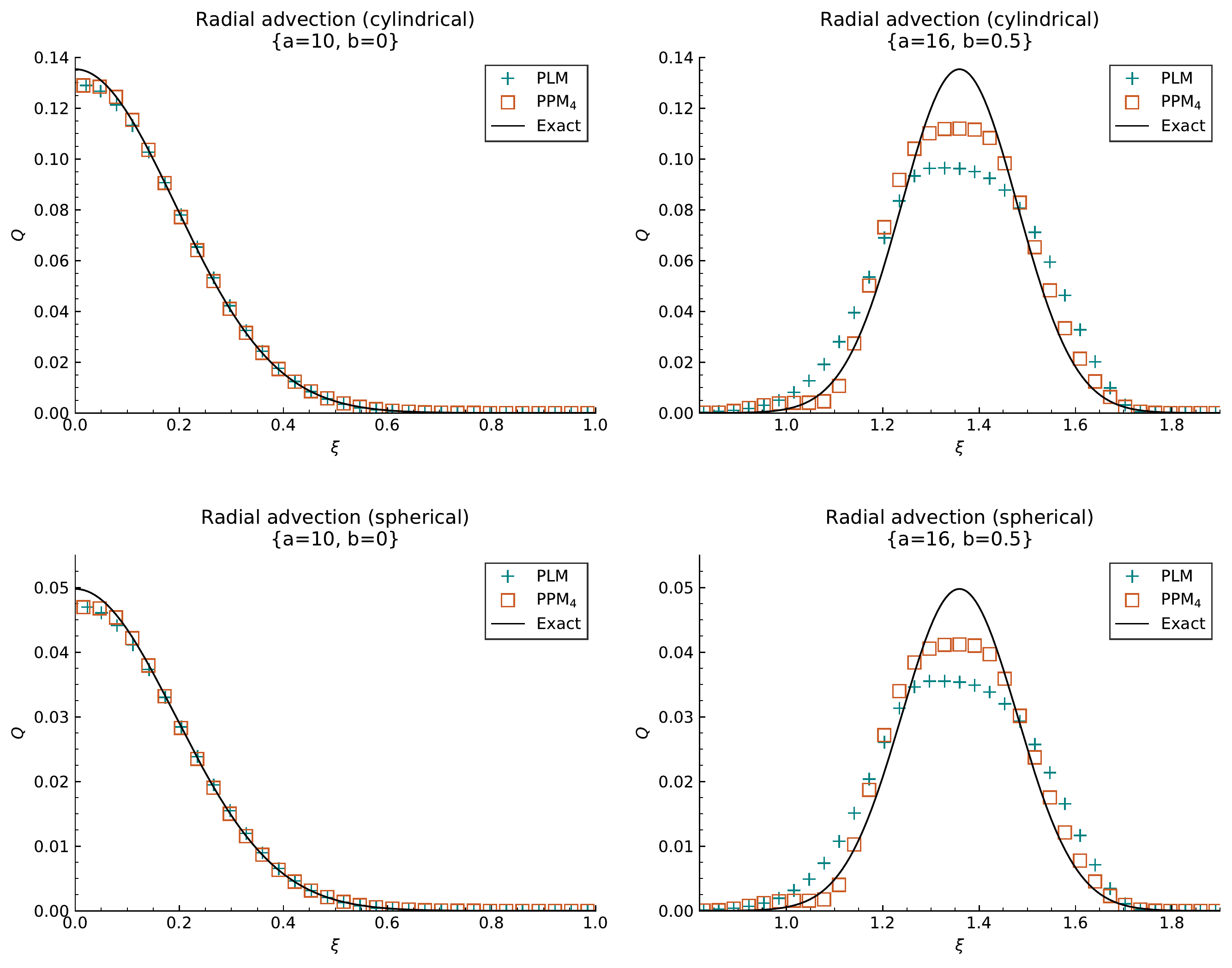}
\caption{Profiles of the 1D radial advection problems in curvilinear coordinates at \(t=1\) and \(N=64\). The top row plots show cylindrical coordinates solutions, and the bottom row plots show the results in spherical-polar coordinates. The left column is advection of monotonic initial data, while the right column is advection of a non-monotonic profile. Compare to \citet[][figure~2]{Mignone2014}. \label{fig:mignone_fig2}}
\end{figure*}

\begin{figure*}[htb!]
\centering 
\includegraphics[width=2.0\columnwidth]{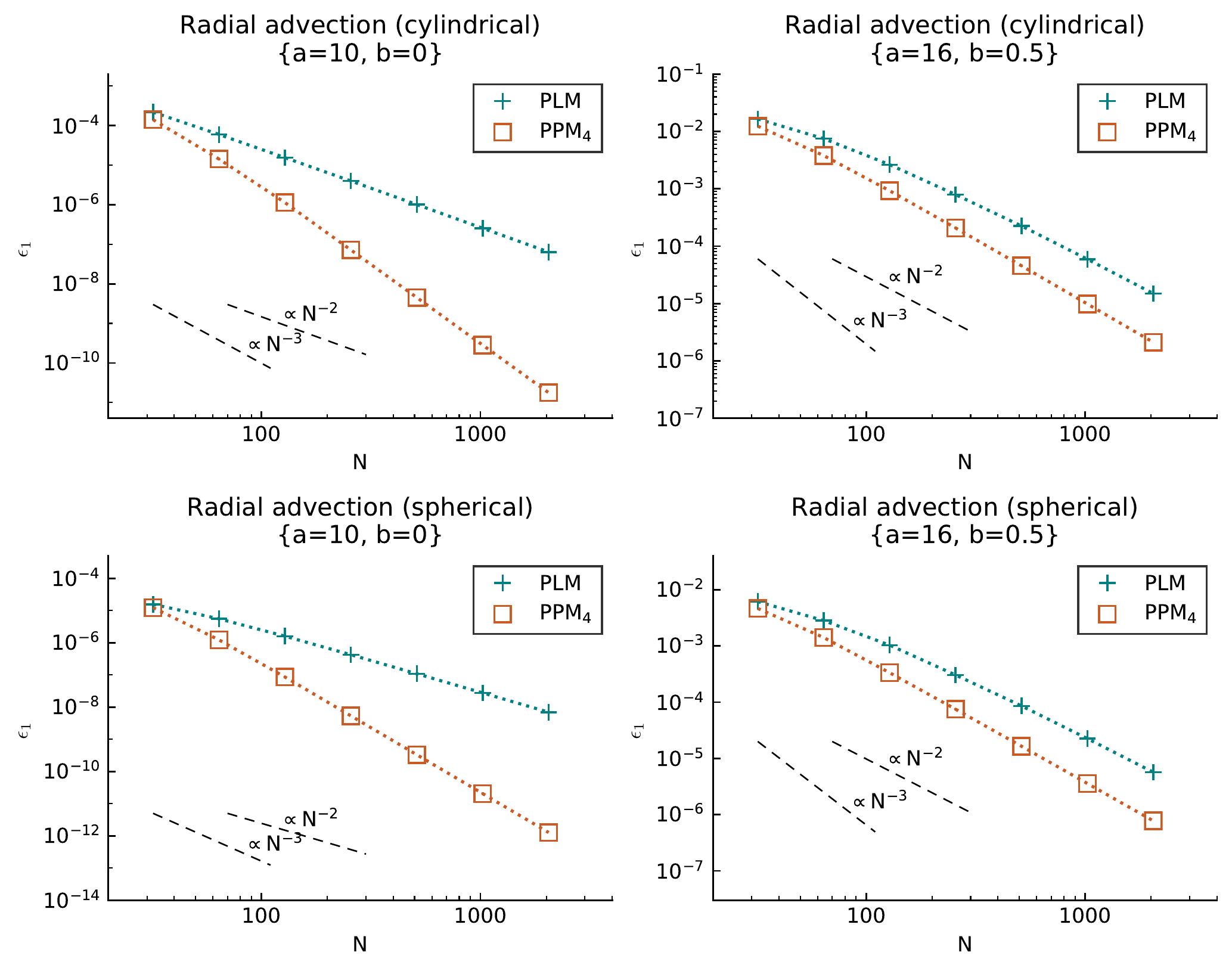}
\caption{Convergence of the \(L_1\) errors of the 1D radial advection problems:\ cylindrical and spherical-polar coordinates, monotonic and non-monotonic data. Compare to \citet[][figure~3]{Mignone2014}. \label{fig:mignone_fig3}}
\end{figure*}
\begin{figure*}[htb!]
\centering 
\includegraphics[width=2.0\columnwidth]{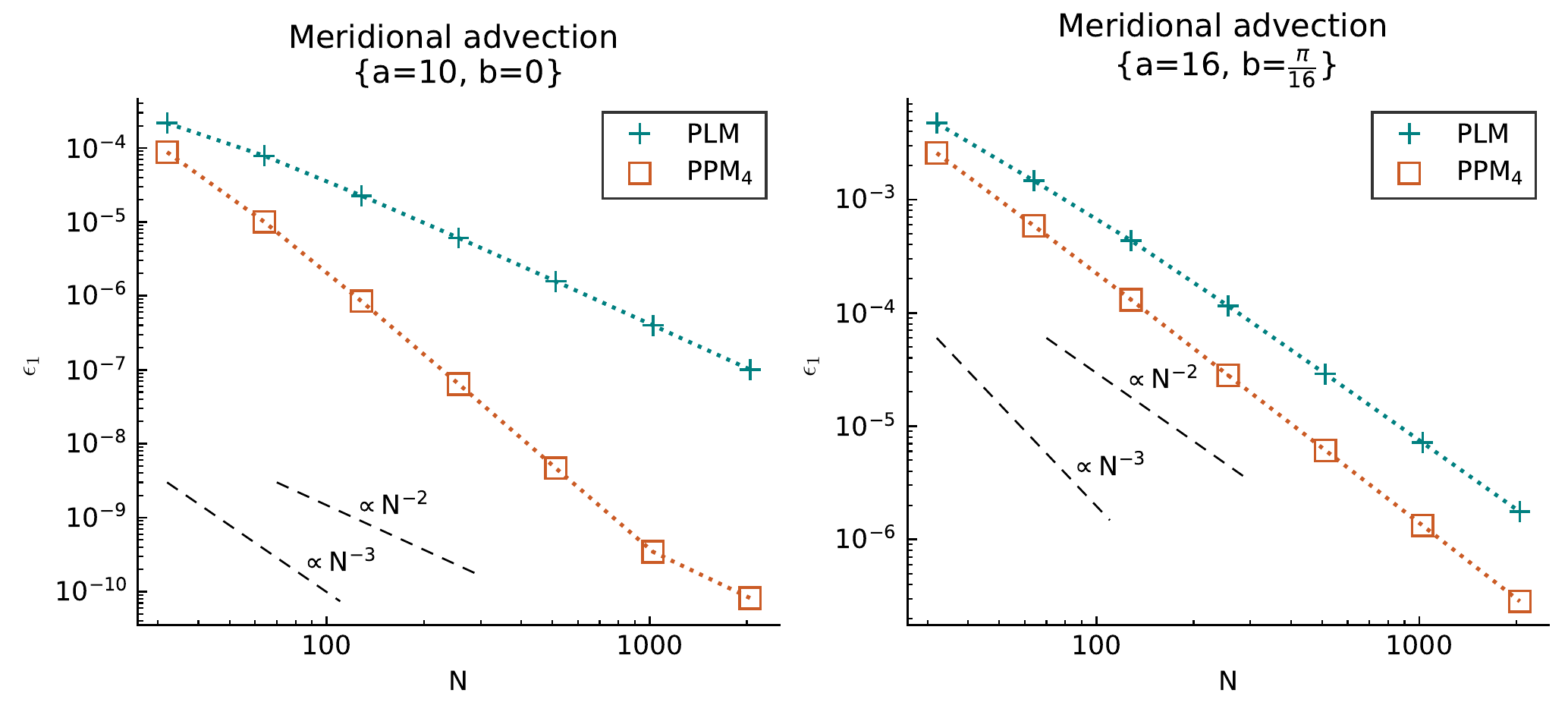}
\caption{Convergence of the \(L_1\) errors of the polar advection problems in spherical-polar coordinates. Compare to \citet[][figure~4]{Mignone2014}. \label{fig:mignone_fig4}}
\end{figure*}

Figure~\ref{fig:mignone_fig2} plots the profiles of the \app{} solutions to the radial 1D advection problem of \citet[][Section 5.1.1]{Mignone2014} in cylindrical and spherical-polar coordinates for both PLM and PPM. For these tests, a passive scalar is initialized with a Gaussian profile and advected with a linear velocity field. The parameters \(a\) and \(b\) in the plot labels control the width of the Gaussian and location of the curve's center, respectively.

Because a variant of the original PPM limiter is used with curvilinear coordinates in \app{}, the smooth extrema in the right column solutions are clipped; this does not occur with the more advanced limiter used in Cartesian coordinates. Future work will consider extending the curvilinear corrections to the smooth extrema preserving PPM limiter. 

Figures~\ref{fig:mignone_fig3} and~\ref{fig:mignone_fig4} show the convergence of the \(L_1\) error in the radial and a 2D counterpart of the meridional \citep[see][Section 5.1.2]{Mignone2014} scalar advection problems, respectively. They demonstrate the formal second- and fourth-order convergence of PLM and PPM reconstruction in \app{}. There are slight differences between the PLM results shown here and those shown in the original reference \citep{Mignone2014}; their choice of a modified monotonized central (MC) limiter for PLM results in lower errors than the van~Leer limiter for the non-monotonic tests and higher errors in the monotonic tests.  Overall, these plots demonstrate the fidelity of the solvers in curvilinear grids.

\subsubsection{Field Loop Advection Through the Pole\label{subsubsec:field-loop-pole}}

Near coordinate singularities (such as the poles in a 3D spherical polar mesh), numerical discretizations generally have nonuniform truncation error which can imprint visible features in solutions.  Moreover, flow through the pole requires special boundary conditions that load the ghost cells of MeshBlocks that overlap regions across the pole with data from the appropriate azimuthal angle. To test the implementation of curvilinear coordinates at poles, the results for an advection of flow through the pole is presented.

The problem consists of a uniform parallel velocity field $v_x=1$ that is represented on a spherical polar mesh, with the poles perpendicular to the flow velocity.  A passive magnetic field loop is then initialized and advected through the poles. Following \citet{gs05}, the magnetic fields are initialized with a vector potential of the form
\begin{multline}
    A_z = B_0\exp{\left[-\frac{(z-z_0)^2}{\sigma^2}\right]} \\
        \times \max\left(R-\sqrt{(x-x_0)^2+(y-y_0)^2},0\right) ,    
\end{multline}
where $(x_0,y_0,z_0) = (-\sqrt{2}/2,0,\sqrt{2}/2)$ is the initial center of the loop, $B_0$ the magnetic field strength, $R=0.5$ the radius of the loop, and $\sigma=0.2$ the thickness of the loop, respectively. The field strength $B_0$ is set so that $\beta = 2p/(B_0^2) = 10^5$ at the mid-plane of the loop. PLM reconstruction, the HLLD approximate Riemann solver, and an adiabatic EOS with $\gamma=5/3$ are used. The computational domain is $0.1<r<2.0$, $0<\theta<\pi/2$, $0<\phi<2\pi$ and the resolution is $160\times80\times160$ using logarithmic spacing in the $r$-direction.

\begin{figure*}[htb!]
\centering
\includegraphics[width=\textwidth]{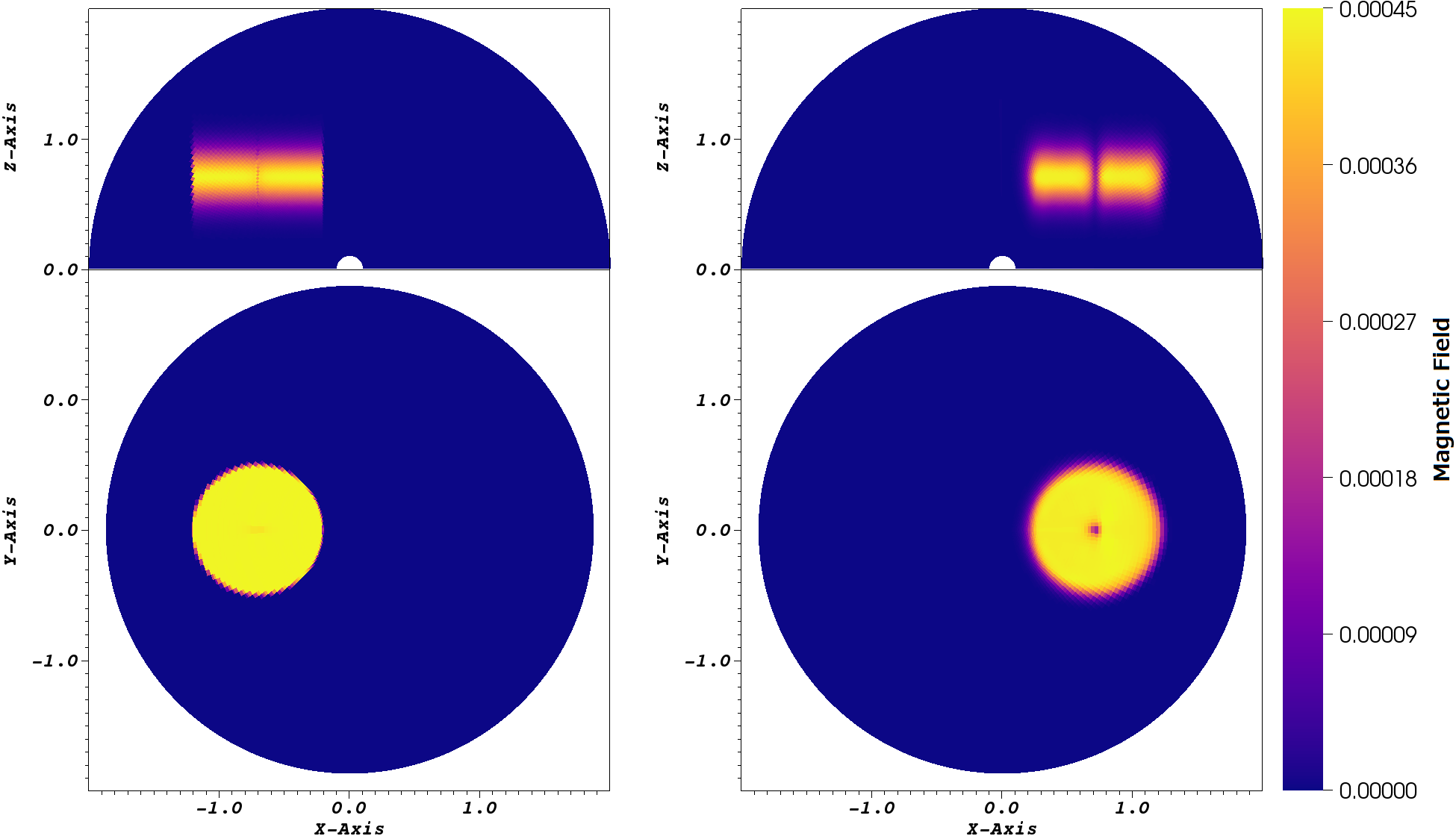}
\caption{Slices of the magnetic field strength for the field loop advection test in spherical-polar coordinates. The left panels indicate the initial condition, while the right panel is the result at $t=\sqrt{2}$. The top panels are vertical cross sections through the $y=0$ plane, while the bottom panels are horizontal cross sections through the $z=\sqrt{2}/2$ plane.  Despite having passed directly through the coordinate singularity at the pole, the loop at the final time is symmetric.\label{fig:field_loop_pole}}
\end{figure*}

Figure~\ref{fig:field_loop_pole} shows the magnetic field strength on slices through the computational mesh at the center of the field loop in the initial and final states.  The loop shows evidence for numerical diffusion, especially at the center where oppositely directed field lines are closely spaced.  However, the structure is well preserved even after advection through highly anisotropic coordinates and the coordinate singularity. While this test is perhaps artificial (spherical polar grids are not a good representation of the initial flow geometry), it nevertheless demonstrates the robustness of our finite-volume scheme in curvilinear coordinates.

\subsubsection{Blast Wave Test in Spherical-Polar Coordinates}

To demonstrate AMR in curvilinear coordinates, the same blast wave tests detailed in section~\ref{subsubsec:cartesian-blast} were run using spherical-polar coordinates. The problem domain is \(0.5 < r < 1.5\), \(\pi/6 < \theta < \pi/2\), and \(-\pi/5 < \phi < \pi/5\). The grid is nonuniformly spaced along the \(r\)-direction so that the aspect ratio of the cells remains close to unity everywhere. The other parameters are the same as the problem in Cartesian coordinates. For the MHD model, the magnetic field is initially uniform along the pole.

\begin{figure}[htb!]
\centering
\includegraphics[width=\columnwidth]{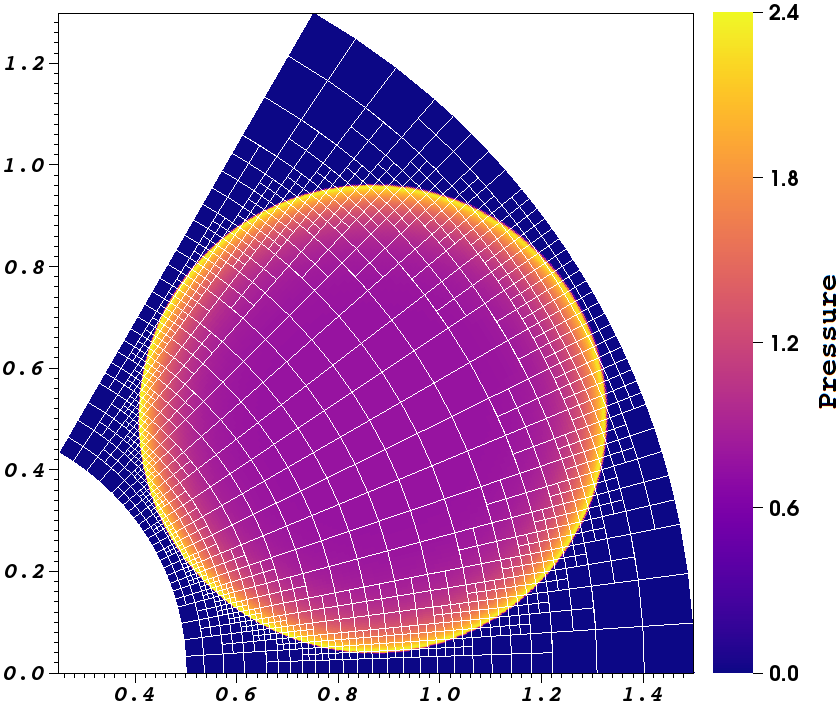}
\caption{Pressure on a slice at \(\phi=0\) in the hydrodynamic blast wave test at \(t=0.1\) in spherical polar coordinates with AMR. The MeshBlock distribution is superimposed as white boxes. \label{fig:hd_sphblast}}
\end{figure}

\begin{figure}[htb!]
\centering
\includegraphics[width=\columnwidth]{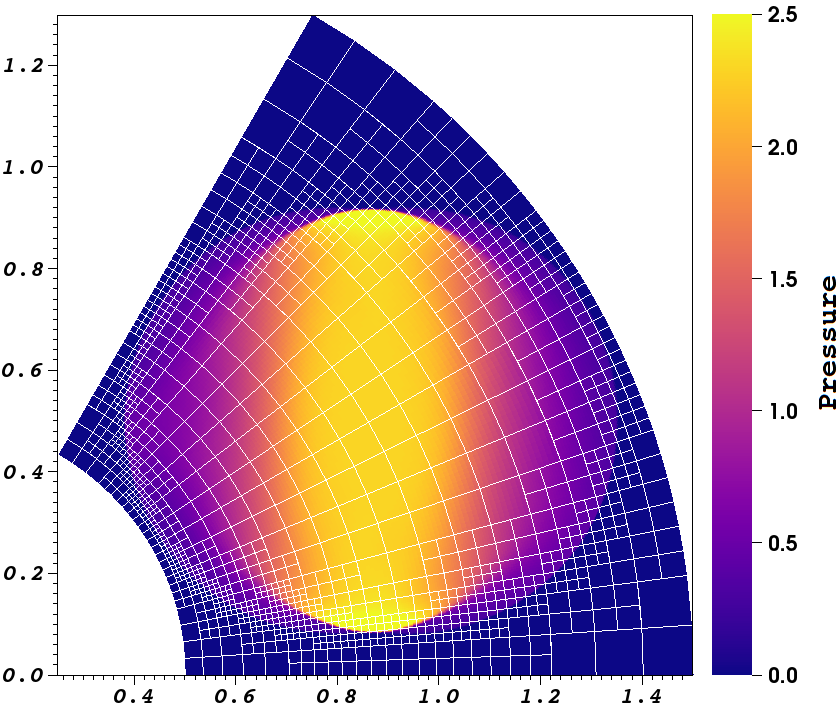}
\caption{Same as figure~\ref{fig:hd_sphblast} but for the MHD model at \(t=0.08\). \label{fig:mhd_sphblast}}
\end{figure}

The results for the hydrodynamic test are shown in figure~\ref{fig:hd_sphblast}.  Note that the blast remains spherically symmetric even on the curvilinear mesh.  The plot shows excellent agreement with figure~\ref{fig:hd_blast}.  The results for the MHD test are shown in figure~\ref{fig:mhd_sphblast}.  Again, there is excellent agreement with the previous results found for a Cartesian grid and shown in figure~\ref{fig:mhd_blast}.

\begin{figure}[htb!]
\centering
\includegraphics[width=\columnwidth]{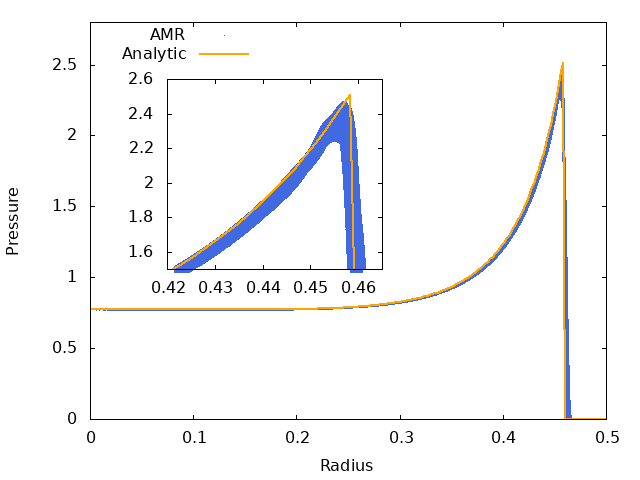}
\caption{Same as figure~\ref{fig:sedov} but for the blast wave test in spherical-polar coordinates.\label{fig:sedov_sph}}
\end{figure}

Finally, figure~\ref{fig:sedov_sph} plots the pressure as a function of radial position from the center of the blast for the hydrodynamic problem shown in figure~\ref{fig:hd_sphblast}, along with the analytic solution for the Sedov-Taylor blast wave.  The results can be compared to figure~\ref{fig:sedov}, which used a Cartesian grid.  The peak of the pressure curve at the location of the blast wave is slightly smeared in the spherical polar grid solution. However, this phenomenon is explained by the use of a nonuniform radial grid that has larger cells at larger radii.  Otherwise, excellent agreement is obtained.

\subsection{Performance and Scaling of the MHD Solver}

Most of the scientific applications of \app{} require multi-dimensional calculations at high resolution.  Performance of the solver is often a rate-limiting step for progress, and therefore we have spent considerable effort trying to maximize the performance and scaling of the MHD solver.  For example, the initial design of the C++ classes used in \app{} resulted from extensive performance benchmarking of the core computational kernel of the algorithm. The design was continually compared against the highest performance achieved for raw C code that implemented the same steps.  Only once the design of the C++ mocked classes met or exceeded the performance of the raw C code was this design used to implement the full code.  In this way, we ensure that none of the abstractions of the object-orientated design inhibit performance optimization by the compiler.  In this section, we report the performance and scaling we have achieved for the MHD solver in the \app{} AMR framework.

\subsubsection{Single-Core Performance}

\begin{deluxetable}{cccccc}  % does not accept p{} alignment
% deluxetable preamble commands:
\tablewidth{\columnwidth}
\tabletypesize{\small}
\tablecolumns{6}
 \tablecaption{\app{} single-node performance: single-core\label{tbl:single-core_performance}}
 \tablehead{   &     &      & \multicolumn{3}{c}{MZone-cycles/sec} \\
&   &   & \colhead{\thead{Xeon Phi\\KNL 7250}} & \colhead{\thead{Broadwell\\E5-2680 v4}} & \colhead{\thead{Skylake-SP\\Gold 6148}}}
%&   &   & \colhead{\thead{KNL\\7250}} & \colhead{\thead{Broadwell\\2680v4}} & \colhead{\thead{Skylake\\6148}}}
% end deluxetable preamble

\startdata
\multirow{6}{*}{Hydro} & \multirow{3}{*}{PLM} & HLLC &           1.472 &                           3.136 &                           5.227 \\
      &     & HLLE &        1.617 &                               3.346 &                           5.814 \\  
      &     & Roe &           1.520 &                            3.367 &                           5.471 \\   
\cmidrule(lr){2-6}
      & \multirow{3}{*}{PPM} & HLLC &        0.665 &                        1.316 &                           2.527 \\   
      &     & HLLE &        0.689  &                       1.353 &                           2.643 \\                    
      &     & Roe &           0.674    &                  1.352 &                           2.593 \\                     
\cmidrule(lr){1-6}
\multirow{6}{*}{MHD} & \multirow{3}{*}{PLM} & HLLD &              0.754 &                   1.519 &                           2.924 \\   
      &    & HLLE &          0.875 &                                             1.626 &                           2.757 \\              
      &     & Roe &             0.689 &                                         1.294 &                           2.191 \\               
\cmidrule(lr){2-6}
      & \multirow{3}{*}{PPM} & HLLD &         0.381 &                           0.775 &                           1.559 \\   
      &     & HLLE &         0.437 &                           0.799 &                           1.512 \\                    
      &     & Roe &           0.347 &                         0.708 &                           1.323 \\                     
\enddata
\end{deluxetable}
% Version 4abd888 from 2019-03-20 w/ 64^3 MeshBlock

Table~\ref{tbl:single-core_performance} summarizes the performance (averaged across 20 independent trials) using only a single physical core on a single node of three target Intel architectures.  The test is based on a three-dimensional benchmark problem (the blast wave test in section~\ref{subsubsec:cartesian-blast}) for adiabatic hydrodynamics and MHD, and it considers multiple Riemann solvers and primitive variable reconstruction techniques. The default second-order accurate VL2 time integrator is used in all cases, and the problem size is fixed to a single \(64^3\) MeshBlock.  Performance is measured in the number of cells updated per second (the inverse of which is the CPU time required to update a single cell).

For 3D MHD with the HLLD Riemann solver and PLM reconstruction (a typical combination of algorithmic options), we achieve nearly 3 million zone updates per second per core on the Intel Skylake processor.  With PPM reconstruction the performance drops by about a factor of two. For comparison, we have run the same benchmark problem using the same algorithmic choices and the same compiler optimizations for the latest public versions of the \texttt{FLASH}, \texttt{PLUTO}, and \texttt{Enzo} codes, and we find that the per-core performance of \app{} is the highest of all four, in some cases by as much as a factor of ten.
Good performance on modern processors is achieved only through the use a high percentage of vectorized instructions. Using Intel diagnostic tools, we find about 85 percent (based on the CPU time) of the MHD code is vectorized using the AVX/AVX2/AVX512 vector instruction sets.

\subsubsection{Multi-Core Performance}

\begin{deluxetable}{cccccc}  % does not accept p{} alignment
% deluxetable preamble commands:
\tablewidth{\columnwidth}
\tabletypesize{\small}
\tablecolumns{6}
 \tablecaption{\app{} single-node performance: multi-core\label{tbl:multi-core_performance}}
 \tablehead{   &     &      & \multicolumn{3}{c}{MZone-cycles/sec} \\
&   &   & \colhead{\thead{\\Xeon Phi\\KNL 7250}}& \colhead{\thead{(2x)\\Broadwell\\E5-2680 v4}} & \colhead{\thead{(2x)\\Skylake-SP\\Gold 6148}}}
%&   &   & \colhead{\thead{\\KNL\\7250}} &  \colhead{\thead{(2x)\\Broadwell\\2680v4}} & \colhead{\thead{(2x)\\Skylake\\6148}}}
% end deluxetable preamble

\startdata
\multirow{6}{*}{Hydro} & \multirow{3}{*}{PLM} & HLLC &      81.992 &                           49.744 &                              84.769 \\  
      &     & HLLE &            83.110 &                                        51.278 &                              87.877 \\                 
      &     & Roe &            79.129 &                                        51.425 &                              87.754 \\                  
\cmidrule(lr){2-6}
      & \multirow{3}{*}{PPM} & HLLC &           42.554     &                           24.834 &                              49.759 \\  
      &     & HLLE &           42.804      &                          25.183 &                              50.012 \\                   
      &     & Roe &            42.002      &                         25.242 &                              49.875 \\                    
\cmidrule(lr){1-6}
\multirow{6}{*}{MHD} & \multirow{3}{*}{PLM} & HLLD &        37.953 &                     24.624 &                              44.361 \\     
          &     & HLLE &         43.139  &                                  25.480 &                              43.345 \\                  
          &     & Roe &            35.287   &                              22.045 &                              39.853 \\                   
\cmidrule(lr){2-6}                                                                                                                           
          & \multirow{3}{*}{PPM} & HLLD &  21.024           &                                14.624 &                              28.826 \\ 
          &     & HLLE &       24.090 &                                     14.954 &                              28.457 \\                  
          &     & Roe &            19.683  &                               13.657 &                              26.612 \\                   
\enddata
\end{deluxetable}
% Version 4abd888 from 2019-03-20 w/ 64^3 MeshBlock

Table~\ref{tbl:multi-core_performance} summarizes the code's performance when using
all of the cores available on a single node. For this test, both Broadwell (14 cores per socket) and Skylake (20 cores per socket) CPUs configured as dual-socket nodes were used, for a total of 28 and 40 cores respectively. The KNL node possesses a total of 68 physical cores, but we use only 64 cores in order to minimize jitter from the operating system. A single MPI rank is pinned to each core in these tests. In addition, the KNL tests benefit from using 4 OpenMP threads per MPI rank in order to utilize the 4-way Hyper-Threading of the 64 physical (256 logical) cores on these nodes. In this case, each thread owns a MeshBlock of size \(64\times 32\times 32\).

Note that in all cases, the performance per core is significantly less than that reported in table~\ref{tbl:single-core_performance}, typically by a factor of two regardless of the choice of algorithm. Modern Intel processors decrease the overall clock speed when all cores are active and are executing AVX2/AVX512 instructions, which contributes in part to the decrease.  However, most of the decrease is due to memory bandwidth limits and less-than-optimal use of cache.  Generally algorithms with higher arithmetic intensity (ratio of flops to memory accesses) are less affected by memory bandwidth limits.  However, we observe the same decrease in performance when all cores are used independent of which algorithm we adopt.  For example, PPM reconstruction with the HLLD Riemann solver for MHD requires nearly three times the number of floating point operations per cell than PLM reconstruction and the HLLE solver for hydrodynamics, yet both display the same factor of two decrease in performance when all cores are used.  We have observed the same trend even for the complex fourth-order algorithm implemented in \citet{FelkerStone2018}.  This indicates the overall design and implementation of the MHD solver in \app{} is cache-limited.

It is important to note that different algorithmic choices can greatly improve cache-performance.  For example, \citet{Woodward2019} describe an approach for organizing data into small ``mini-briquettes'' that fit entirely into cache and which enables excellent performance for the dimensionally-split PPM algorithm for hydrodynamics.  However, this approach requires special-purpose coding, and it is not clear if it is extensible to the dimensionally unsplit integrators required for MHD that are implemented in \app{}.  Nevertheless, exploring such approaches in the future could be important for achieving further performance increases.

Recently, \citet{Grete19} reported the port of the public version of \app{} to GPUs based on the \texttt{Kokkos} library \citep{Edwards14}.  Figure 3 in their paper explores the efficiency of the implementation on various architectures.  Generally excellent results are obtained, with between 75-90\% architectural efficiency on most processors including both CPUs and GPUs. Figure 4 in their paper compares the performance of the resulting code, called \texttt{K-ATHENA}, on both Intel CPUs and NVIDIA GPUs.  The performance per CPU shown in the right panel of their figure is somewhat lower than the value reported in table~\ref{tbl:multi-core_performance} for the same test (MHD using PLM reconstruction and the Roe Riemann solver), due to recent optimizations that were not available in the public version they used.  Using our values, the ratio of the performance of \texttt{K-ATHENA} on the latest NVIDIA Volta GPU to a single Intel Skylake (20-core) CPU performance is about a factor of five, which is about the same as the ratio of the peak performance for these two architectures. This indicates that despite the limitations of cache performance in \app{} inherent in table~\ref{tbl:multi-core_performance}, overall the code performs extremely well.

\subsubsection{Weak Scaling on Uniform Grids}

\begin{figure}[htb!]
    \centering
    \includegraphics[width=\columnwidth]{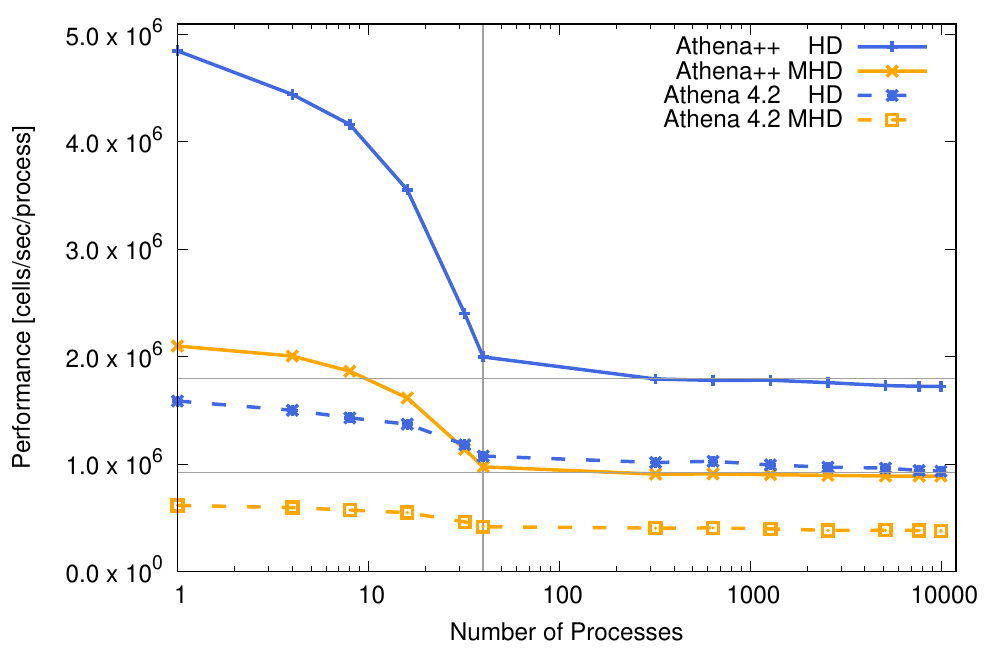}
    \caption{Weak scaling test on Cray XC50 (2x Skylake 6148, 40 cores per node). The vertical line indicates one node. To the left of this delimiter, the performance is limited mainly by memory bandwidth, and beyond this we observe the influence of the network overhead. The gray horizontal lines indicate the \app{} performance with 8 nodes.\label{fig:scaling}}
\end{figure}

On modern architectures, good parallel scaling is essential to make large calculations feasible. Figure~\ref{fig:scaling} shows the results of weak scaling tests on a Cray XC50 machine containing dual Intel Skylake 6148 processors with 40 cores per node.  The test uses a uniform grid with \(64^3\) cells per MeshBlock and one MeshBlock per process.  The test uses up to 250 nodes ($10^4$ cores).  For the hydrodynamic tests, the HLLC Riemann solver is used.  For the MHD tests, the HLLD solver is used, and both use the VL2+PLM integration algorithm.  Performance is again measured in zone updates per CPU second per core.

Note the rapid decrease in performance when scaling from one to 40 processes, as all cores on a node are used and memory bandwidth limits performance.  This behavior reflects the trends already noted in tables~\ref{tbl:single-core_performance} and~\ref{tbl:multi-core_performance} and discussed above.
While improving the cache utilization would likely reduce the memory bandwidth per node limitations evident in figure~\ref{fig:scaling} (as has been achieved in a few other codes, e.g. \citet[][]{Woodward2019}), this will require substantial changes to the implementation.
Once all cores on a node are utilized, the weak scaling of \app{} is essentially perfect.  The parallel efficiencies of the hydrodynamic and MHD simulations between 8 and 250 nodes are about 97\% and 95\% respectively.  Thus only a small fraction of the time for the calculation is used for communication costs.

\begin{figure}[htb!]
    \centering
    \includegraphics[width=\columnwidth]{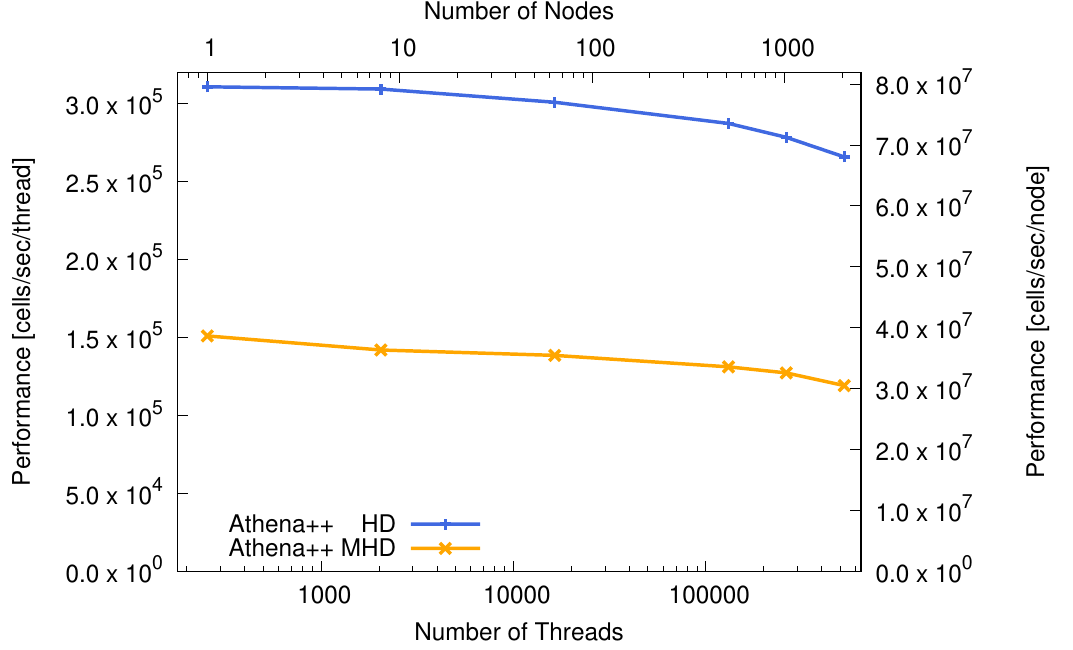}
    \caption{Weak scaling test on Intel Xeon Phi 7250 on the Oakforest-PACS supercomputer. \label{fig:knlscaling}}
\end{figure}

To test the weak scaling and parallel efficiency on even larger core counts, we have performed another set of weak scaling tests on the \mbox{Oakforest-PACS} supercomputer equipped with \mbox{Intel} Xeon Phi 7250 (Knights Landing) multi-core processors. We use 64 cores per node (4 cores are left unused to accommodate the operating system and other tasks), and 4 OpenMP threads per process using the \texttt{COMPACT} affinity.  Each thread owns one MeshBlock consisting of $64 \times 32 \times 32$ cells. The results are shown in figure~\ref{fig:knlscaling}. 

Note that even when using 2048 nodes (equivalent to 524,288 threads), the parallel efficiency compared to 8 nodes is 86\% for hydrodynamics and 84\% for MHD.  This excellent scaling is due in part to the ability of the TaskList to interleave communications and calculations.  The test demonstrates that finite-volume algorithms show excellent scaling up to millions of cores and are highly capable of exploiting emerging resources in the exascale era.

\subsubsection{Strong Scaling with AMR\label{sec:AMRStrongScaling}}

Quantifying the performance of the AMR framework when used with the MHD solver is difficult, because the amount of work per calculation is highly variable and depends on the refinement criteria, the size of the MeshBlocks, and the efficiency of the implementation.  In section~\ref{subsubsec:khi-tests}, we discussed the performance of AMR in terms of reducing the time to solution of some given accuracy compared to a uniform grid for the particular problem of the KHI test.  To further quantify the performance of our AMR framework, we measure strong scaling in this section using a different problem.

\begin{figure}[htb!]
    \centering
    \includegraphics[width=\columnwidth]{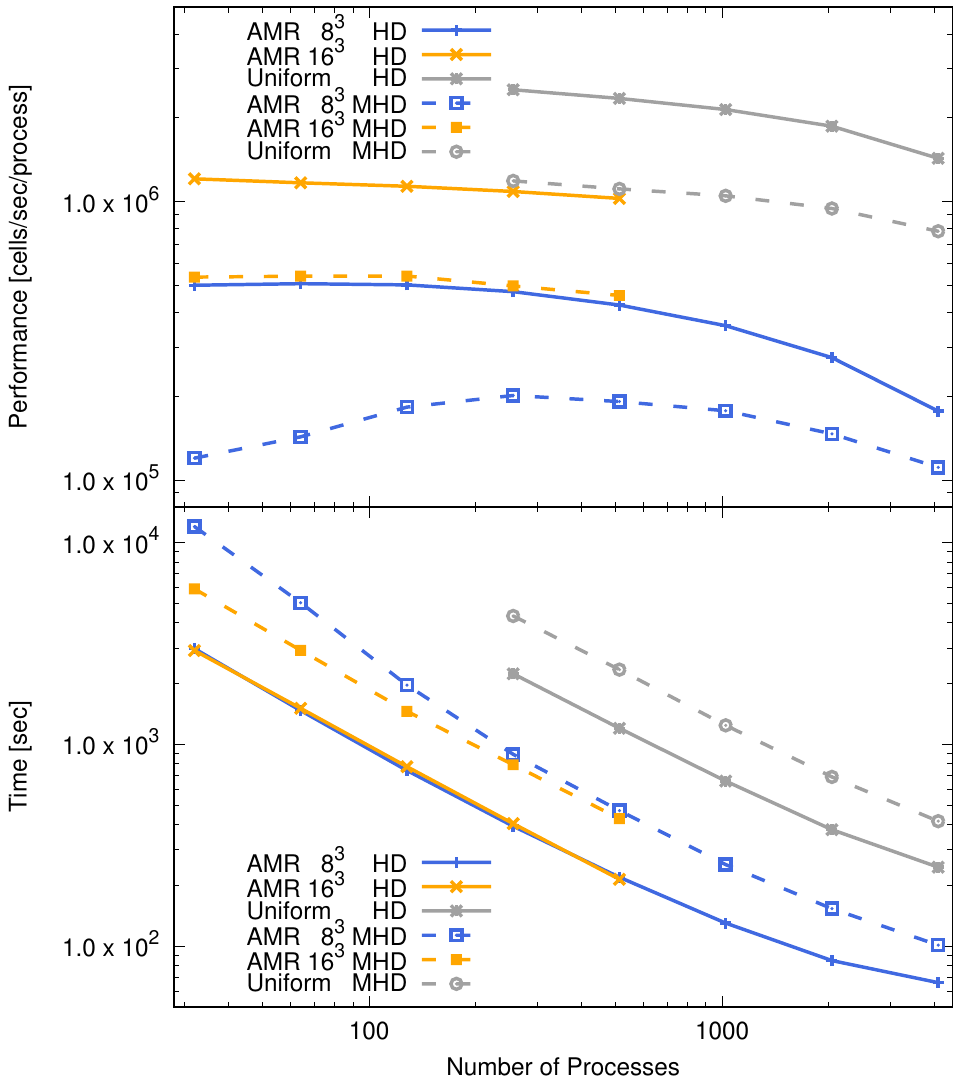}
    \caption{Strong scaling test on Cray XC50 (2x Skylake 6148 per node, using only 32 out of 40 cores per node), including file outputs. \label{fig:amrscaling}} 
\end{figure}

We use the blast wave test discussed in section~\ref{subsubsec:cartesian-blast}.  Our timing measurements include outputs at every \(\Delta t=0.01\). The result is presented in figure~\ref{fig:amrscaling}. The computational throughput of AMR with \(16^3\) MeshBlocks in terms of cell-updates per second is about half of the uniform grid's efficiency, but its time to solution is about five times shorter than that of uniform grid. AMR, when used with relatively small \(8^3\) MeshBlocks, has even higher overhead, but its time to solution is as fast as AMR with \(16^3\) MeshBlocks. The short computing time with AMR is not only due to the reduced number of cells in AMR, but also the larger time step by a factor of \(\sim 2\) because the hot region near the center of the explosion is derefined.

The optimal choice for the refinement parameters depends on many factors such as the volume of the refined regions, the size of the root grid, the number of refinement levels, the number of processes,  etc., and thus is highly problem dependent. While smaller MeshBlocks give more flexibility to adapt to solutions, they are computationally less efficient. It is reasonable to start with MeshBlocks of size $8^3$ or $16^3$, but ultimately the best choice for each problem must be found through experimentation.

\subsubsection{Size of MeshBlocks and Performance\label{sec:MeshBlockSize-Performance}}

\begin{figure}[htb!]
    \centering
    \includegraphics[width=\columnwidth]{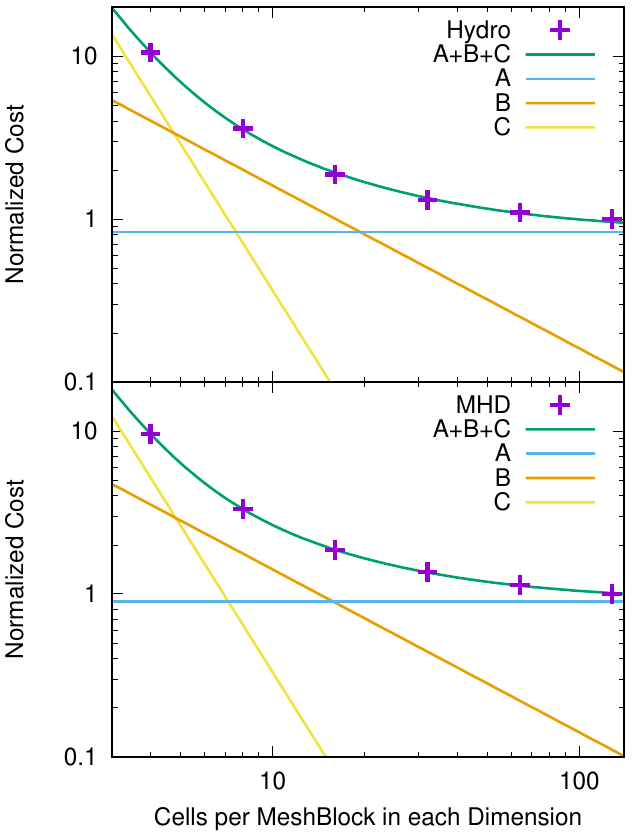}
    \caption{Computational cost for a 3D blast wave test using MeshBlocks of different sizes, normalized to the cost using a single $128^3$ MeshBlock. The purple crosses are measured results, while the three lines labeled $A$ through $C$ are different components of a simple model fit to the data.  See text for details.\label{fig:mbsize}} 
\end{figure}

In order to quantify how the size of MeshBlocks affects performance, we have run a series of tests using the 3D blast wave problem described earlier but with file outputs disabled.  In each case the computational domain is resolved with $128^3$ cells, and the CPU time required for solution is measured with MeshBlocks ranging in size from $4^3$ (32,768 MeshBlocks) to $128^3$ (1 MeshBlock).  All tests were run on a single core of a Skylake 6148 processor.  The results are shown in figure~\ref{fig:mbsize}, with each point normalized to the CPU time required for the run with a single $128^3$ MeshBlock. Similar trends are observed in both hydrodynamic and MHD runs, with the CPU time increasing by nearly an order of magnitude as the MeshBlock size decreases from $128^3$ to $4^3$.

This behavior can be explained by a simple model that assumes there are three contributions to the cost of runs with different sized MeshBlocks. The first part $A$ represents the actual cost to update all the active cells. Because the total number of cells is fixed for all runs, this term does not depend on MeshBlock size.  The second part $B$ represents the cost of communications, and therefore is proportional to the total surface area of MeshBlocks.  Let $x$ be the number of cells in each dimension per MeshBlock.  The total number of MeshBlocks is $(128/x)^3$, while the surface area per MeshBlock scales as $x^2$, therefore the cost of this second part $B(x) \propto x^{-1}$.  Finally, the last part of the model $C$ accounts for the overhead incurred in generating and managing MeshBlocks and therefore is proportional to the total number of MeshBlocks $C(x)\propto x^{-3}$.  The total cost $W$ can be expressed as the sum of these three parts:
\begin{equation}
W(x) = A+B+C = A + b/x + c/x^{3}
\end{equation}
where $b$ and $c$ are constant coefficients.
We plot this model, along with each of the three contributing terms $A$ through $C$, to the measured normalized CPU time shown in figure~\ref{fig:mbsize}. We use a least-squares method to fit the coefficients in each term, since they cannot be predicted analytically. This simple model can explain the observed performance trends very well. As expected, with large MeshBlocks the cost is dominated by the actual computation ($A$). However, as the MeshBlock size decreases, the communication cost ($B$) becomes more important, exceeding $A$ around $16^3$ in both hydrodynamics and MHD. For MeshBlocks as small as $4^3$, the overhead term ($C$) becomes dominant and makes the simulation inefficient. 

Although the actual balance between these cost components depends on many factors (including the size of the simulation, physics modules in use, CPU and memory performance, parallelization, use of AMR, etc.) this result clearly demonstrates that small MeshBlocks are computationally inefficient. Therefore users must choose the optimum MeshBlock size specific to their problem. For uniform grid simulations, larger MeshBlocks are obviously better. For AMR, it is not trivial to balance performance and flexibility, but somewhere between $8^3$ and $16^3$ should be reasonable as discussed in the previous section.

\section{A Relativistic MHD Solver\label{sec:RelativisticMHD_solver}}

The details of the SR and GR methods have already been presented in \citet{WhiteStone2016}. Here we summarize the important equations and highlight salient differences from Newtonian MHD. In this section we use units with \(c = 1\).

\subsection{Equations and Discretization\label{subsec:GR_equations}}

The differential equations of relativistic MHD can be written in a form similar to those of Newtonian MHD. In SR the primitive variables are fluid-frame density $\rho$, fluid-frame gas pressure $\pgas$, spatial part of lab-frame fluid 4-velocity $\mathbf{u}$, and lab-frame magnetic field $\mathbf{B}$. The Lorentz factor is $\gamma = (1 + \mathbf{u}^2)^{1/2}$, and the 3-velocity is $\mathbf{v} = \mathbf{u}/\gamma$. The magnetic pressure is
\begin{equation}
  \pmag = \frac{1}{2} \paren[\bigg]{\frac{1}{\gamma^2} \mathbf{B}^2 + (\mathbf{v} \cdot \mathbf{B})^2}
\end{equation}
and the total enthalpy is
\begin{equation}
  w = \rho + \frac{\Gamma}{\Gamma-1} \pgas + 2 \pmag.
\end{equation}
Here $\Gamma$ is the adiabatic index, taken to be constant. The analogues of equations~\eqref{eq:cons} are then
\begin{subequations} \label{eq:sr_cons} \begin{align}
  \frac{\partial D}{\partial t} + \nabla \cdot (D \mathbf{v}) & = 0, \label{eq:sr_cons:mass} \\
  \frac{\partial\mathbf{M}}{\partial t} + \nabla \cdot \mathbf{S} & = 0, \label{eq:sr_cons:momentum} \\
  \frac{\partial E}{\partial t} + \nabla \cdot \mathbf{M} & = 0, \label{eq:sr_cons:energy} \\
  \frac{\partial\mathbf{B}}{\partial t} - \nabla \times (\mathbf{v} \times \mathbf{B}) & = 0. \label{eq:sr_cons:induction}
\end{align} \end{subequations}
Here the conserved variables include the lab-frame density, energy, and momentum given by
\begin{subequations} \begin{align}
  D & = \gamma \rho, \\
  E & = \gamma^2 w - \gamma^2 (\mathbf{v} \cdot \mathbf{B})^2 - (\pgas + \pmag), \\
  \mathbf{M} & = (E + \pgas + \pmag) \mathbf{v} - (\mathbf{v} \cdot \mathbf{B}) \mathbf{B},
\end{align} \end{subequations}
and the stress tensor is
\begin{align}
  \mathbf{S} & = \gamma^2 w \mathbf{v} \mathbf{v} - \frac{1}{\gamma^2} \mathbf{B} \mathbf{B} - (\mathbf{v} \cdot \mathbf{B}) (\mathbf{v} \mathbf{B} + \mathbf{B} \mathbf{v}) \notag \\
  & \quad \qquad - \gamma^2 (\mathbf{v} \cdot \mathbf{B})^2 \mathbf{v} \mathbf{v} + (\pgas + \pmag) \mathbf{I}.
\end{align}
Given the forms of equations~\eqref{eq:sr_cons} are the same as those for Newtonian MHD, the same discretization scheme applies, with cell-centered volume averages and face-centered fluxes of hydrodynamical quantities and with face-centered area averages and edge-centered fluxes of magnetic fields.

With GR, all of our equations acquire a dependence on the metric $\mathbf{g}$. The primitive variables in the GRMHD module of \app{} are fluid-frame density $\rho$, fluid-frame gas pressure $\pgas$, normal-frame spatial velocity components $u^{i'}$, and coordinate-frame magnetic field $B^i$. The primitive velocities are related to the coordinate-frame velocity components via $u^0 = \gamma / \alpha$ and $u^i = u^{i'} - \beta^i \gamma / \alpha$, where $\alpha = (-g^{00})^{-1/2}$ is the lapse, $\beta^i = \alpha^2 g^{0i}$ is the shift, and
\begin{equation}
  \gamma = \paren[\big]{1 + g_{ij} u^{i'} u^{j'}}^{1/2}
\end{equation}
is the Lorentz factor in the normal frame (the frame with time direction orthogonal to surfaces of constant time). The contravariant magnetic field $b^\mu = u_\nu (*F)^{\nu\mu}$ has components
\begin{subequations} \begin{align}
  b^0 & = u_i B^i, \\
  b^i & = \frac{1}{u^0} (B^i + b^0 u^i),
\end{align} \end{subequations}
and with this a magnetic pressure
\begin{equation}
  \pmag = \frac{1}{2} b_\mu b^\mu
\end{equation}
and total enthalpy
\begin{equation}
  w = \rho + \frac{\Gamma}{\Gamma-1} \pgas + 2 \pmag
\end{equation}
can be defined.

The equations of GRMHD are simply
\begin{subequations} \begin{align}
  \nabla_\mu (\rho u^\mu) & = 0, \\
  \nabla_\mu \tensor{T}{^\mu_\nu} & = 0, \\
  \nabla_\mu (*F)^{\nu\mu} & = 0,
\end{align} \end{subequations}
where the stress-energy tensor has components
\begin{equation}
  \tensor{T}{^\mu_\nu} = w u^\mu u_\nu - b^\mu b_\nu + (\pgas + \pmag) \delta^\mu_\nu
\end{equation}
and the electromagnetic field tensor can be written $(*F)^{\mu\nu} = b^\mu u^\nu - b^\nu u^\mu$. Put into a more useful form, the equations solved by \app{} are
\begin{subequations} \begin{align}
  \partial_t(\sqrt{-g} \rho u^0) + \partial_{j}(\sqrt{-g} \rho u^j) & = 0, \label{eq:gr_cons_mass} \\
  \partial_t(\sqrt{-g} \tensor{T}{^0_\mu}) + \partial_{j}(\sqrt{-g} \tensor{T}{^j_\mu}) & = \frac{1}{2} \sqrt{-g} (\partial_\mu g_{\alpha\beta}) \tensor{T}{^{\alpha\beta}}, \label{eq:gr_cons_momentum} \\
  \partial_t(\sqrt{-g} B^i) + \partial_{j}(\sqrt{-g} (*F)^{ij}) & = 0, \label{eq:gr_induction}
\end{align} \end{subequations}
where $g = \det\mathbf{g}$. The conserved variables are $\rho u^0$, $\tensor{T}{^0_\mu}$, and $B^i$. Again, the equations have the same form and can be discretized as before, as long as the volumes, areas, and lengths used account for the appropriate factors of $\sqrt{-g}$.

Note the source term on the right-hand side of \eqref{eq:gr_cons_momentum} also appears in \eqref{eq:cons_momentum} when expressing the divergence operator in terms of partial derivatives in non-Cartesian coordinate systems. By choosing the free index in \eqref{eq:gr_cons_momentum} to be lowered, the source term vanishes for ignorable coordinates, as noted in \citet{Gammie2003}. In practice, this often means the global energy and $z$-angular momentum are easily conserved to machine precision.

\subsection{Numerical Algorithms}

\subsubsection{Reconstruction in Relativistic MHD}

In SR and GR, reconstruction is only allowed on the primitive variables. This avoids the numerical expense and potential variable inversion failures associated with characteristic reconstruction. 
Note also that the choice of primitive velocities ensures there is a unique, physically admissible (i.e.\ subluminal) state of the fluid for any finite real numbers $u^i$ (SR) or $u^{i'}$ (GR). This would not be true in general were $v^i$ (SR) or $u^i$ (GR) to be used. \app{} as described in \citet{WhiteStone2016} originally used 3-velocities for SR, but we have found the change to spatial 4-velocity components makes the code more robust.

\subsubsection{Relativistic Riemann Solvers}

\app{} includes relativistic versions of the HLLE Riemann solver for both pure hydrodynamics and MHD. It also includes the relativistic HLLC solver for hydrodynamics \citep{Mignone2005} and HLLD solver for MHD \citep{Mignone2009}. The latter two solvers are designed for SR only, but can be used in GR via the local frame transformations described in \citet{WhiteStone2016}.

\subsubsection{Variable Inversion}

The highly nonlinear, tightly coupled nature of the primitive-conserved variable relations in relativity make finding primitives both expensive (requiring iterative solvers that are difficult to vectorize) and prone to failure, such as when a conserved state has no corresponding subluminal primitives. 
\Citet{Noble2006} catalog six root-finding procedures used for variable inversion---four one-dimensional, one two-dimensional, and one five-dimensional, with the lower-dimensional versions solving for an enthalpy-like variable and/or a velocity-like variable, and with some of them only working for select equations of state. In practice, different methods find use in modern relativistic codes. For example, \texttt{GENESIS} \citep{Aloy1999} and \texttt{Enzo} \citep{Bryan2014,Wang2008} perform a 1D iteration on pressure, the initial implementation of \texttt{HARM} uses the 5D method \citep{Gammie2003}, \texttt{ECHO} \citep{DelZanna2007} uses a velocity-based 1D method, \texttt{RAMSES} \citep{Teyssier2002,Lamberts2013} uses a modified enthalpy-based method from \citet{Mignone2007b}, \texttt{PLUTO} \citep{Mignone2007a,Mignone2014} uses the enthalpy-based 1D method of \citet{Mignone2006}, and \texttt{BHAC} uses both an enthalpy-based 1D method and a 2D method \citep{Porth2017}.

Early versions of \app{} employed an enthalpy-based 1D method \citep{WhiteStone2016}. However, we have found inversion to be more robust by adapting the algorithm presented in \citet{Newman2014}, which involves a one-dimensional root-find operation and guarantees that a solution will be found if it exists.

Additionally, robustness relies on the ability to impose appropriate floors and ceilings depending on the problem being solved. The relativistic modules not only put floors on $\rho$ and $\pgas$ (in a position-dependent way, if desired), but also employ ceilings on $\gamma$, $\beta^{-1} = \pmag / \pgas$, and $\sigma = 2 \pmag / \rho$. In the latter two cases, the field components $B^i$ are never altered by ceilings, but rather these constraints are interpreted as additional field- and velocity-dependent floors on $\rho$ and $\pgas$.  We are also exploring the use of a first-order flux correction step, as implemented by \citet[][see the appendix]{lemaster2009}.

\subsection{Tests of the Relativistic MHD Module}

In the following subsections, we present several tests of the relativistic MHD module in \app{}, focusing especially on the use of mesh refinement with both SR and GR.

\subsubsection{Relativistic Shock Tube}

Relativistic Riemann problems can be challenging because very thin features can be formed \citep[][section 6.1]{RAMcode} which are hard to resolve with a uniform mesh.  To demonstrate the use of AMR with relativistic MHD, we present results for a strong shock tube problem using the initial conditions from \citet[][section 6.1]{Mignone2012}:
\begin{equation}
    (\rho, \pgas, B^y, B^z) =
    \begin{cases}
        (1, 1000, 7, 7), & x < 0.5, \\
        (1, 0.1, 0.7, 0.7), & x > 0.5,
    \end{cases}
\end{equation}
with \(B^x = 10\) and \(v^i = 0\) everywhere and with \(\Gamma = 5/3\). The root grid consists of \(400\) cells divided into MeshBlocks of \(16\) cells each. The VL2 integrator with PLM reconstruction and the HLLD Riemann solver are used. The CFL number is set to \(0.6\). The refinement criterion is the maximum value of the curvature on the MeshBlock,
\begin{equation}
    g = \max\paren[\bigg]{\frac{\abs{q_{i-1} - 2 q_i + q_{i+1}}}{q_i}},
\end{equation}
where
\begin{equation}
    q = \frac{(B^y)^2 + (B^z)^2}{\gamma \rho}.
\end{equation}
The refinement and derefinement thresholds are set to \(10^{-3}\) and \(10^{-4}\), and up to \(6\) levels of refinement beyond the root grid are allowed.

\begin{figure}[htb!]
    \centering
    \includegraphics[width=\columnwidth]{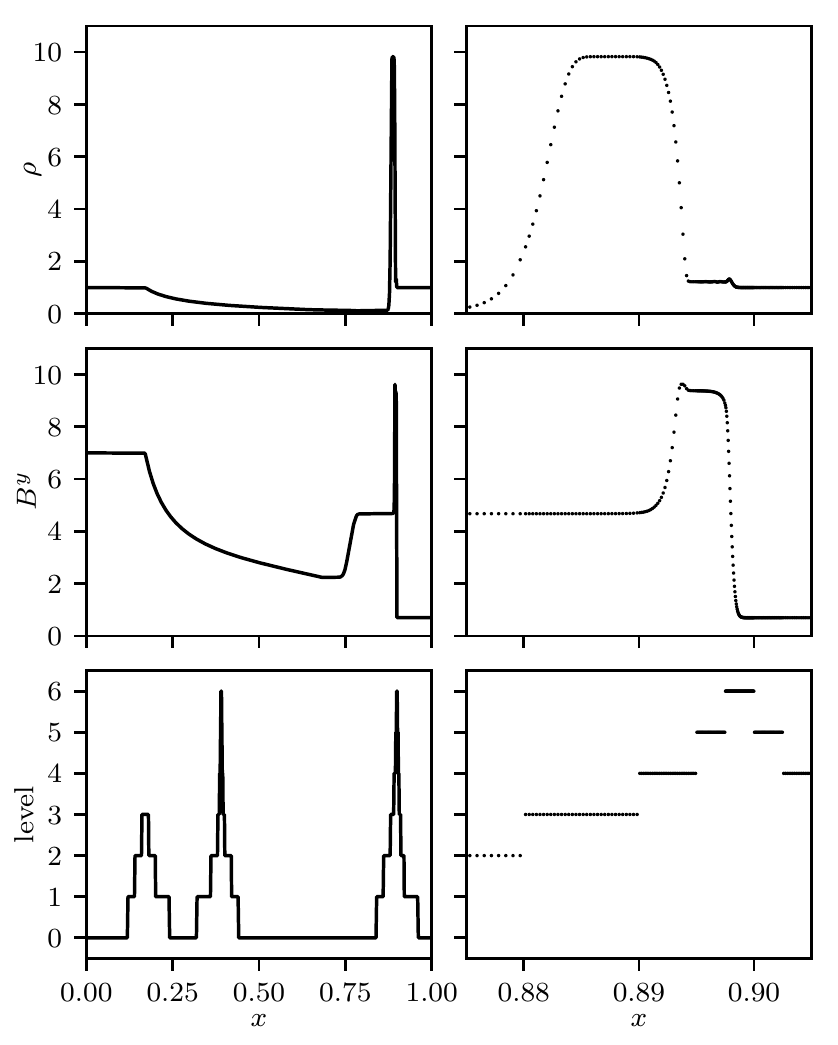}
    \caption{Density, single component of transverse magnetic field, and refinement level for the relativistic shock tube. Each cell is represented by a single point on the right. The thin shell consisting of multiple shocks is captured by high levels of refinement, as are the steepest parts of the rarefaction zones. \label{fig:sr_shock}}
\end{figure}

Figure~\ref{fig:sr_shock} shows the results for this test at time \(t = 0.4\). AMR naturally refines the very thin shell propagating to the right, as well as the steep parts of the rarefaction fans.  The results compare favourably to those presented in \citet[][figure 23]{Mignone2012}. When run on \(4\) cores of a Skylake (Xeon 8160) node, this simulation takes \(22.8\) core-seconds. The same simulation done at a uniform resolution of \(25{,}600\) cells takes \(520\) core-seconds, so the use of AMR results in a speedup of a factor of \(23\) for this test. While this speedup from mesh refinement is slightly lower than that reported for the \texttt{PLUTO} code \citep[][table 3]{Mignone2012}, this is in part because the run time on a uniform mesh is significantly lower using \app{}.

\subsubsection{Relativistic Kelvin--Helmholtz Instability}

The ability to simulate relativistic MHD problems with AMR is illustrated in two dimensions with a magnetized KH problem. The same primitive state as in \citet{Mignone2012} is initialized:\ the domain has constant values \(\rho = 1\), \(\pgas = 20\), \(v^z = 0\), \(B^x = \sqrt{2/5}\), \(B^y = 0\), and \(B^z = 10 \sqrt{2/5}\), and the in-plane velocity has the perturbed shear profile
\begin{subequations} \begin{align}
    v^x & = \frac{1}{4} \tanh(100 y), \\
    v^y & = \frac{1}{400} \sin(2\pi x) \exp(-100 y^2).
\end{align} \end{subequations}
Here \(\Gamma = 4/3\) is used in the equation of state. Note, \citeauthor{Mignone2012} use the Taub--Mathews equation of state instead, though the differences are small:\ the initial enthalpy is \(w = 81\) in our case, and \(w \approx 80.02\) in theirs.

The domain spans \(0 \leq x \leq 1\) and \(-0.25 \leq y \leq 0.25\), with periodic boundary conditions in \(x\) and outflowing conditions in \(y\). The root grid consists of \(64 \times 32\) cells in MeshBlocks of size \(16^2\). Up to \(5\) levels of refinement beyond the root grid are allowed. Refinement is based on the curvature of the conserved energy in each dimension:
\begin{subequations} \label{eq:sr_kh_curvature} \begin{align}
    g & = \max(g_x + g_y), \\
    g_x & = \frac{\abs{E_{i-i,j} - 2 E_{i,j} + E_{i+1,j}}}{E_{i,j}}, \\
    g_y & = \frac{\abs{E_{i,j-1} - 2 E_{i,j} + E_{i,j+1}}}{E_{i,j}}.
\end{align} \end{subequations}
The refinement and derefinement thresholds are set to be \(10^{-2}\) and \(10^{-3}\), respectively.

\begin{figure*}[htb!]
    \centering
    \includegraphics{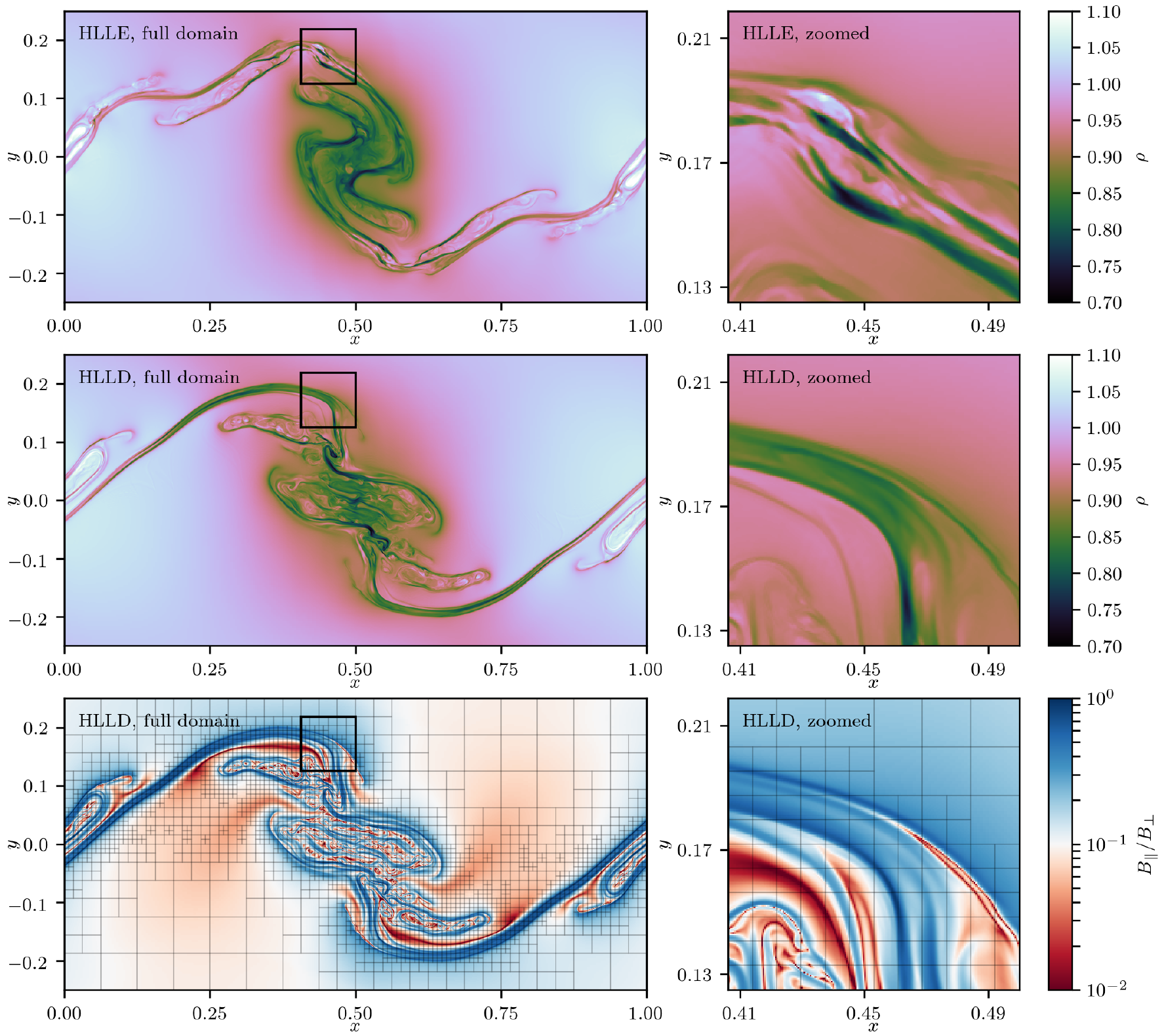}
    \caption{Density and ratio of \(B_\parallel = ((B^x)^2 + (B^y)^2)^{1/2}\) to \(B_\perp = B^z\) at \(t = 5\) in the relativistic magnetized KH test. The grid lines denote MeshBlocks consisting of \(16^2\) cells. The top row shows results with the HLLE Riemann solver, while the lower rows use HLLD. The left panels show the full domain with refinement levels \(2\) through \(5\) present, while the right panels zoom in to the region with a black border, with refinement levels \(4\) and \(5\). \label{fig:sr_mhd_kh}}
\end{figure*}

The VL2 integrator, PPM reconstruction, and, separately, the HLLE and HLLD Riemann solvers are used. The simulation is run to a time of \(t = 5\), using a CFL number of \(0.4\). The density and ratio of in-plane to perpendicular magnetic field strength at the end of the simulation are shown in figure~\ref{fig:sr_mhd_kh}, where the refined regions can be seen to track the locations of small-scale structures.

\begin{figure}[htb!]
    \centering
    \includegraphics{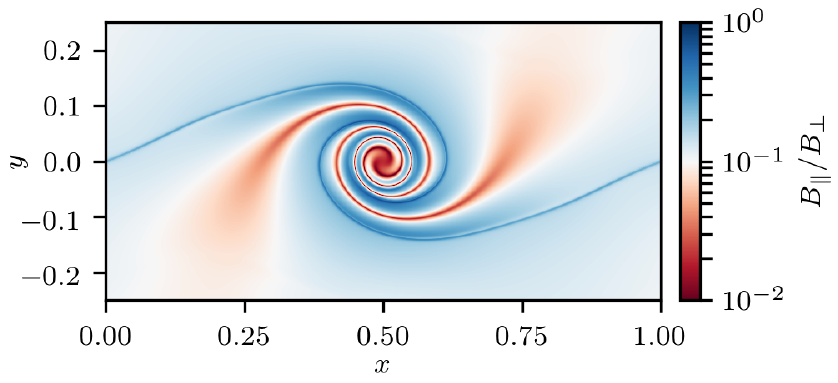}
    \caption{Ratio of \(B_\parallel = ((B^x)^2 + (B^y)^2)^{1/2}\) to \(B_\perp = B^z\) at \(t = 5\) in the relativistic magnetized KH test on a uniform grid.  Compare to the bottom panel in figure~\ref{fig:sr_mhd_kh}.
    \label{fig:sr_mhd_kh_high}}
\end{figure}

There are some differences between figure~\ref{fig:sr_mhd_kh} and figure~31 of \citet{Mignone2012}. While some of these may be attributable to the different equations of state, we also note that the result at the end of the simulation depends strongly on details of the numerical algorithms employed. For example, when the same test is performed with two different Riemann solvers but all else equal, the locations and shapes of even the largest KH rolls shift (e.g., compare the top and middle panels in figure~\ref{fig:sr_mhd_kh}). In fact, many of the short-wavelength features in the solution are introduced by changes in resolution at fine/course boundaries.  For example, figure~\ref{fig:sr_mhd_kh_high} shows the ratio of the parallel and perpendicular components of the magnetic field in the same problem run on a uniform grid with a resolution of
$4096\times2048$ (twice the effective resolution at the highest refinement level in the AMR calculation).  In this case, the vortex produced by the instability is smooth.  Therefore we conclude that most of the complex features visible in figure~\ref{fig:sr_mhd_kh} are due to the AMR boundaries, and this in part contributes to the difference between these solutions and those shown in \citet{Mignone2012}.

In the AMR runs, the refinement criterion partially refines the root grid before time evolution begins, and so the run begins with \(176\) MeshBlocks. Over the course of the HLLD simulation \(1174\) MeshBlocks are refined and \(177\) coarser blocks are created from finer ones. The AMR simulation takes \(32.2\) core-hours to run on \(2\) KNL nodes (Xeon Phi 7250, \(68\) cores each), while the same problem run on a uniform \(2048 \times 1024\) grid takes \(251\) core-hours. Thus using AMR gives a speedup of \(7.8\).

\subsubsection{Relativistic Magnetized Blast Wave}

A further test of the relativistic MHD module in the code is provided by the evolution of a magnetized blast wave. Variations on this test are commonly used to test the propagation of strong shocks in relativistic MHD codes, for example in \citet{Komissarov1999,Leismann2005,DelZanna2007,Beckwith2011,Mignone2012}.

We first run a strongly magnetized blast in two dimensions on a Cartesian grid, with the magnetic field not aligned with the grid. On a domain \([-6, 6]^2\), we have initial values \(v^i = 0\), \(B^x = 1/\sqrt{20}\), \(B^y = 1/\sqrt{20}\), and \(B^z = 0\). The density \(\rho\) is \(10^{-2}\) within a distance \(r = 0.8\) of the origin, \(10^{-4}\) outside \(r = 1\), and it varies linearly with radius between these circles. \(\pgas\) varies from \(1\) to \(5 \times 10^{-3}\). \(\Gamma = 4/3\) for this test.

The simulation is evolved to a time of \(t = 4\) using a CFL number of \(0.25\), PLM reconstruction, and the HLLD Riemann solver. We use both a uniform grid with \(1536^2\) cells and an AMR grid with \(48^2\) cells at root level. In both cases MeshBlocks with \(16^2\) cells are used. The AMR grid can have up to \(5\) additional levels of refinement. Refinement is triggered with the same curvature condition in equation~\eqref{eq:sr_kh_curvature} as in the previous test, except using conserved density \(D\) instead of energy. The thresholds are set to be \(0.025\) and \(0.005\).

\begin{figure}[htb!]
    \centering
    \includegraphics{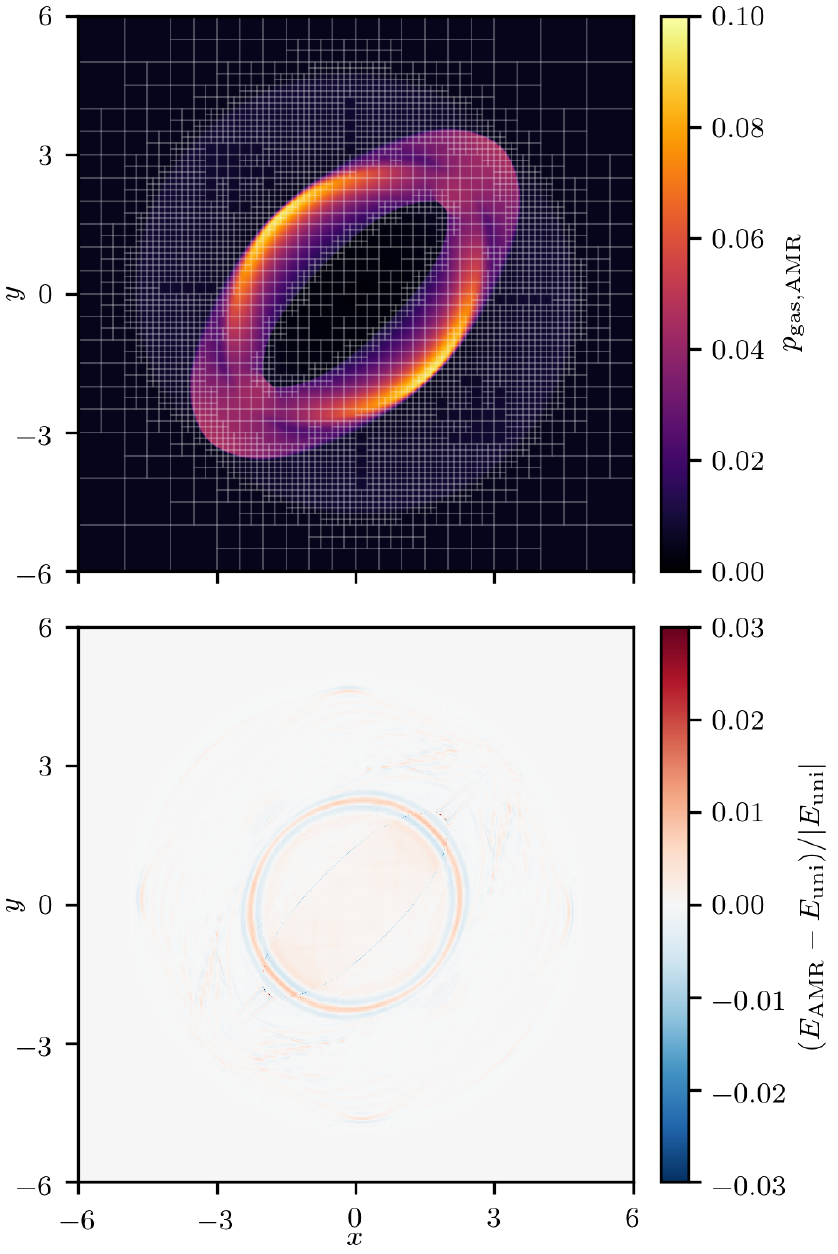}
    \caption{Top:\ gas pressure in the 2D strongly magnetized relativistic blast test using AMR. The grid lines denote MeshBlocks consisting of \(16^2\) cells, and refinement levels \(2\) through \(5\) are present. Bottom:\ relative difference in conserved energy between the AMR grid and a uniform grid. \label{fig:sr_mhd_blast_2d}}
\end{figure}

The upper panel of figure~\ref{fig:sr_mhd_blast_2d} shows the gas pressure at the end of the AMR simulation. Refinement tracks the shock fronts that are directed by the magnetic field. The lower panel shows the relative difference in conserved energy between the simulations. In most of the volume, the agreement is better than \(1\%\).

The uniform simulation takes \(24.7\) core-hours on \(2\) KNL nodes (Xeon Phi 7250, 68 cores each), while the AMR simulation takes \(6.18\) core-hours. Thus AMR gives us a factor of \(4.0\) speedup.

We next run a similar but spherical test in three dimensions, using the same physical parameters as in \citet{Mignone2012}. The domain is \([-6, 6]^3\), with initial values \(v^i = 0\), \(B^x = 1/\sqrt{200}\), \(B^y = 1/\sqrt{200}\), and \(B^z = 0\). Density and pressure are the same as in the 2D case, and again we have \(\Gamma = 4/3\).

The simulation is evolved to a time of \(t = 4\) using a CFL number of \(0.25\), PLM reconstruction, and the HLLD Riemann solver, using both a uniform grid with \(768^3\) cells and an AMR grid with \(48^3\) cells at root level. In both cases MeshBlocks with \(16^3\) cells are used. The AMR grid can have up to \(4\) additional levels of refinement. The curvature condition for refinement is extended naturally to 3D, with thresholds set to be \(0.15\) and \(0.03\).

\begin{figure}[htb!]
    \centering
    \includegraphics{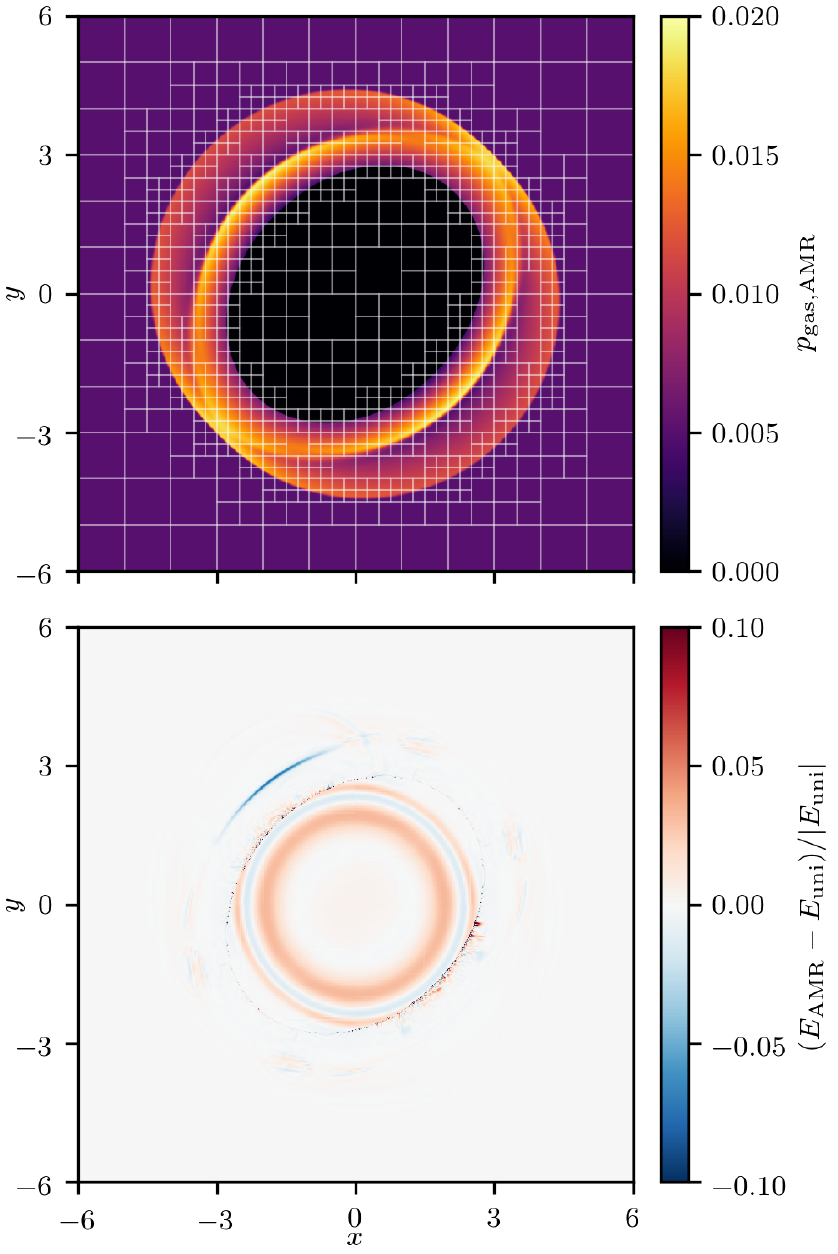}
    \caption{Top:\ midplane \(z = 0\) slice of gas pressure in the 3D relativistic magnetized blast test using AMR. The grid lines denote MeshBlocks consisting of \(16^3\) cells. Refinement levels \(2\) through \(4\) are present. Bottom:\ relative difference in conserved energy between the simulations. \label{fig:sr_mhd_blast_3d}}
\end{figure}

Figure~\ref{fig:sr_mhd_blast_3d} shows the gas pressure in the \(z = 0\) slice at the end of the simulation. As in the 2D case, we see refinement tracking the shock fronts. Again, the agreement is better than \(1\%\) over most of the volume, with most of the relative error in the interior, which has been evacuated to near the density and pressure floors.

The uniform simulation takes \(2720\) core-hours on \(16\) KNL nodes, while the AMR simulation takes \(316\) core-hours. Thus AMR gives us a factor of \(8.6\) speedup.

\subsubsection{Black Hole Accretion}

As a demonstration of the general relativistic capabilities of \app, we show the evolution of a weakly magnetized, hydrostatic equilibrium torus around a spinning black hole. The initial conditions are those of \citet{Fishbone1976} with dimensionless spin $a = 0.9$, inner edge at $r = 15 \rg$, and pressure maximum at $r = 25 \rg$, where $\rg = GM/c^2$ is the characteristic length scale of a black hole of mass $M$. The magnetorotational instability \citep{Balbus1991} is seeded with a single magnetic field loop in the poloidal plane, normalized such that the mass-weighted average of $\beta^{-1}$ is $0.01$.

We evolve the torus in horizon-penetrating, spherical Kerr--Schild coordinates. Our root grid has $56 \times 32 \times 44$ cells in $r$, $\theta$, and $\phi$. Cells are spaced logarithmically in radius, from $r = 1.329 \rg$ to $r = 100 \rg$, and they are uniform in both angles. Static mesh refinement adds three successive refinement levels away from the polar axis, achieving an effective resolution of $448 \times 256 \times 352$ everywhere within $50.625^\circ$ of the midplane while still keeping cells from being unnecessarily small near the axis.

\begin{figure}[htb!]
    \centering
    \includegraphics{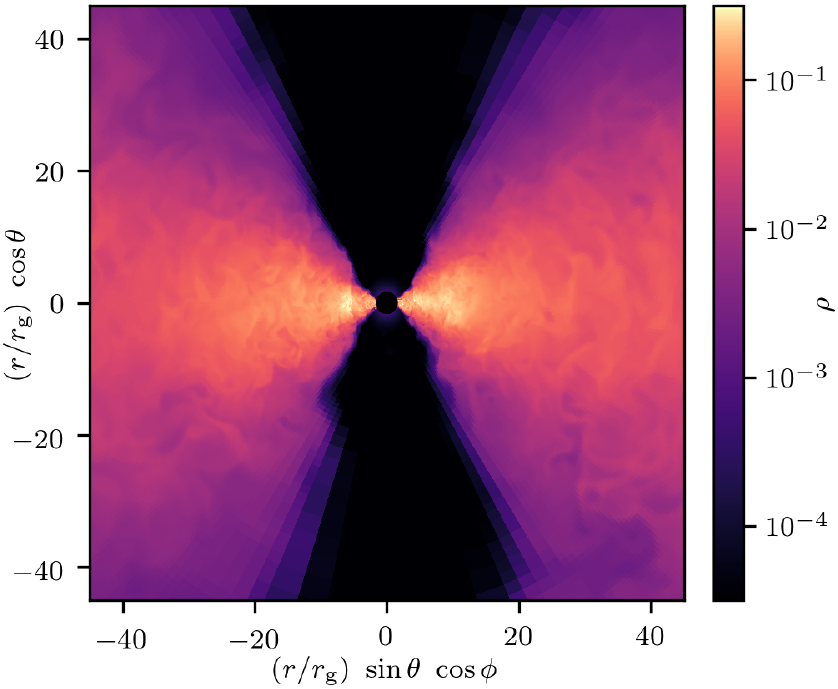}
    \caption{Poloidal slice of density in the GR torus demonstration. Turbulence is fully developed after $10{,}000$ gravitational times $\rg/c$. \label{fig:gr_torus}}
\end{figure}

The density after a time of $10{,}000 \rg/c$ is shown in figure~\ref{fig:gr_torus}. By this point the turbulence has saturated, and inflow equilibrium has been achieved in the inner parts of the thick disk that has formed. The evolution of accretion flows such as this, at similar resolutions, is ubiquitous in the black hole modeling community, and it is used as a test of codes' GRMHD capabilities \citep[see][including a comparison of \app{} with other codes]{Porth2019}.

\section{Additional Physics}

%Our flexible framework design makes it possible to easily add new physics.  New variables can be added as $n$-dimensional arrays to the {\tt MeshBlock} class, and any steps associated with the new algorithm can be enrolled through the TaskList.  The framework includes methods for applying simple boundary conditions to any variable (such as periodic), as well as methods for MPI communication of ghost cells between processors and prolongation and restriction algorithms for AMR.  Only in the case of special conditions unique to the new physics are extensions needed to any of these components.

In this paper, we have described in detail modules for non-relativistic and relativistic MHD that have already been implemented in the AMR framework.  These modules include additional physics for MHD, including non-ideal MHD, a general EOS, the shearing box approximation, and orbital advection.  Below we describe some of the additional physics that will be available in new modules in the future.

{\em Self-gravity.}  Two different methods are implemented for self-gravity.  The first solves the Poisson equation using fast Fourier transforms (FFTs), following the method implemented in \athena.  This module is included in the public version.  There is also a new implementation of self-gravity based on the solution of the Poisson equation using the full multigrid (FMG) method, which is more efficient and scalable than FFTs (FMG is $O(N)$ while FFTs are $O(N\log N)$). The new FMG solver is being extended to work with AMR (Tomida et al.\ in prep.).  In addition, the FMG solver is designed to be flexible so that it can be used for a variety of applications, for example solving implicit discretizations of the radiation transfer moment equations as in \citet{Jiang2012}.

{\em Radiation transfer.} A variety of modules for incorporating radiation transfer into MHD calculations are being developed.  The time-dependent radiation transport algorithm described in \citet{Jiang2014} has already been implemented and has been used to study a variety of problems in radiation-dominated accretion disks \citep{Jiang2014b} and massive stars \citep{Jiang2018}.  This module is being extended to full GR.  Moment-based methods such as flux-limited diffusion and the variable Eddington tensor (VET) method \citep{Davis2012,Jiang2012} will also be implemented.  For post-processing calculations that compute synthetic images and spectra, a Monte Carlo-based radiation transfer solver is also under development (Davis et al.\ in prep.).  A method for following radiation from point sources using adaptive ray tracing has been implemented in \athena{} \citep{JGKim}, and will be re-implemented in \app{}.

{\em Reaction networks.} A module to solve chemical reaction networks is implemented using a publicly available sparse matrix solver (Gong et al.\ in prep.). Nuclear reaction networks are also being implemented using the same algorithmic infrastructure (Halevi et al.\ in prep.).  When coupled with the passive scalar capabilities described in section~\ref{subsubsec:additional-phys}, \app{} becomes capable of solving chemo-hydrodynamics.  When these solvers are coupled with the radiation transfer module and the general EOS capabilities, they enable new studies of a diverse set of problems from the dynamics of the multi-phase interstellar medium (ISM) to the merger of compact objects.

{\em Dust Particle Dynamics:} To simulate particle motions coupled with hydrodynamics, a general particle module is under development (Yang et al.\ in prep.), based on the methods implemented in \athena{} and described in \citet{Bai-particles}.  The module will enable calculations of the dynamics of dust particles in planet formation, the kinematics of tracer particles to diagnose flow, and the use of sink particles to represent stars and compact objects.

%for modeling turbulence and feedback in the ISM, and even particle kinetics in weakly collisional plasmas similar to the use of the particle module in \athena{} \citep{pegasus,athena-particle-cr}. 

\section{Summary and Conclusion}

In this paper, we have described a new framework for AMR as implemented in the \app{} code.  This framework adopts a block-based AMR design, with blocks organized into a tree data structure, for improved performance, scalability, and ease of implementation.  It can be used with any logically rectangular coordinates, and with nonuniform mesh spacing.  We also describe a dynamic execution model based on a simple design we call a task list.  This model is capable of overlapping communication with computation on distributed memory parallel systems, which helps improve the parallel efficiency and scalability of the algorithms on very large numbers of processors.  Moreover, different combinations of physics can be included in calculations by simply adding new steps to the task list.  Finally, since different regions of the calculation can have different task lists, it is even straightforward to implement multi-physics calculations that include different physics in different locations (such as kinetic MHD in dense regions of a plasma that are weakly collisional, and particle-in-cell methods in diffuse regions that are collisionless).  The task list could also be used to solve different physics on different physical cores in a parallel calculation; for example the Poisson equation for self-gravity could be solved on different cores from those dedicated to hydrodynamics or MHD.

We have also described two physics modules that have been implemented in this framework, for non-relativistic and relativistic MHD respectively.  These modules are based on the numerical algorithms for MHD developed in the \athena{} code \citep{Stone+2008} using a finite volume discretization combined with the constrained transport algorithm to enforce the divergence-free constraint on the magnetic field.  They have been updated with new algorithmic extensions, such as higher-order reconstruction in curvilinear and/or nonuniform meshes, new higher-order time integrators based on a method of lines approach, and diffusive terms that can be updated using new Runge--Kutta--Legendre super-time-stepping methods.  Most importantly, these modules for MHD work effectively with AMR.  A variety of test problems were presented to show the accuracy and fidelity of the MHD algorithms with AMR; see \citet{Stone+2008} and \citet{WhiteStone2016} for a more comprehensive list of tests we have used to validate the algorithms.

A significant aspect of this new framework is excellent performance and parallel scaling.  The MHD solvers have been highly optimized to exploit vector instructions on modern processors.  
Based on tests run with the public versions of several MHD codes built using the same compilers and optimizations, the performance of the \app{} MHD module is amongst the highest of any publicly available astrophysical MHD code of which we are aware, with only the \texttt{DISPATCH} code being similar \citet{DISPATCH}.  Using all cores on a single Intel Skylake CPU, the performance is only about $5\times$ slower than the same algorithm implemented on the latest NVIDIA Volta GPU \citep{Grete19} (as expected given the ratio of peak performance for these two devices).
On up to 500,000 threads, the MHD module shows excellent weak scaling, with 84\% parallel efficiency.  Thus, the finite volume algorithms implemented in \app{} are clearly capable of exploiting the new hardware emerging in the exascale era.

A variety of new physics modules are under development, including self-gravity, radiation transfer, chemical and nuclear reaction networks, and particles coupled to the fluid.  In addition, improvements to the algorithms in existing modules is planned.  For example, a fully fourth-order accurate algorithm for MHD has been implemented in the \app{} framework \citep{FelkerStone2018}, and is currently being tested and compared with existing algorithms in the code on astrophysical applications to determine the relative advantages and disadvantages of each.  Finally, a performance-portable version of the entire \app{} AMR framework is being built using the \texttt{Kokkos} library (Dolence et al, private communication), and will be released as open source in the near future.

The \app{} is publicly available through a GitHub repository, and is distributed under the BSD open source license.  Once new modules are thoroughly tested and deemed reliable, they will also be made publicly available.  While the code has been developed primarily to enable scientific applications by the core members of the development team, it is hoped that others will find the code useful.

\acknowledgments

We thank the many contributors to the \app{} code project, especially Matt Coleman, Shane Davis, Munan Gong, Goni Halevi, Yan-Fei Jiang, Chang-Goo Kim, Alwin Mao, Patrick Mullen, Tomohiro Ono, Ji-Ming Shi, Bei Wang, Chao-Chin Yang, and Zhaohuan Zhu. We also thank the referee for comments which improved the manuscript.

JMS was supported by National Science Foundation, grant number AST-1715277. KT was supported by Japan Society for the Promotion of Science (JSPS) KAKENHI Grant Numbers 16H05998, 16K13786, 17KK0091, 18H05440. KT also acknowledges support by Ministry of Education, Culture, Sports, Science and Technology (MEXT) of Japan as ``Exploratory Challenge on Post-K Computer'' (Elucidation of the Birth of Exoplanets [Second Earth] and the Environmental Variations of Planets in the Solar System). CJW was supported in part by the National Science Foundation, grant number PHY-1748958. KGF was supported by the Department of Energy Computational Science Graduate Fellowship (CSGF), grant number DE-FG02-97ER25308.  The initial design and development of \app{} was undertaken while JMS was a Simons Distinguished Visiting Scholar at the KITP, University of California Santa Barbara, in 2014; the support of the KITP is gratefully acknowledged.

The simulations presented in this article were performed partly on computational resources managed and supported by Princeton Research Computing, a consortium of groups including the Princeton Institute for Computational Science and Engineering (PICSciE) and the Office of Information Technology at Princeton University. We thank David Luet for his support of the Jenkins server at PICSciE. This work also used the Oakforest-PACS supercomputer at the Joint Center for Advanced High Performance Computing through the HPCI System Research Project (Project IDs:\ hp180128, hp190088); the Cray XC50 supercomputer at Center for Computational Astrophysics, National Astronomical Observatory of Japan; and the Extreme Science and Engineering Discovery Environment (XSEDE) Stampede2 at the Texas Advanced Computing Center (allocation:\ AST170012).

\newpage
\bibliography{main}

\end{document}